\documentclass[10pt]{iopart}
\usepackage{iopams}  
\usepackage{setstack} 
\usepackage{xcolor}
\usepackage{graphicx}
\usepackage{soul}
\usepackage{placeins}
\usepackage{cite}

\usepackage{enumerate,subcaption} 

\usepackage{svg}

\newcommand{\fulltableh}[1]{\begin{table}[h]
  \caption{#1}
  \footnotesize
  \lineup
  \begin{tabular*}{\textwidth}{@{}l*{15}{@{\extracolsep{0pt plus 12pt}}l}}}
\def\endfulltableh{\end{tabular*}\end{table}\normalsize}

\begin{document}

\title{The role of mobility in epidemics near criticality}

\author{Beatrice Nettuno$^*$, Davide Toffenetti$^*$, Christoph Metzl$^*$, Linus Weigand$^*$, Florian Raßhofer, Richard Swiderski, and Erwin Frey}
\address{Arnold Sommerfeld Center for Theoretical Physics and Center for NanoScience, Department of Physics, Ludwig-Maximilians-Universit\"at M\"unchen, Theresienstra\ss e 37, D-80333 Munich, Germany}

\address{$^*$ These authors contributed equally. B Nettuno and D Toffenetti performed the analytical work, C Metzl and L Weigand did the numerical one.  }
\ead{frey@lmu.de}
\vspace{10pt}
\begin{indented}
\item[]\today
\end{indented}

\begin{abstract}
 
The \textit{general epidemic process} (GEP), also known as \textit{susceptible-infected-recovered model} (SIR), describes how an epidemic spreads within a population of susceptible individuals who acquire permanent immunization upon recovery. 
This model exhibits a second-order absorbing state phase transition, commonly studied assuming immobile healthy individuals. 
We investigate the impact of mobility on disease spreading near the extinction threshold by introducing two generalizations of GEP, where the mobility of susceptible and recovered individuals is examined independently. 
In both cases, including mobility violates GEP's rapidity reversal symmetry and alters the number of absorbing states. 
The critical dynamics of the models are analyzed through a perturbative renormalization group approach and large-scale stochastic simulations using a Gillespie algorithm.
The renormalization group analysis predicts both models to belong to the same novel universality class describing the critical dynamics of epidemic spreading when the infected individuals interact with a diffusive species and gain immunization upon recovery. 
At the associated renormalization group fixed point, the immobile species decouples from the dynamics of the infected species, dominated by the coupling with the diffusive species. 
Numerical simulations in two dimensions affirm our renormalization group results by identifying the same set of critical exponents for both models. 
Violation of the rapidity reversal symmetry is confirmed by breaking the associated hyperscaling relation.
Our study underscores the significance of mobility in shaping population spreading dynamics near the extinction threshold.

\end{abstract}

\noindent

\newpage

\section{Introduction}

The recent COVID-19 pandemic has sparked an increased interest in the mathematical modeling of disease and virus spreading~\cite{Chinazzi.et.al:2020,Kissler.et.al:2020,Arenas.et.al:2020}.
Various quantitative models have been successfully applied to aid vaccination and containment strategies~\cite{Bauer.et.al:2021,Dehning.et.al:2020,Ferretti.et.al:2020,Gros.et.al:2021}.
This underscores the critical importance of understanding epidemic processes.
However, realistic modeling efforts remain challenging: Many factors such as human mobility~\cite{Muley.et.al:2020} and social behavior~\cite{Epstein.et.al:2008, Weitz.et.al:2020} as well as spatial heterogeneity~\cite{Lloyd.et.al:1996,Pastor-Satorras.et.al:2015} are known to play pivotal roles in shaping the course of epidemics.

Naturally, this calls for exploring simpler spreading models of epidemics that may be applied to an even broader class of phenomena, including opinion dynamics~\cite{Holley.Liggett:1975,Castellano.et.al:2009}, forest fires~\cite{Drossel.Schwabl:1992,Albano:1995} and even catalytic surface reactions~\cite{Ziff.et.al:1986,Jensen.et.al:1990}.
While inherently less intricate, these conceptual models are crucial in identifying common themes and providing qualitative insights~\cite{Marro.Dickman:1999,Odor:2004,Tauber:2014}.
Typically, one considers lattice-gas models, where active agents~($B$), such as a burning tree or an infected individual, activate nearby susceptible agents~($A$)~\cite{Tauber.et.al:2005,Henkel.et.al:2008}.
In its most basic form, this may be accounted for by a \textit{local} reaction 
\begin{eqnarray}
    A+B \overset{\sigma}{\rightarrow} B + B \, \nonumber
\end{eqnarray}
at constant rate $\sigma$. Over time, agents lose their activity, e.g.~by recovery of infected individuals, with deactivation being either permanent or transitory.
If the activation process proves stronger, activity (e.g.~disease or fire) may persist and spread throughout the system.
Conversely, if deactivation dominates and there is no spontaneous creation of active or susceptible agents, the system approaches an inactive \textit{absorbing} state where active individuals are absent. 
The transition between these two phases is coined an absorbing state phase transition~\cite{Henkel.et.al:2008, Hinrichsen:2000}, which is a class of non-equilibrium phase transitions.

Like their equilibrium counterparts, the near-critical behavior of absorbing state phase transitions can be categorized into a limited number of universality classes, the most important ones being \textit{directed percolation}~(DP)~\cite{Hinrichsen:2000, Janssen:1981, Janssen.Tauber:2005} and \textit{dynamical percolation}~(DyP), here referred to as \textit{general epidemic process}~(GEP)~\cite{Grassberger:1983, Janssen:1985}.
The \textit{contact process}~(CP)~\cite{Harris:1974} gives one particular realization of the DP universality class, which acts as a minimal model of disease spreading when there is no immunization.
It is formulated in terms of local reactions
\begin{eqnarray}
  A_{i}+B_{j} \overset{\sigma}{\rightarrow} B_{i} + B_{j} \, , \quad
    B_{i}\overset{\zeta}{\rightarrow} A_{i} \, , \nonumber
\end{eqnarray}
with at most one agent per site and~($i,j$) denoting adjacent lattice sites.
Due to the absence of immunization, it is often referred to as the \textit{susceptible-infected-susceptible}~(SIS) model.
Whenever immunization is permanent, one reverts to the \textit{susceptible-infected-recovered}~(SIR) model~\cite{Kermack.McKendrick:1927} which falls into the GEP universality class~\cite{Grassberger:1983, Janssen:1985}. 
Its associated reaction scheme reads
\begin{eqnarray}
     A_{i}+B_{j} \overset{\sigma}{\rightarrow} B_{i} + B_{j} \, , \quad
    B_{i}\overset{\zeta}{\rightarrow} C_{i} \, , \nonumber
\end{eqnarray}
with immune (recovered) individuals being represented by an additional species~$C$.
The crucial difference between these two reaction schemes is that for the CP/SIS dynamics, an epidemic can survive locally. In contrast, in the GEP/SIR model, activity may only persist as a traveling wave spreading through the population.
A generalized model that allows for reinfection at a varying rate contains both of these scenarios as limiting cases~\cite{Grassberger:1997}.

Hitherto, one crucial aspect has been omitted in these classical conceptual models: 
In general, activity may not only propagate due to the infection dynamics but also strongly depend on the mobility of susceptible and infected agents.
In its most elementary form, this can be accounted for by each agent undergoing \textit{diffusive} motion.
It is known that more long-ranged forms of mobility, such as L\'{e}vy-flights~\cite{Janssen.Stenull:2008} or chemotactic motion~\cite{Kolk.et.al:2023} can affect the behavior close to a non-equilibrium phase transition.
In this work, we exclusively focus on diffusive motion, which alters the local environment of active agents.

Previous studies show that the inclusion of diffusion alone can significantly change the observed critical behavior:
While the ordinary \textit{pair contact process}~(PCP)~\cite{Jensen:1993} falls into the DP universality class~\cite{Munoz.et.al.1998, Silva.Dickman:1999}, the \textit{pair contact process with diffusion}~(PCPD)~\cite{Grassberger:1982, Howard.Tauber:1997} shows deviating critical behavior.
However, the nature of its universality class remains a subject of ongoing discussions~\cite{Henkel.Hinrichsen:2004, Deng.et.al:2020}.
Moreover, numerical simulations indicated that SIR dynamics with symmetric exchange hopping culminates in a small albeit significant deviation from the known GEP universality class~\cite{Deng.Odor:2023}.

For mobile agents, it is further necessary to revisit the occupation number constraint:
If each site can be occupied by at most one agent, diffusion may only occur in the form of two-agent exchange hopping, as illustrated in reference~\cite{Deng.Odor:2023}; it is impossible to account for the individual mobility of different species.
Further, in a coarse-grained description, nearest neighbor (n.n.)\@~infections give rise to gradient terms akin to a classical diffusion law for the active species~\cite{Jensen.Dickman:1999}. 
Thus, a clear distinction between the two modes of activity spreading (individual mobility and nearest-neighbor infections) cannot be made on a mesoscopic level.
This rationalizes why the contact process (SIS model) with exchange hopping of infected and susceptible agents and a constrained agent number remains in the DP universality class~\cite{Jensen.Dickman:1999}.

In contrast, relaxing the exclusion constraint and, thus, allowing for distinct mobilities of susceptible and infected agents gives rise to the \textit{diffusive epidemic process}~(DEP)~\cite{Wijland.et.al:1998}.
Its emergent critical behavior is only partially understood and crucially depends on the relative mobilities of susceptible and recovered agents:
The limit of immobile susceptible agents was shown to fall into the universality class of quenched \textit{Edwards-Wilkinson} growth by an exact mapping~\cite{Doussal.Wiese:2015, Janssen.Stenull:2016}.
Field-theoretical studies have identified two new classes of critical behavior~\cite{Wijland.et.al:1998, Kree.et.al:1989, Janssen:2001} for infected agents diffusing faster or equally as fast as the susceptible ones.
When diffusion of susceptible agents is faster, neither perturbative~\cite{Wijland.et.al:1998} nor non-perturbative methods~\cite{Tarpin.et.al:2017} could identify an entirely conclusive fixed point structure.
DEP was even speculated to undergo a fluctuation-induced first-order transition~\cite{Oerding.et.al:2000}.
However, follow-up numerical studies have questioned this possibility, reporting a continuous transition for the latter regime~\cite{Maia.Dickman:2007,Dickman.Maia:2008,Argolo.et.al:2019}.
The ongoing debate about the nature of the phase transition and the associated critical exponents was addressed in a recent publication for one-dimensional systems~\cite{Polovnikov.et.al:2022}.
By unambiguously showing that the transition is continuous, this numerical study supports the hypothesis of a strong coupling fixed point and further reported sub-diffusive activity spreading.

The above examples illustrate how unraveling the role of mobility in epidemic spreading remains a field of ongoing research, with a conclusive answer yet to be found.
Therefore, in this study, we investigate how the mobility of susceptible and recovered agents affects critical behavior in a generalized version of the SIR/GEP model.
For that purpose, we employ a field-theoretical Renormalization Group (RG) analysis~\cite{Tauber.et.al:2005} as well as extensive numerical simulations. 

Our modifications aim to explore different possibilities of coupling infected individuals with a diffusive species.
In particular, an additional caregiving reaction is introduced.
This enables us to consider two limiting cases in which the role of mobile susceptible and recovered agents can be studied independently. 
Moreover, we exclusively consider systems without occupation number constraints, allowing for a clear distinction of different mobilities for different species and a consistent field-theoretical description.
Regardless of how the GEP model is generalized by introducing additional diffusivity of the different species, our results show a consistent critical behavior that deviates from the conventional GEP paradigm and thus establishes a new universality class.

Our theoretical prediction of a new universality class is further corroborated through exact stochastic simulations. 
Not considering any exclusion constraint, our numerical approach complements and shows remarkable agreement with earlier numerical results obtained in~\cite{Deng.Odor:2023}.
This enables us to state beyond doubt that mobility is a critical perturbation to the important GEP universality class.
Moreover, the presented analysis offers a clear understanding of how mobile agents alter the universality class and thereby may shed light on analogous cases such as DEP or PCPD.

This paper is organized as follows:
In section~\ref{sec:models}, we introduce our modifications of the SIR/GEP reaction scheme and discuss their phenomenological implications. 
Particular emphasis is put on limiting cases that we consider to address our questions. We further introduce the observables analyzed in this paper. 
Section~\ref{sec:mean_field} provides a mean-field analysis to gain intuition on our models' absorbing state phase transitions. 
We elucidate how diffusing agents influence deterministic dynamics, specifically examining the emergence of diffusive currents close to the propagating front of infected agents in the active phase.
Section~\ref{sec:field.theory} presents a perturbative renormalization group analysis, providing analytical evidence for a new universality class in the form of a new IR-stable fixed point. 
In addition, we discuss under which circumstances the mobility of either susceptible or recovered individuals violates an intrinsic symmetry of GEP and alters the absorbing state in the model, ultimately modifying the RG flow functions. 
Section~\ref{sec:numerics} concisely describes the algorithm used in the simulations and presents the key numerical results, including critical scaling exponents and their estimated error. 
We compare our results to the literature where possible and conclude that the numerical analysis underpins all our analytical predictions of the new universality class.
The last section summarizes our findings and discusses their implications with possible future research avenues.
Further, we provide detailed appendices that include a derivation of the field-theoretical action, a complete summary of the perturbative RG calculations, a description of our numerical approach and error estimates, and additional data.

\section{Reaction-diffusion models of epidemic processes} 
\label{sec:models}

The epidemic models considered in this paper are formulated as reaction-diffusion models on a bosonic $d$-dimensional lattice—wherein an arbitrary number of individuals can occupy each site. We consider a population consisting of three species: $A$ represents the individuals susceptible to the disease, $B$ the infected, and $C$ the recovered members immune to the disease. 
We refer to susceptible and recovered individuals collectively as healthy individuals.
The general class of epidemic models  we consider is given by the stochastic reaction scheme:\\
\begin{minipage}[b]{.48\textwidth}
\label{eq:chemical.rates}
\numparts
\begin{eqnarray}
    \fl \mbox{infection} \qquad &&A_{i}+B_{i}\overset{\sigma}{\rightarrow} 2 B_{i}, \label{eq:chemical.rates1} \\
    \fl \mbox{recovery}\qquad &&B_{i}\overset{\zeta}{\rightarrow} C_{i}, \label{eq:chemical.rates2} \\
    \fl \mbox{caregiving} \qquad &&C_{i}+B_{i}\overset{\gamma}{\rightarrow} 2 C_{i} \label{eq:chemical.rates3}
\end{eqnarray}
\endnumparts
\end{minipage} 
\hfill
\begin{minipage}[b]{.48\textwidth}
\numparts
\label{eq:diffusion.rates}
\begin{eqnarray}
    \fl  \mbox{diffusion of }A \qquad &&A_{i} \overset{D_A}{\rightarrow}A_{j},
    \label{eq:diffusion.rates1} \\
   \fl \mbox{diffusion of }B \qquad  &&B_{i} \overset{D_B}{\rightarrow}   B_{j},
    \label{eq:diffusion.rates2} \\
   \fl \mbox{diffusion of }C \qquad  &&C_{i}\overset{D_C}{\rightarrow}C_{j}.
    \label{eq:diffusion.rates3}
\end{eqnarray}
\endnumparts
\end{minipage}\\

\begin{figure}[t]
\centering
\includegraphics[width=13cm]{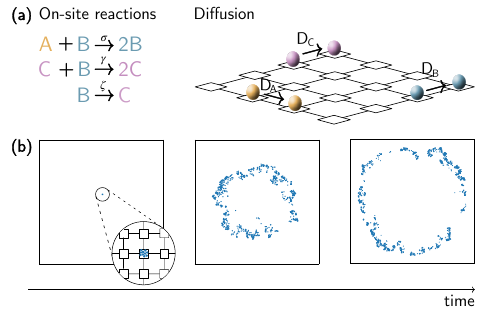}
\caption{\textbf{Models and phenomenology.}
(a) Reactions in the general model: On-site reactions~\eref{eq:chemical.rates1}--\eref{eq:chemical.rates3} (left) and diffusion~\eref{eq:diffusion.rates1}--\eref{eq:diffusion.rates3}(right) as a stochastic nearest-neighbor hopping process on a 2D lattice (right). 
The reaction scheme of the studied sub-models can be obtained in the following limits: GEP ($D_C=D_A=\gamma=0$), DGEP ($D_C=\gamma=0$), CGEP ($D_A=0$). (b) Exemplary stochastic simulation of an epidemic with recovery in the active phase, starting from a single seed (left-most panel) and showing an annular spreading of infected individuals depicted as blue dots (other species not shown).
The inset shows the 2D square lattice on which the simulations were performed. In this simulation, ten infected individuals $B$ are placed initially at the center of the system with a homogeneous density of susceptible individuals $A$ (not shown) and no recovered individuals~$C$. }
\label{fig:1}
\end{figure}

The lower indices $i,j$ indicate the lattice sites with $(i,j)$ being neighboring sites. 
The first reaction~\eref{eq:chemical.rates1} describes the infection of a healthy individual~($A$) by a sick individual~($B$) with rate $\sigma$.  
The second reaction~\eref{eq:chemical.rates2} represents the recovery from the disease with rate~$\zeta$. After recovery, full immunization~($C$) is assumed, implying that each individual can be sick only once.
Compared to the standard GEP reaction scheme, we introduce an additional ``caregiving" reaction~\eref{eq:chemical.rates3}, where immune individuals~($C$) help the infected~($B$) to recover with a rate~$\gamma$.
Finally, equations~\eref{eq:diffusion.rates1}--\eref{eq:diffusion.rates3} represent the mobility of each of the three species $A$, $B$ and $C$, which can hop from a site \textit{i} to a nearest-neighbor site \textit{j} with rates $D_A$, $D_B$, and $D_C$, respectively.
While the reactions~\eref{eq:chemical.rates1}--\eref{eq:chemical.rates3} are on-site processes, the nearest-neighbor hopping~\eref{eq:diffusion.rates1}--\eref{eq:diffusion.rates3} introduces a non-trivial spatial dependence. 
All the reactions described above are stochastic, meaning that the reaction rates represent the probability per unit of time for the corresponding reaction to occur. 
This general model exhibits an absorbing state phase transition between active and inactive phases:
In the active phase, the disease has a non-zero probability of spreading through the whole system, and the mean number of infected individuals increases. 
In contrast, in the inactive phase, the epidemic subsides, and the system reaches the absorbing state, where infected individuals are no longer present. 
The spatiotemporal distribution of the three types of individuals in this system is dictated by the dynamics of the infected population, denoted as $B$. 
Therefore, the density of $B$ functions akin to an order parameter in phase transitions, and we will adopt this terminology throughout this paper. However, in the active phase, the density of $B$ does not uniformly reach a non-zero value across space. 
Instead, it is non-zero along an expanding annulus since the epidemic spreads through the system as a traveling wave, as depicted in \fref{fig:1}. 

This paper analyzes how accounting for the healthy species' diffusivity affects the order parameter's critical dynamics. 
We study three limiting cases of the general reaction scheme~\eref{eq:chemical.rates1}--\eref{eq:diffusion.rates3}, all exhibiting a second-order absorbing state phase transition: the \textit{general epidemic process}~(GEP), the \textit{diffusive general epidemic process}~(DGEP) and the \textit{general epidemic process with caregiving}~(CGEP).
The reaction scheme of the GEP consists of~\eref{eq:chemical.rates1},~\eref{eq:chemical.rates2} and~\eref{eq:diffusion.rates2}, i.e., it assumes immobile healthy individuals and disregards the caregiving reaction (${D_A=D_C=\gamma=0}$).
This model was first proposed in 1927~\cite{Kermack.et.al:1927} but was formulated in terms of stochastic processes only much later in~\cite{Grassberger:1983}. 
At first, the critical properties of the GEP were believed to be different from other known universality classes (see, e.g.~\cite{Cardy:1983}). Still, later, it was established that the GEP transition lies in the universality class of Dynamical Percolation (DyP)~\cite{Cardy.Grassberger:1985,Janssen:1985}, even though the properties of these two systems far away from criticality differ~\cite{Tome.Ziff:2010}.

GEP is the paradigmatic model for disease spreading with immunization, but it disregards the mobility of healthy individuals.
To study the role of mobility of the healthy species $A/C$, we consider two distinct limiting cases of the general reaction scheme~\eref{eq:chemical.rates1}--\eref{eq:diffusion.rates3}. In these, mobility of susceptible and recovered individuals is examined independently.
To account for the mobility of susceptible individuals, we introduce a model variant of GEP with a finite diffusion constant $D_A \neq 0$, consisting of the reactions~\eref{eq:chemical.rates1},~\eref{eq:chemical.rates2},~\eref{eq:diffusion.rates1} and~\eref{eq:diffusion.rates2};  we refer to it as DGEP. Note that in both GEP and DGEP, the recovered individuals $C$ can effectively be treated as dead individuals since they do not interact with other species; thus, the mobility of $C$ cannot influence the dynamics of $B$.

To account for the mobility of recovered individuals, we introduce the CGEP model (\eref{eq:chemical.rates1}--\eref{eq:chemical.rates3},~\eref{eq:diffusion.rates2} and~\eref{eq:diffusion.rates3}), where a ``caregiving" reaction~\eref{eq:chemical.rates3} is taken into account, coupling the infected population with the recovered one. 
We consider the case where the recovered population is mobile (${D_C\neq 0}$), while the healthy population is immobile (${D_A= 0}$). 
This results in infected individuals ($B$) interacting with both mobile individuals ($C$) through caregiving (see equation~\eref{eq:chemical.rates3}) and immobile individuals ($A$) through infection (see equation~\eref{eq:chemical.rates1}). Consequently, CGEP allows us to investigate whether the dynamics at criticality are predominantly influenced by coupling with a diffusing or non-diffusing field.

\subsection*{Observables}
   Epidemic models intend to describe how a disease spreads through a population of healthy individuals. 
To quantitatively characterize the spreading, it is sufficient to consider a finite set of observables defined as an average over many stochastic realizations. Here we introduce three observables, which are typically considered for epidemic spreading:
The average number of infected individuals $\langle N\rangle(t)$ at time $t$, the survival probability $P_{\rm s}(t)$ of the epidemic, which represents the probability of having at least one infected individual left at time $t$, and the mean square radius $\langle R^2 \rangle (t)$ of the cluster of infected individuals. 
The qualitative behavior of these observables depends on whether the system is in the active or inactive phase. 
In the active phase, where the disease has a non-zero probability of spreading through the whole system, $\langle N\rangle(t)$ and $\langle R^2 \rangle (t)$ grow indefinitely, while $P_{\rm s}(t)$ approaches a non-zero value for $t\rightarrow\infty$. 
\Fref{fig:1}(b)
shows the typical behavior of an epidemic process with immunization in the active regime. 
In contrast, infected individuals go extinct in the inactive phase, such that both $\langle N\rangle(t)$ and $P_{\rm s}(t)$ approach zero for long times, and the mean square radius is not defined anymore.
At the phase transition, the system exhibits scale invariance, with observables following a power-law scaling~\cite{Tauber:2014, Henkel.et.al:2008}. 
Consequently, the three observables introduced previously take the following {asymptotic form~\cite{Tauber.et.al:2005}:  
\begin{eqnarray}
    \langle N \rangle (t,\sigma) 
    &=& t^{\theta} \hat{N}\left( (\sigma-\sigma_c)^{\nu_{\parallel}} t \right), 
    \label{eq:observable1}\\
    P_{\rm s}(t,\sigma) 
    &=& t^{-\delta} \hat{P}_{\rm s}\left((\sigma-\sigma_c)^{\nu_{\parallel}} t\right) \, , 
    \label{eq:observable2}
    \\
    \langle R^2 \rangle (t,\sigma) 
    &=& 
    t^{\frac{2}{z}} \hat{R}^2\left((\sigma-\sigma_c)^{\nu_{\parallel}} t\right)
    \, .
    \label{eq:observable3}
\end{eqnarray}
where $\sigma_c$ is the critical infection rate at which the system exhibits the absorbing state phase transition.
The four exponents $\theta, \delta, \nu_{\parallel}$, and  $z$---called critical exponents---characterize a universality class and are common to all phase transitions that belong to it. The functions $\hat{N}( x )$, $\hat{P}_{\rm s}(x)$ and $\hat{R}^2(x)$---called scaling functions---contain both a universal as well as a model-specific part; they are, however, difficult to infer analytically. Therefore, they are not the object of study in this paper.

\section{Mean-field analysis}
\label{sec:mean_field}

To gain some intuition on the various epidemic models and the role of diffusion, we first discuss the dynamics in continuous space without considering stochastic effects.
\begin{figure}
    \centering
    \includegraphics[width=13cm]{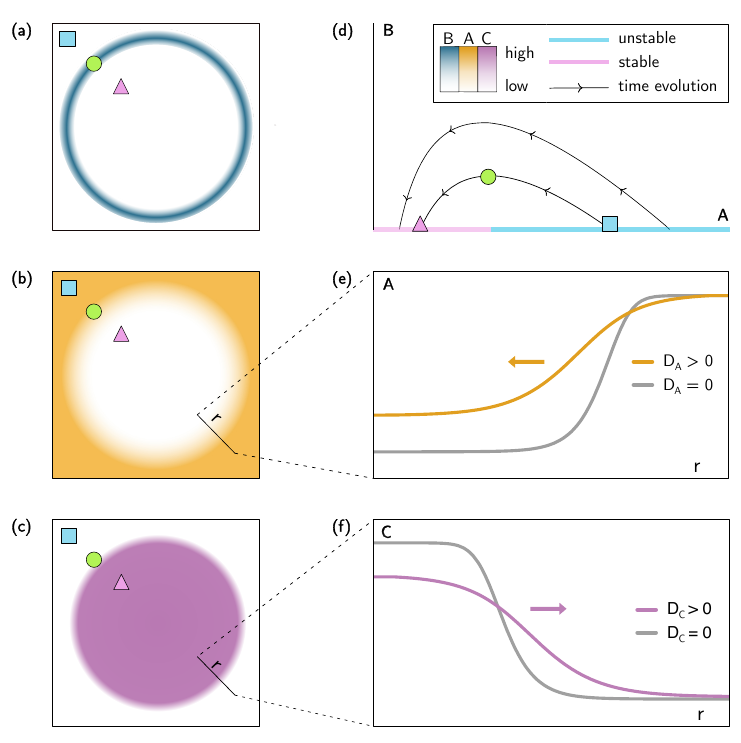}
    \caption{\textbf{Mean-field analysis.} The spreading of the density of infected $B$ (a), susceptible $A$ (b) and recovered $C$ individuals (c). The data was obtained through a finite element method (FEM) simulation of equations~\eref{eq:deterministicequations1}--\eref{eq:deterministicequations3} in a parameter regime corresponding to the active phase and using \textit{seed} initial conditions. The blue squares, yellow circles, and purple triangles in (a)--(d) correspond to phase space points where the wave of $B$ respectively has not yet passed, it is passing or has passed. (d) The well-mixed phase space of the densities $A(t)$ and $B(t)$ of equations~\eref{eq:deterministicequations1} and~\eref{eq:deterministicequations2}, in the ${C=0}$ plane. The dynamics exhibit a fixed line ${B^*=0}$, changing stability at the threshold value ${A^{ \rm mf}_c=\zeta/\sigma}$ at the boundary between the pink and blue line. (e)--(f) The influence of mobility on the profile of the densities $A$ and $C$ as a function of the radius $r$. The value approached at a large radius by $A$ and $C$ is determined by the initial conditions and is therefore equal for the diffusive and non-diffusing case. In (e), profiles with and without diffusivity for the susceptible species are shown in orange and gray, respectively, while in (f), they are shown in purple and gray. The orange (e) and purple (f) arrows show the direction of diffusive currents of susceptible and recovered individuals. }
     \label{fig:2}
\end{figure}
We use a mean-field theory where the system's description is no longer based on the number of individuals at each lattice site, 
but formulated in terms of the three scalar densities $A(\bi{r},t),B(\bi{r},t)$, and $C(\bi{r},t)$.
The corresponding mass-conserving reaction-diffusion equations~\cite{Frey.Brauns:2018} for the general reaction scheme~\eref{eq:chemical.rates1}--\eref{eq:diffusion.rates3} are given by:
\begin{eqnarray}
     \partial_{t}A (\bi{r} ,t) 
     &=& 
     D_A \nabla^2 A -\sigma A B \, , \label{eq:deterministicequations1} \\
     \partial_{t}B (\bi{r} ,t) 
     &=& 
     D_B \nabla^2 B +\sigma A B- \gamma B C-\zeta B \, ,\label{eq:deterministicequations2}\\
     \partial_{t}C (\bi{r} ,t) 
     &=& 
     D_C \nabla^2 C+ \gamma B C+\zeta B \, . \label{eq:deterministicequations3}
\end{eqnarray}
The equations of motion for the different models introduced above are obtained by setting specific parameters to zero: for the GEP, ${D_A = D_C = \gamma = 0}$; for the DGEP, ${D_C = \gamma = 0}$; and the CGEP, ${D_A = 0}$. 

\subsection*{Well-mixed system}
To better understand the type of phase transition exhibited by these epidemic models, we first focus on the well-mixed case, disregarding any spatial dependence and gradients in equations~\eref{eq:deterministicequations1}--\eref{eq:deterministicequations3}. 
Since the total number of individuals ${A+B+C}$ is conserved, it is sufficient to consider the dynamics of densities $A(t)$ and $B(t)$ only. 
We choose the initial conditions such that no recovered individuals are present, $C(0)=0$, imitating the initial conditions used later in stochastic simulations (see e.g.~\fref{fig:1}(b)). 
\Fref{fig:2}(d) illustrates the dynamics of the well-mixed system in the phase space spanned by $A$ and $B$. 
Independent of their initial values, the dynamics ultimately reach a stationary state where infected individuals are extinct (${B^* = 0}$).
This is because an epidemic with immunization cannot survive \textit{in loco}, i.e., in a well-mixed analysis, but has to spread through space to survive. In particular, the line ${B=0}$ is unstable for ${A^*> \zeta / \sigma}$ (light blue line in \fref{fig:2}(d)) and stable for $A^*< \zeta / \sigma$ (pink line in \fref{fig:2}(d)).
Stated differently, an epidemic transiently emerges in a well-mixed system only when ${A^*> \zeta / \sigma}$, as illustrated by the flow lines shown in \fref{fig:2}(d).
This scenario differs from an epidemic lacking permanent immunization, such as the \textit{diffusive epidemic process}~(DEP) defined by the reactions ${A+B \to 2 B}$ and ${B \to A}$~\cite{Polovnikov.et.al:2022}. 
In that case, a well-mixed system exhibits a transcritical bifurcation marked by the appearance of a stable active fixed point (${B^{*} \neq 0}$), allowing the disease to persist \textit{in loco}. 

\subsection*{Spreading dynamics}
Next, we performed 
simulations of the spatially extended model ~\eref{eq:deterministicequations1}--\eref{eq:deterministicequations3} using a Finite Element Method (FEM)~\cite{COMSOL}.
We are interested in the spreading of a disease through a population of susceptible individuals. Therefore, we chose initial conditions with a spatially uniform distribution of susceptible individuals ${A(\bi{r},0) = A_0}$, no recovered agents ${C(\bi{r},0)=0}$ and a small density of infected individuals---a \textit{seed}---placed in the center of the system. In continuous space the \textit{seed} is represented by a Gaussian distribution ${B(\bi{r},0)= B_0 \exp(- |\bi{r}|^2 / \rho^2)}$, where $B_0$ and $\rho$ are small compared to the other model's parameters. 
We term this type of initial conditions \textit{seed} initial conditions. 

As suggested by the phase space flow of the well-mixed system (\fref{fig:2}(d)), the spreading dynamics are qualitatively different depending on the initial homogeneous density $A_0$ and the values of the infection and recovery rates, $\sigma$ and $\zeta$, which are the tuning parameters of the deterministic phase transition. 
Suppose the system is initialized with $A_0$ above the threshold density ${A_0 > A^{\rm mf}_c = \zeta/\sigma}$. In that case, the seed grows, and the infection spreads throughout the system as an annular traveling wave, as depicted in figure~\ref{fig:2}(a). This happens because the system, except for the location of the \textit{seed},  is initialized at uniform density values $(A_0,0,0)$ corresponding to an unstable fixed point in the well-mixed analysis (depicted as light blue squares in figures~\ref{fig:2}(a)--(d)); thus, enabling the perturbation to grow.
The susceptible individuals \(A\) at the leading edge of the wavefront are gradually converted by infected \(B\) individuals.
After the wave has passed, these infected individuals recover such that the density of the susceptible individuals drops below the threshold, ${A < A^{\rm mf}_c}$} (pink triangle in figures~\ref{fig:2}(a)--(d)), resulting in the extinction of the epidemic behind the trailing edge of the wavefront. 
This spreading scenario corresponds to the \textit{active} phase. 
Conversely, if the initial concentration of
susceptible individuals is below the threshold, ${A_0 < A^{\rm mf}_c}$, the perturbation induced by the \textit{seed} of sick individuals decays and the epidemic goes extinct:
The system is initialized at uniform density values $(A_0,0,0)$, corresponding to a stable fixed point in the well-mixed analysis so that any perturbation decays.
The model is the \textit{inactive} phase in this scenario. The absorbing state phase transition is located on the active and inactive phase boundary at the deterministic threshold ${A_0 = A^{\rm mf}_c}$. 

\subsection*{Diffusive currrents}
Last, we discuss how the mobility of healthy individuals ($A$/$C$) influences the spreading dynamics in the generalized model. 
\Fref{fig:2}(e) and (f) illustrate the density profiles of susceptible ($A$) and recovered ($C$) individuals in the spatially extended system.
The density of susceptible individuals is low where the wave of infected individuals has already passed and high where it has not since the susceptible individuals get infected by the $B$ individuals in the wave. 
On the other hand, the spatial profile of the recovered individuals ($C$) is inverted compared to the spatial profile of susceptible individuals ($A$): 
High-density regions in $A$ coincide with low-density areas of $C$ and vice versa. 
This is because infected individuals are converted into recovered ones by the caregiving and recovering reactions, increasing their number after the wave of infected individuals has passed.
A comparison between \fref{fig:2}(a),(b), and (c) shows that the boundary between the low/high-density regions of $A$ and $C$ is located at the wavefront of infected individuals $B$.
If $A$ or $ C$ are mobile, the difference in density induces a diffusive current from regions of high to low density (see figures~\ref{fig:2}(e) and~(f)).
These currents have two notable effects: 
(i) The diffusivity of~$A$ and $C$ tends to minimize gradients by smearing out the interface between high and low-density phases.
(ii) The currents transport individuals from high-density regions to low-density ones, resulting in a higher density of $A$ and a reduced density of $C$ after the wave has passed as compared to the case where $A/C$ individuals are immobile (see \fref{fig:2}(e) and (f)). 

In the DGEP model, susceptible individuals are diffusing. Therefore, currents of $A$ individuals are present. Conversely, the CGEP model exhibits currents of $C$ individuals. 
The currents of $A$ and $C$ individuals in DGEP and CGEP, respectively, have the same orientation but reversed direction due to the inverted configuration of the densities.
Likewise, the impact on infected individuals resulting from the coupling with $A$ and $C$ is also reversed: Susceptible individuals transform into infected ones (${A + B \rightarrow 2B}$), whereas recovered individuals aid in the recovery of infected ones (${C + B \rightarrow 2C}$). 
The overall effect of the two currents on the spreading of the infection might, however, be similar as there is compensation for the opposite direction of currents and opposite reactive effects.
In the next section, we demonstrate through an RG analysis that the DGEP and CGEP systems lie in the same universality class, which could be linked to the similar effects of the respective diffusive fluxes.

\section{Field-theoretical analysis}
\label{sec:field.theory}
In the previous section, we discussed the spreading of infected individuals within the mean-field approximation of the stochastic epidemic model. We observed that, above a threshold for the initial density of susceptible individuals, the seed of infected individuals spreads throughout the entire system. Below this threshold, infected individuals initially spread, but eventually, the epidemic goes extinct, reaching the absorbing state. 
A similar phenomenology, typical of epidemic models featuring an absorbing state phase transition (see for example~\cite{Polovnikov.et.al:2022}), persists when stochasticity is considered. 
Stochastic effects become increasingly important near criticality when the system is close to the extinction threshold, and demographic noise cannot be neglected.
To analyze the stochastic critical dynamics, we use a field-theoretical description, also termed Fock space formulation, in which the stochastic lattice gas model given by~\eref{eq:chemical.rates1}--\eref{eq:diffusion.rates3} is formulated in terms of a master equation for the probability distribution of individuals' densities and then converted into a path integral representation~\cite{Tauber.et.al:2005, Weber.Frey:2017}. 
This enables the use of standard perturbative renormalization group tools to determine the universality class and critical exponents of the GEP, DGEP, and CGEP models~\cite{Tauber:2014}.

\subsection*{Action of the field theories}\label{subsec:action}
The actions of the three analyzed models containing only relevant terms are derived in appendix A 
and can be expressed as limiting cases of the following general action: 
\begin{eqnarray}
    \fl \mathcal{S}= \int_{\bi{q}} 
    \tilde{B}(-\bi{q}) \, 
    \bigg[
    \rmi\omega+D_{B}k^{2}+\tau 
    \bigg] \, 
    B(\bi{q}) - \frac{U}{2} \int_{\bi{q}_1,\bi{q}_2} \!\!\! B(\bi{q}_1+\bi{q}_2)\tilde{B}(-\bi{q}_1)\tilde{B}(-\bi{q}_2) \nonumber
   \\  
   +\int_{\bi{q}_1,\bi{q}_2} 
    \Bigg[ \frac{V \, B(\bi{q}_1)}{\rmi\omega_1+D k_1^2} + \frac{W \, B(\bi{q}_1)}{\rmi\omega_1} \Bigg] 
    B(\bi{q}_2)
    \tilde{B}(-\bi{q}_1-\bi{q}_2) \, ,
    \label{eq:action}
\end{eqnarray}   
where we denote frequency $\omega$ and momentum $\bi{k}$ (${|\bi{k}|=k}$) by a 4-vector ${\bi{q}=(\omega, \bi{k})}$ and write the integral as ${\int_{\bi{q}}:=\int_{\mathbb{R}^{d+1}}   \rmd \omega \, \rmd^{d}\bi{k}} /(2 \pi)^{d+1}$ . 
Here, ${B(\bi{q})= B(\omega, \bi{k})}$ denotes the Fourier transform of the density of infected individuals. In contrast, ${\tilde{B} (\bi{q}) = \tilde{B} (\omega, \bi{k})}$ is its dual response field; for an interpretation on its physical meaning see~\cite{Tauber:2014,Weber.Frey:2017}. 

The first term contains the inverse propagator for the field $B$, where the critical tuning parameter $\tau$ plays the role of an effective infection rate, encoding the dynamical interplay of infection and recovery. 
At criticality, one expects that these processes balance out such that $\tau= 0$ and the propagator contains no intrinsic length scale. 
From the Fock space formulation, one finds that $\tau=\zeta-A_0 \sigma$, which can be tuned to zero by adjusting the values of $\zeta$, $\sigma$ and $A_0$ such that the deterministic threshold is reached $A^{\rm mf }_c=A_0=\zeta / \sigma$, as found in section~\ref{sec:mean_field}. 
Note, however, that the actual critical value of $\tau$ is shifted with respect to ${\tau = 0}$ when stochasticity is considered. 
The interaction term
proportional to $U$ originates from the demographic noise in the model and has the same functional form as in the DP and GEP actions~\cite{Janssen:1981, Janssen:1985}. 
The two terms proportional to $V$ and $W$ arise from the non-linear contributions of the infection and caregiving reactions mediated by $A$ and $C$ individuals. 
Notably, the details on how the healthy species ($A, C$) interact with the infected one ($B$) do not enter these two terms, but only whether the interacting species is diffusing or immobile. 
Indeed, the two terms contain an explicit momentum contribution---${(\rmi\omega_1+D k_1^2)^{-1}}$ for $V$ and $(\rmi\omega_1)^{-1}$ for $W$---corresponding to the propagator of a diffusing field and a non-diffusing field, respectively. Which of the fields $A$ or $C$ corresponds to which propagators depends on the model. Accordingly, $D$ corresponds to $D_A$ (DGEP) or $D_C$ (CGEP) depending on the considered model. Table~\ref{table:1} summarizes the values of~$U$,~$V$,~$W$, and~$D$ expressed in terms of the microscopic parameters for the different models. 

For the GEP model, where only a coupling with a non-diffusing field ($A$) is present, $V$ equals zero, and $W$ describes the infection reaction. 
In the DGEP model, where only a coupling with a diffusing field ($A$) is present, $W$ equals zero, and the infection reaction is described by $V$. 
Finally, in the CGEP model, the infection reaction is described by $W$ (coupling with a non-diffusing species $A$), and the caregiving one is described by $V$ (coupling with a diffusing species $C$). 
Note that in the limit of ${D=0}$, the term proportional to $V$ becomes identical to the one proportional to $W$, and thus, it describes a coupling with a non-diffusing field. 
Following this reasoning, we recover the GEP action from the DGEP and CGEP models in the limit ${D=0}$. 

\Table{\textbf{Interaction vertices in terms of microscopic parameters.} Values of the interaction vertices in the action~\eref{eq:action} in terms of physical parameters for the three models. For a derivation of those values, please see appendix A. Note that the parameter $D$ is not defined in the GEP model, as no other diffusing species other than the infected one is present. } 
\br
Vertex 
& GEP  & DGEP  & CGEP   \\
\mr
$U$ & $2 \sigma A_0$ & $2 \sigma A_0$ & $2 \sigma A_0$  \\
$V$ & $0$ & $\sigma^2 A_0$ &  $\gamma \zeta$\\
$W$ & $\sigma^2 A_0$ & $0$ & $\sigma^2 A_0$  \\
$D$ & $-$ & $D_A$ & $D_C$ \\
\br \label{table:1}
\endTable

\subsection*{Rapidity reversal symmetry}
The action of the GEP model, (equation~\eref{eq:action} with ${V=0}$), is invariant under the following transformation~\cite{Janssen:1985}:
\begin{eqnarray}
   B( \bi{q}) 
   \rightarrow 
   -\frac{U}{W} \, \rmi\omega \,  
   \tilde{B}(\bi{q})
   \, , 
   \qquad 
   {\rm and}
   \qquad 
    \tilde{B}(\bi{q}) 
    \rightarrow 
    -\frac{W}{U}\frac{B(\bi{q})}{\rmi\omega}
    \, .
    \label{eq:rapidity.reversal.symmetry}
\end{eqnarray}
This symmetry is known as \textit{rapidity reversal symmetry}.
Generally, symmetries constrain the degrees of freedom of physical systems. In the context of criticality, this happens by reducing the number of independent critical exponents through a hyperscaling relation~\cite{Henkel.et.al:2008}
\begin{eqnarray}
    \theta=\frac{d}{z}-2\delta-1, \label{eq:hyperscaling.relationII}
\end{eqnarray}
where $\theta$, $\delta$, and $z$ were defined in equations~\eref{eq:observable1}--\eref{eq:observable3} and $d$ is the spatial dimension of the system. 
The number of independent exponents in the GEP universality class is reduced to three by the above hyperscaling relation.

Because of the vertex proportional to $V$, encoding the coupling of the field $B$ with a diffusive field, action \eref{eq:action} violates the rapidity reversal symmetry, similarly as in \cite{Deng.Odor:2023}. Consequently, the critical exponents of models featuring a coupling with a diffusing species are not linked by the hyperscaling relation~\eref{eq:hyperscaling.relationII} and the four critical exponents appearing in~\eref{eq:observable1}--\eref{eq:observable3} are all independent. 
This suggests a scenario in which at least one of the critical exponents deviates from the corresponding GEP, making DGEP and CGEP belong to a separate universality class. 

\subsection*{Number of absorbing states }
Another significant difference between coupling infected individuals with a diffusing and with a non-diffusing species lies in the number and type of absorbing states the system exhibits. This is an essential property because it is generally believed that the nature of the absorbing states has a decisive influence on the critical dynamics and the universality class of a model (see, e.g.~\cite{Janssen.Tauber:2005, Tauber.et.al:2005}).
A famous example is the DP conjecture, according to which all absorbing state phase transitions described by a single order parameter decoupled from any slow variable and with a single absorbing state belong to the same DP universality class~\cite{Grassberger:1982, Janssen:1981}. 

In the GEP model, where the infected individuals are coupled to a non-diffusing species only, the entire dynamics of the system are encoded in the movement of the infected individuals.
Once these entities are extinct (i.e., ${B(\bi{r},t)=0}$), the dynamics come to a halt, resulting in a perpetual frozen state for the system. 
Consequently, each arrangement of the non-diffusing coupled fields represents a distinct absorbing state, reached when no remaining $B$ individuals are left. In the particular case of GEP, the distribution of the frozen-in cluster of individuals $A$ at criticality is described by the universality class of isotropic percolation~\cite{Janssen:1985}. 
The situation changes if the infected individuals are coupled to a mobile species instead, as in the DGEP model. In this case, the susceptible species can still diffuse freely after the extinction of the infected individuals and approach a homogeneous distribution in the limit ${t \rightarrow \infty}$. Therefore, in the presence of mobility, the absorbing state only exists in a state with homogeneous density in $A$, reached by diffusive currents. Note that the distribution of recovered individuals doesn't influence the dynamics of the infected population. Therefore, its configuration in the absorbing state can be disregarded. 
The classification is more subtle in the CGEP model since it features absorbing states where the recovered individuals are mobile and reach a homogeneous distribution while the susceptible individuals are frozen in time.
We perform a perturbative renormalization group analysis to examine if and how these differences affect the critical behavior of the different models.

\subsection*{Renormalization group analysis and flow equations}
The renormalization group 
theory is applied to study how the parameters of a theory change as the typical length scale on which the system is observed is increased by a coarse-graning procedure~\cite{Wilson:1975, Wilson.Fisher:1972}. This scale dependence is encoded in a set of differential equations termed flow functions~\cite{Fisher:1974}. 
Scale-invariant behavior in equilibrium systems is typically observed at the critical points of second-order phase transitions or in the low-temperature phase of systems featuring Goldstone modes~\cite{Tauber:2014}. In non-equilibrium systems, scale invariance is obtained in the presence of Goldstone modes as for the KPZ equation~\cite{Frey.Tauber:1994} and---fairly generically---also in reaction-diffusion systems with absorbing states~\cite{Henkel.et.al:2008, Tauber.et.al:2005}. In all these cases, the critical behavior is described by a fixed point of the flow equations, such that the theory's parameters remain invariant at different length scales. This fixed point is reached at large length scales and is therefore denoted an IR-stable fixed point.

Here, we utilize a perturbative field-theoretical ($Z$-factor) 
RG approach to calculate these flow functions to one-loop order. The expansion parameter is $\epsilon = d_c - d$, the deviation from the upper critical dimension of the theory, which in our case is $d_c = 6$. 
The $Z$-factors are determined through dimensional regularization in a minimal subtraction scheme (for a review of these methods, see~\cite{Weinberg:1995,Tauber:2014}).
To perform a perturbative RG study, it is crucial to identify the effective couplings, which are combinations of the theory's coupling parameters.
In appendix~B, we show that one possible choice of effective couplings is given by
\begin{eqnarray}
    \fl
    \frac{UW}{(4 \pi D_B)^3} \, \mu^{-\epsilon}=:g_1 \, ,   
    \qquad 
    \frac{UV}{(4 \pi D_B)^3} \,  \mu^{-\epsilon}=:g_2 \, ,
    \qquad {\rm and}
    \quad 
    \frac{D}{D_B}=:g_D \, . \label{eq:CGEP.eff.coupling}
\end{eqnarray}
The renormalized momentum scale $\mu^{-\epsilon}$ is inserted to ensure that the effective couplings $g_1$~and$g_2$ are dimensionless at ${d=6-\varepsilon}$ dimensions. They encode the interplay between
the demographic noise (vertex $U$) and the non-linear reactions, which couples the infected species with a diffusing (vertex $V$) or non-diffusing species (vertex $W$). As previously discussed, the information on how the healthy species $A$ and~$C$ are coupled to the infected one does not enter the effective couplings. It only matters whether the coupled species is diffusing (effective coupling $g_2$) or immobile (effective coupling $g_1$). Finally, the effective coupling $g_D$ describes the mobility of diffusing healthy individuals 
relative to the mobility of infected ones.
In appendix~B, a detailed derivation of the RG flow functions for the action~\eref{eq:action} is given. We ensure those flow functions describe the system dynamics at criticality by restricting the analysis to the critical plane $\tau=0$.
We find: 
\begin{eqnarray}
\fl  \beta_{g_{1}}(g_{1},g_{2},g_{D})
:=\mu \frac{d}{d\mu}g_{1}=-\epsilon g_{1}+g_{1}\left(\frac{7g_{1}}{2}+g_{2}\frac{5+5g_{D}+2g_{D}^2}{2(1+g_{D})^3}\right) \, , 
\label{eq:flow.function.CGEP1}
 \\
\fl  \beta_{g_{2}}(g_{1},g_{2},g_{D})
:=\mu \frac{d}{d\mu}g_{2}= -\epsilon g_{2}+g_{2}\left(\frac{5g_{1}}{2}+g_{2}\frac{5+5g_{D}+2g_{D}^2}{2(1+g_{D})^2}\right) \, , 
\label{eq:flow.function.CGEP2}
\\
\fl  \beta_{g_{D}}\left(g_{1},g_{2},g_{D}\right) :=\mu \frac{d}{d\mu}g_{D}=-g_{D}\left(\frac{7g_{1}}{12}+g_{2}\frac{7+4g_{D}+g_{D}^2}{12(1+g_{D})^3}\right) \, . 
\label{eq:flow.function.CGEP3}
\end{eqnarray}

\begin{figure}[htb]
\centering
\includegraphics[width=12cm]{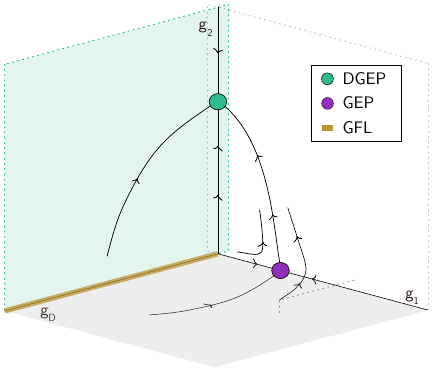}
\caption{\textbf{Renormalization group flow.} 
A sketch of the RG flow of the effective couplings $g_i$ (solid black lines with arrows) obtained from the RG flow equations~\eref{eq:flow.function.CGEP1}--\eref{eq:flow.function.CGEP3} where the direction of the arrows denotes the flow towards larger length scales. Note that we consider only the physical region of the parameter space where all effective couplings are positive.
The flow exhibits a Gaussian fixed line  (brown shaded line) at ($g_1=g_2=0$), the GEP fixed point (purple dot) at (${g_1^*=2\epsilon/7, g_D^*=g_2^*=0}$) and the new DGEP fixed point (green dot) at (${g_1^*=0=g_D^*}$, ${g_2^*=2\epsilon/5 }$).
In GEP, the infected
individuals ($B$) are exclusively coupled to immobile susceptible individuals via the infection reaction, such that the model's RG
flow is restricted to the $g_D$-$g_1$ plane (gray-shaded plane). It approaches the GEP fixed point. Here, $g_D$ represents the mobility of an uncoupled species, and
the flow is indicated by a curved line as the dynamics of $g_1$ is independent of $g_D$, but the one of $g_D$ depends on $g_1$ (see~\eref{eq:flow.function.CGEP1}--\eref{eq:flow.function.CGEP3}).
On the contrary, in DGEP, the infected individuals are only coupled to diffusing susceptible individuals via the infection reaction, restricting the model's RG flow to the $g_2$-$g_D$ plane (green shaded plane). The flow reaches the DGEP fixed point along this plane. 
In CGEP, the infected individuals are coupled to both immobile susceptible individuals via the infection reaction, and mobile recovered individuals via the caregiving reaction. The RG flow starts from ($g_1\neq 0, \ g_2 \neq 0, \ g_D \neq 0$) (coordinates of the point indicated by gray dashed lines) and approaches the DGEP fixed point.
}
\label{fig:3}
\end{figure}

\Fref{fig:3} illustrates the RG flow resulting from~\eref{eq:flow.function.CGEP1}--\eref{eq:flow.function.CGEP3} for dimensions below ${d_c = 6}$ (${\epsilon>0}$).
It exhibits three different fixed points: the Gaussian fixed line, the well-known GEP fixed point, and a new fixed point we term the DGEP fixed point.

The Gaussian fixed line (${g_1=g_2=0}$) corresponds to a model where the noise amplitude is effectively zero ${U=0}$, i.e., the mean-field limit, or equivalently where both the infection and caregiving reactions are absent (${V=W=0}$), effectively decoupling the dynamics of the infected individuals $B$ from that of the healthy ones. Note that at the Gaussian fixed line (brown line in \fref{fig:3}), ${g_2}$ is equal to zero, indicating that the effective coupling $g_D$ no longer enters in action~\eref{eq:action}, since it represents the mobility of a species which does not interact with the infected one, like the recovered individuals \(C\) in the CGEP model.

In GEP, where infected individuals only interact with a non-diffusing species, the RG flow starts from ${g_2=0}$ and is constrained to the $g_1$--$g_D$ plane (the gray plane in \fref{fig:3}). The flow reaches the GEP fixed point (${g_1^*=2\epsilon/7,g^*_2=g^*_D=0}$) as identified by~\cite{Janssen:1985}. Since ${g_2=0}$, $g_D$ represents the mobility of a species whose dynamics are not coupled to that of \(B\) and any $g_D$-dependence vanishes in $\beta_{g_1}$. Therefore, $g_1$ is the only effective coupling in the GEP model, and its value approaches the GEP fixed point under the RG flow given by $\beta_{g_1}(g_1,0,0)$, which is the same flow function as found in~\cite{Janssen:1985}. 
The GEP fixed point is stable in the $g_1$ direction but unstable along the $g_2$ direction. 
For finite $g_2$, i.e., when the infected individuals are coupled to a diffusing field, the RG flow approaches the new DGEP fixed point (${g_1^*=0=g_D^*}$, ${g_2^*=2\epsilon/5 }$). This fixed point is stable along all three directions $g_1$,~$g_2$ and~$g_D$. Indeed, both the flow corresponding to DGEP---described by $\beta_{g_2}(0,g_{2},g_{D})$ and $\beta_{g_D}(0,g_{2},g_{D})$---and CGEP---which requires all three effective couplings---approach this new IR-stable fixed point. 
Thus, both models belong to the same universality class, different from GEP, whose critical exponents are given in table~\ref{table:2}.
This universality class describes the critical dynamics close to the extinction threshold of models where the infected individuals are coupled to a diffusing species, as in the DGEP and CGEP models.
The surprising fact is that, at criticality, the interactions with a non-diffusing species are entirely suppressed by the coupling with a diffusive field since at the DGEP fixed point ${g_1=0}$. 

In summary, the universality class of the critical dynamics is altered if the infected population is coupled to at least one diffusing species, suppressing all other interactions with non-diffusing fields. Previous studies found similar results \cite{Costa.et.al:2007}, suggesting that the same principle can be valid in other generic epidemic models where the coupling between a diffusive and a non-diffusive species is present.
\fulltable{\textbf{Critical exponents.} The value of the critical exponents~\eref{eq:observable1}--\eref{eq:observable3} of the GEP and DGEP universality classes calculated both with an $\epsilon$-expansion below the upper critical dimension $d_c=6$ (theory) and through stochastic simulations of the GEP (sim.~GEP), DGEP (sim.~DGEP) and CGEP (sim.~CGEP) models in dimension $d=2$.  The obtained exponents are compared with the numerical~\cite{Munoz.et.al:1999} and theoretical~\cite{Janssen:1985} literature values of the GEP universality class.}
 & \multicolumn{3}{c}{\textbf{GEP}} & & \multicolumn{3}{c}{\textbf{DGEP}} \\[1.5mm]
 & \text{theory} & \text{lit.~\cite{Munoz.et.al:1999}} & \text{sim. GEP} &  &\text{theory} & \text{sim. DGEP} & \text{sim. CGEP} \\
\mr
$z$ & $2-\frac{1}{6}\epsilon$ & $1.1295$ & $1.13 \pm 0.004$& &$2-\frac{7}{30}\epsilon$ & $1.14 \pm 0.03$ & $1.128 \pm 0.004$ \\[1.5mm]
$\delta$ & $1-\frac{5}{28}\epsilon$ & $0.092$ & $0.093 \pm 0.001$& & $1-\frac{1}{4}\epsilon$ & $0.091 \pm 0.009$ & $0.093 \pm 0.003$ \\[1.5mm]
$\theta$ & $\frac{3}{28}\epsilon$ & $0.586$ &  $0.582 \pm 0.003$ & &$\frac{3}{20}\epsilon$ & $0.54 \pm 0.02$ & $0.554 \pm 0.005$ \\[1.5mm]
$v_{\parallel}$ & $1+\frac{1}{28}\epsilon$ & $1.506$ &  $1.5 \pm 0.1$ & & $1+\frac{1}{20}\epsilon$ & $1.4 \pm 0.1$ & $1.4 \pm 0.1$ \\[0.75mm]
\label{table:2}
\endfulltable

\subsection*{Vanishing diffusivity implies additional divergences}

At first sight, it may appear counterintuitive that at the DGEP fixed point, the effective coupling $g_D$ is equal to zero since this is the value it takes when the coupled species is immobile. 
To clarify this, we look at the limit ${D\rightarrow 0}$, which helps to elucidate how this limit leads to a different universality class and why $g_D$ is zero at both the GEP and DGEP fixed point. 

In the minimal subtraction scheme used in this work, the divergences arising in the perturbative calculation of the $n$-point correlation functions completely determine the flow functions and, thus, the fixed point structure of the model. In appendix~B, we show that
almost all divergences arising from the vertices $V$ are $W$ are symmetric, meaning that divergences 
proportional to $g_1$ can be obtained from the ones proportional to $g_2$ in the limit ${D\rightarrow 0}$.
The only exception stems from a single diagram (diagram (A), in appendix B), which turns out to have a divergent contribution only proportional to $g_1$ and not $g_2$. This divergence is responsible for the asymmetry between the flow functions $\beta_{g_1}$ and~$\beta_{g_2}$~\eref{eq:flow.function.CGEP1}--\eref{eq:flow.function.CGEP2}.
In appendix~B we evaluate this diagram, which results in two different contribution:~$g_2 V D^3_B  I(D)+g_1 W D^3_B I(D=0)$, where:
\begin{eqnarray}
   \fl I(D):= \int  
    \frac{\rmd^d \bi{p}}{(2 \pi)^d} 
    \frac{1}{(D-D_B) p^2 +\tau} \, \frac{1}{\big[\rmi\omega_1 +(D_B+D)p^2 +\tau\big]^2} \, \frac{1}{\rmi\omega_1 +2 D p^2}
    \, .  \label{eq:anonamlous.divergent .diagram}
\end{eqnarray}
The integral over the momentum $ \bi{p}$ is evaluated at dimension ${d=6-\epsilon}$, as dimension regularization requires. The key point is that only $I(D=0)$ is divergent, not $I(D\neq 0)$. The reason can be inferred by simple power counting. If ${D\neq0}$, the denominator scales for large momentum as $p^{8}$ and the integral converges. If ${D=0}$, the scaling changes to $p^{6}$ and the integral diverges. 
Consequently, the only divergent contribution of diagram (A), calculated through dimensional regularization, is given by ${g_1 W D^3_B I(0)=g_1 W/(\rmi\omega_1 \epsilon)+\mathcal{O}(\epsilon^0)}$.
This divergence has the same structural form as the vertex $W$ in action~\eref{eq:action}; thus, it 
has to be reabsorbed by the $Z$-factor of the vertex $W$ in the renormalization procedure. The implication is that the $Z$-factor of the vertex $W$
and $V$ differ, leading to different flow functions for the effective couplings $g_1$ and $g_2$. In particular, the additional divergence arising from diagram (A) shifts the $7g^2_1 /2$ in~\eref{eq:flow.function.CGEP2} to $5g^2_2 /2$ in~\eref{eq:flow.function.CGEP1} and creates an asymmetry. The asymmetry in the flow functions of $g_1$ and $g_2$ is responsible for changing the value of the GEP fixed point to
(${g_1^*=2\epsilon/7,g^*_2=g^*_D=0}$) as compared to the DGEP one  (${g_2^*=2\epsilon/5,g^*_1=g^*_D=0}$), leading to a different universality class.

We believe that a similar mechanism by which some Feynman diagrams acquire additional divergences in the limit of no diffusivity can equally explain why many models featuring absorbing state phase transitions change universality class depending on the diffusivity of the various species. One example is the DEP model with immobile healthy species, which lies in a different universality class than the diffusive DEP model~\cite{Pastor-Satorras.Vespignani:2000,Janssen:1997}.

Having explained how setting ${D = 0}$ leads to an additional divergence, we can now resolve the seeming ambiguity that the effective coupling ${g_D=D / D_B}$ vanishes at the DGEP fixed point.  
The important difference between the GEP and DGEP fixed points is how $g_D$ reaches its zero value.
When infected individuals are coupled to a diffusing species, like in DGEP, $g_D$ approaches zero by $D_B$ flowing to infinity under the RG flow. In contrast, $D$ does not flow and remains fixed at its non-zero initial value. 
On the other hand, if $B$ is coupled to an immobile species, like in the GEP model, the effective coupling $g_D$ is zero due to the value of the diffusion constant ${D=0}$. 
However, we showed before that the flow functions acquire an additional contribution only if ${D=0}$, independent of the value of $D_B$. Therefore, even if $g_D = 0$ at the DGEP fixed point, we have $D\neq0$, thus guaranteeing that the DGEP flow functions differ from the GEP ones. 

\section{Stochastic simulations of epidemic models} 
\label{sec:numerics}
In the preceding section, we discerned a novel universality class through an RG analysis by identifying a relevant modification of the GEP model. 
However, since we are using a perturbative method expanding around the upper critical dimension~${d_c=6}$, the quantitative predictions concerning the values of the critical exponents may not be accurate in physical dimensions~${d\leq 3}$. 
Therefore, to determine the values of the critical exponents, we carried out stochastic simulations. 
Specifically, we used a Gillespie algorithm based on an algorithm published in previous work~\cite{Polovnikov.et.al:2022} to study the critical dynamics in the vicinity of the absorbing state phase transition for two-dimensional lattice models as defined in section~\ref{sec:models}. 

\subsection*{Gillespie algorithm for spatially extended systems}

Most previous studies implement non-diffusive processes on a regular lattice with an exclusion constraint and simulate the dynamics by standard Monte Carlo updates~\cite{Henkel.et.al:2008,Grassberger:1983, Munoz.et.al:1999, Grassberger:1989, Hinrichsen:2000,  Argolo.et.al:2011, Souza.et.al:2011}, where two-individual reactions are implemented as nearest-neighbor interactions~\cite{Grassberger:1979, Grassberger:1983}.
The exclusion constraint considers steric interactions between the individuals and allows, at most, a single occupancy for each lattice site. 
With this constraint, the mobility of individual agents can only be implemented by pairwise exchange processes~\cite{Deng.Odor:2023}. This, however, implies a symmetry in the mobilities of different species and, thus, disallows exploring asymmetric diffusion constants. Especially, the interesting limiting cases with certain immobile species cannot be implemented without further effort. Beyond that, the above class of Monte Carlo algorithms does not bear an intrinsic notion of time. 
Rather, time is associated with the number of updating steps.
These limitations can be overcome by resorting to a different class of stochastic updating schemes.

Here, we employ a modified version of the \textit{stochastic simulation algorithm} (SSA), commonly referred to as  \textit{Gillespie} algorithm~\cite{Gillespie:1976, Gillespie:2007}. In this approach, the underlying lattice structure of the models is retained while the exclusion constraint is abandoned. Thus, multiple individuals can occupy a single site. 
Two-individual reactions occur on a single site, and diffusion is implemented as nearest-neighbor hopping. This decouples the individual species' mobility and allows their hopping rates to differ. In particular, to account for the complexity of the stochastic many-body dynamics, we resort to the \textit{hybrid next reaction scheme}~\cite{Gibson.Bruck:2000, Polovnikov.et.al:2022} which, in principle, allows to solve the problem in sub-polynomial time.
The benefit of this choice is that it generates \textit{exact} stochastic trajectories with a clear notion of time and does not require any exclusion constraint. We are specifically building on an approach that has recently proven successful in resolving a longstanding debate regarding the critical dynamics of the DEP model~\cite{Polovnikov.et.al:2022}. While the hybrid next reaction scheme provides certain benefits, as outlined above, it also requires substantially more computational resources than standard Monte Carlo updates on a lattice with exclusion constraint.

\subsection*{Initial conditions and units.}
As outlined in section~\ref{sec:mean_field}, we study the spreading dynamics starting from a single seed of infected individuals. To that end, we initialize a square lattice with a homogeneous density of~$A$ individuals and an infectious seed of~$B$ individuals placed on the central site. 
For all of our simulations, the initial density of~$A$ individuals is one individual per lattice site, and the infectious seed consists of ten~$B$ individuals. Additionally, finite-size effects can be largely eliminated with single seed initial conditions~\cite{Henkel.et.al:2008}. Finite-size effects arise mostly from the interaction of~$B$ individuals with the boundary of a finite lattice. In all the simulations, we ensure that no such finite-size effects occur by choosing a large enough system size~$L$, such that no $B$ individual ever reaches the boundary during the finite simulation time. Note that this is only made possible by the algorithm's sub-linear scaling with system size. Lastly, time is measured in units of~$10$ times the recovery time, setting ${\zeta = 0.1}$. 
Length scales are measured in multiples of the lattice spacing~${a = 1}$.

\subsection*{GEP model as a benchmark}

\begin{figure}[t]
    \includegraphics[width=\columnwidth]{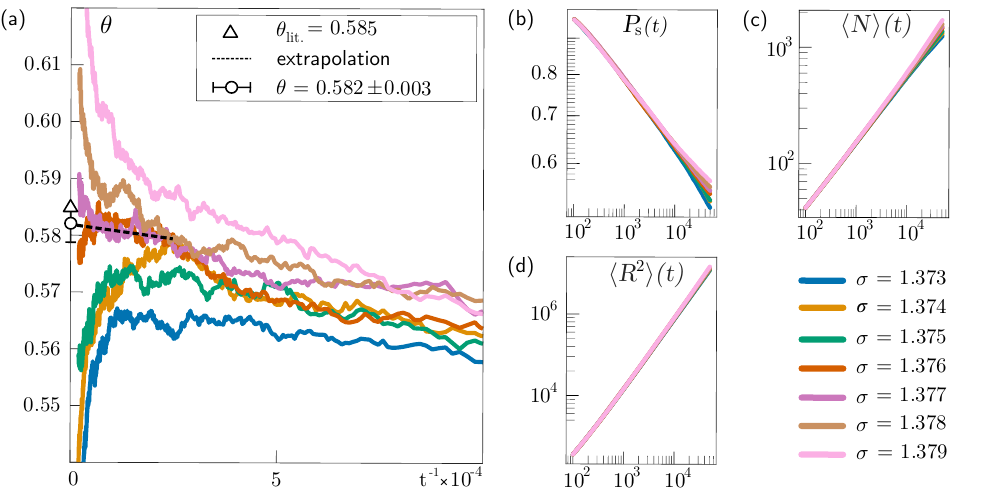}
    \caption{
    {\bf Critical exponents of GEP model.}
    (a) Averaged local slope for the number of infected individuals for a set of values of the infection rate~$\sigma$ as indicated in the legend. Based on these results, we estimate the critical point to lie at~$\sigma_c = 1.3765$ (between the two lines at~$\sigma=1.376$ and~$\sigma=1.377$) and the value of~$\theta=0.582 \pm 0.003$ as indicated by the circular marker and error bars.
    The triangular marker indicates the current best literature estimate for~$\theta = 0.585$~\cite{Henkel.et.al:2008}.
    For each value of~$\sigma$, results were obtained by averaging over 24,000 stochastic trajectories starting from a single seed at the origin. 
    For each realization, the dynamics were monitored for 50,000-time units in a 2D system with linear expansion of 10,000 lattice sites. Other parameters were chosen as~${D_A=D_C=0}$,~${D_B=0.1}$,~${\zeta=0.1}$, and~${\gamma=0}$.
    (b)--(d) Double logarithmic plots for the key observables of the spreading dynamics for a range of infection rates indicated in the graph:
    (b) survival probability~$P_{\rm s} (t)$, 
    (c) mean number of infected individuals~$\langle N \rangle (t)$, and (d) mean square radius of the infected cluster~$\langle R^2 \rangle (t)$.
    }
    \label{fig:4}
\end{figure}
To validate the chosen algorithm and to establish a benchmark for the critical exponents and their errors, we first reinvestigate the GEP model to test how well our code reproduces the GEP critical exponents from the literature~\cite{Munoz.et.al:1999, Argolo.et.al:2011, Souza.et.al:2011, Deng.Odor:2023}.
Figures~\ref{fig:4}(b)--(d) show double logarithmic plots of the three key observables defined in equations~\eref{eq:observable1}--\eref{eq:observable3} of the spreading dynamics for a range of values of the infection rate~$\sigma$. 
Inferring the critical value of~$\sigma$ from these plots amounts to finding the curves that do not bend upwards or downwards, i.e., are straight and thus represent power laws. 
Given the proximity of the curves, achieving high accuracy proves difficult. 
Therefore, a common approach is to use the \textit{averaged local slope}~\cite{Grassberger:1979, Grassberger:1989}. It is defined as follows for the mean number $\langle N\rangle(t)$ of infected individuals and a local approximation~$\theta(t)$ of the corresponding critical exponent,
\begin{eqnarray}
    \theta(t) 
    = \frac{\ln[\langle N \rangle(t) ] - \ln[\langle N \rangle (t/\Delta t) ]}{\ln[\Delta t]} 
    \,  ,
\label{eq:local.slope.theta}
\end{eqnarray}
and analogously for all other observables and exponents. 
Here, the time interval~$\Delta t$ must fulfill the condition~${\Delta t > 1}$.~\ref{appendix:time.interval} elaborates on our choice of~${\Delta t=10}$, which is comparable to previous studies~\cite{Grassberger:1989, Jensen.et.al:1990, Jensen:1993, Voigt.Ziff:1997, Deng.et.al:2020}. 

The intercept with the ${1/t=0}$ axis---which signifies the measured critical exponent---is estimated by a linear extrapolation of~$\theta(t)$ for our best estimate of the critical infection rate~\cite{Henkel.et.al:2008, Grassberger:1989}. It corresponds to the line of the infection rate, for which a linear extrapolation is possible and thus results in a finite intercept.
Figure~\ref{fig:4}(a) shows the averaged local slope for the mean number of infected individuals. In appendix~E, we explain how we determined the error of the exponents. 
Comparing our results for the critical exponents with literature values~\cite{Munoz.et.al:1999, Argolo.et.al:2011, Souza.et.al:2011, Deng.Odor:2023} we find excellent agreement, cf.\@~table~\ref{table:2}.
The larger size of our error bars is due to using a different simulation algorithm~\cite{Polovnikov.et.al:2022} combined with our cautious and conservative approach to error estimation as outlined in appendix~E.
Our simulation algorithm convincingly reproduces well-established results from prior studies~\cite{Munoz.et.al:1999, Argolo.et.al:2011, Souza.et.al:2011, Deng.Odor:2023}, providing a robust benchmark for evaluating the outcomes of the DGEP and CGEP models in the subsequent section.

\subsection*{Simulation results for the DGEP and CGEP models}
\begin{figure}[ht]
\centering
\includegraphics[width=\columnwidth]{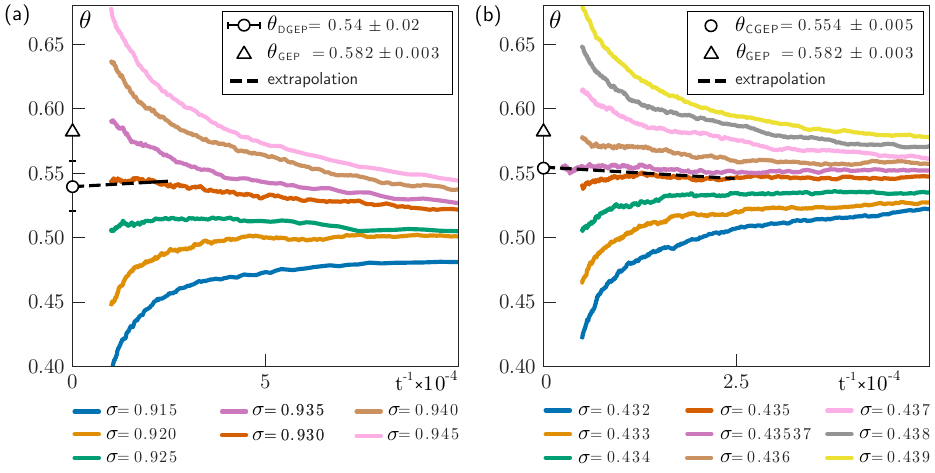}
\caption{
{\bf Critical behavior of the DGEP and CGEP models.} 
(a) Averaged local slope for the number of infected individuals for the DGEP model for a set of values for the infection rate~$\sigma$ indicated below the graph.
Based on these results, we estimate the critical point to lie at~${\sigma_c = 0.93}$ and the value of~${\theta=0.54 \pm 0.02}$ as indicated by the circular marker with error bars. The triangle shows our estimate for GEP for comparison, where we omitted the error bars because they are smaller than the marker. 
For each value of~$\sigma$, results were obtained by averaging over $20,000$ stochastic trajectories starting from a single seed at the origin. The dynamics were monitored for $10,000$ time units in a 2D system with linear expansion of $2,600$ lattice sites for each realization. Other parameters were chosen as~${D_C=0}$,~${D_A=D_B=0.1}$,~${\zeta=0.1}$, and~${\gamma=0}$. 
(b) Averaged local slope for the number of infected individuals for the CGEP model for a set of values for the infection rate~$\sigma$ indicated below the graph.
Based on these results, we estimate the critical point to lie at~${\sigma_c = 0.43537}$ and the value of~${\theta=0.554 \pm 0.005}$ as indicated by the circular marker. A triangular marker represents our estimate for GEP. Both results' errors are smaller than the respective marker.
For the critical value~$\sigma_c$, results were obtained by averaging over $34,000$ stochastic trajectories starting from a single seed at the origin. The dynamics were monitored for $40,000$ time units in a 2D system with linear expansion of $8,800$ lattice sites for each realization.
For each other value of~$\sigma$, results were obtained by averaging over 34,000 stochastic trajectories starting from a single seed at the origin. The dynamics were monitored for $20,000$ time units in a 2D system with linear expansion of $4,400$ lattice sites for each realization.
In both cases, the other parameters were chosen as~${D_A=0}$,~${D_B=D_C=0.1}$,~${\zeta=0.05}$, and the rate~${\gamma=0.1}$ for the new caregiving reaction. 
}
    \label{fig:5}
\end{figure}
\begin{figure}[t]
\centering
\includegraphics[width=\columnwidth]{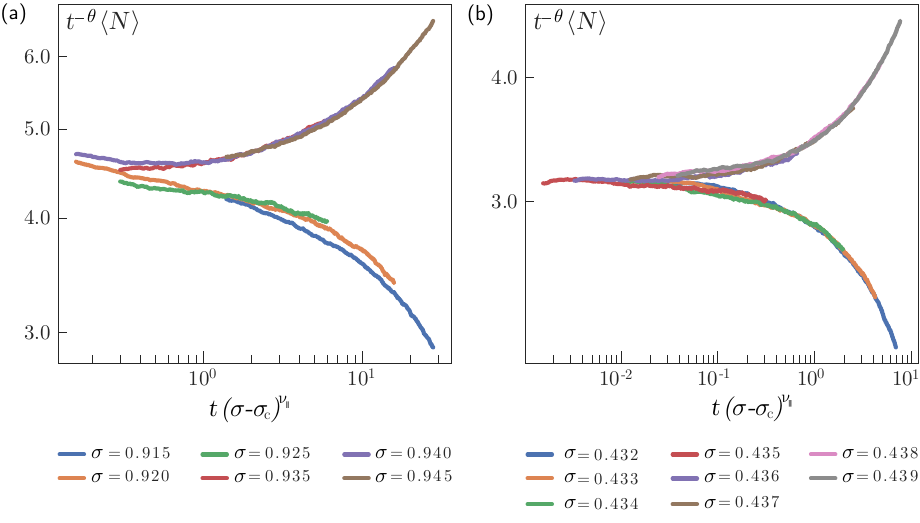}
\caption{
     {\bf Data collapse for the DGEP and CGEP models.} (a) Rescaled number of infected individuals in the DGEP simulations. The data was collapsed using the critical~${\sigma_c=0.93}$,~${\theta=0.54}$, and~${\nu_\parallel=1.4}$. (b) Rescaled number of infected individuals in the CGEP simulations. The data was collapsed using the critical~${\sigma_c=0.43537}$,~${\theta=0.55}$, and~${\nu_\parallel=1.4}$. The underlying data and simulation parameters are identical to the data shown in figure~\ref{fig:5}.}
    \label{fig:6}
\end{figure}
After benchmarking our stochastic simulation algorithm using the GEP model, we next performed stochastic simulations of the spreading dynamics of the DGEP and CGEP models.
The corresponding results for the averaged local slope of the number of infected individuals are shown in figure~\ref{fig:5}. For the figures of the other observables, we refer the reader to appendix~F.
Using the methods outlined above and in appendix~E, we determined the critical exponents~$\theta$,~$\delta$ and~$z$ of the three key observables defined in equations~\eref{eq:observable1}--\eref{eq:observable3}.
They are summarized in table~\ref{table:2} together with their error estimates. 

There are two key observations: 
Our numerical estimates of the exponents~$z$ and~$\delta$---the critical exponents related to the temporal scaling of the mean squared radius and the survival probability---match their respective values of our GEP simulations within error bars as well as those of prior work~\cite{Munoz.et.al:1999, Argolo.et.al:2011, Souza.et.al:2011, Deng.Odor:2023}. 
However, the numerical estimate of the critical exponent~$\theta$---related to the temporal scaling of the mean number of infected individuals---of the DGEP~($\theta = 0.54 \pm 0.02$)  and CGEP~($0.554 \pm 0.005$) models distinctly differ from the corresponding GEP~($0.582 \pm 0.003$) value. Importantly, they lie well outside of the error bars.
Secondly, the numerical estimates of the three critical exponents for DGEP and CGEP are identical within error bars. 
Taken together, the numerical results affirm our analytical findings from the RG analysis that the critical dynamics of the DGEP and CGEP models belong to the same universality class, which differs from the GEP universality class.

Remarkably, we find very good agreement between our RG predictions of the exponents in dimension~${d=2}$ (i.e., ${\epsilon = 4}$) and our numerical estimates. 
The analytic prediction for $\theta$ of~${\theta_{\rm theory} = 0.6}$ deviates less than~$10 \%$ from the numerical estimates of~${\theta_{\rm num} = 0.54 \pm 0.02}$~(DGEP) and~${\theta_{\rm num} = 0.554 \pm 0.005}$~(CGEP). 
This is also the case for the dynamical exponent $z$. 
Here, the RG calculation predicts ${z_{\rm theory} = 1.0\overline{6}}$ compared to the numerical estimates of ${z_{\rm num} = 1.14 \pm 0.03}$ (DGEP) and ${z_{\rm num} = 1.128 \pm 0.004}$ (CGEP). 
For $\delta$, the analytical prediction is ${\delta_{\rm theory} = 0}$, which deviates from the numerical estimates of~${\delta_{\rm num} = 0.091 \pm 0.009}$~(DGEP) and~${\delta_{\rm num} = 0.093 \pm 0.003}$~(CGEP). Nevertheless, the absolute deviation is still minimal considering the value of~${\epsilon =4}$. 
However, our simulations can neither confirm different values for the~$\delta$ and~$z$ exponents of the DGEP universality class when compared to the GEP one nor the analytic prediction of~${\theta_{\rm DGEP} > \theta_{\rm GEP}}$.

Moreover, using our simulation results, we confirm that the hyperscaling relation~\eref{eq:hyperscaling.relationII} is indeed violated for the CGEP model:
\begin{eqnarray}
   0.554 \pm 0.005 = \theta \neq \frac{d}{z}-2\delta-1 = 0.59 \pm 0.02 \, .
   \label{eq:broken.hyperscaling.relation.CGEP}
\end{eqnarray}
Conversely, for the DGEP model, our carefully estimated errors do not permit a definite conclusion.
The absence of a rapidity reversal symmetry~\eref{eq:rapidity.reversal.symmetry} enforces the need to infer one additional independent critical exponent to fully characterize the universality class~\cite{Henkel.et.al:2008}. 
Figure~\ref{fig:6} illustrates the estimation of the independent critical exponent~$\nu_\parallel$ via a data collapse~\cite{Tauber:2014, Hinrichsen:2000} for both models. 
The specifics of the methodology and the error estimation are discussed in appendix~E.  
This analysis confirms the analytical result that the DGEP and CGEP models exhibit identical values for the $\nu_\parallel$ exponent as indicated in table~\ref{table:2}. 
The RG calculation in two dimensions predicts ${\nu_{\parallel, {\rm theory}} = 1.2}$, while the numerical estimate for both models is ${\nu_{\parallel, {\rm num}} = 1.4 \pm 0.1}$. 
This is the largest absolute deviation between the two methods, though it is still smaller than $20\%$ in relative terms.

Moreover, our results show excellent agreement with the ones recently reported by Deng and Odor~\cite{Deng.Odor:2023} for a version of the GEP model with symmetric exchange hopping and an exclusion constraint, which is a different microscopic implementation of the DGEP model.
Generally reporting smaller error bars, even a significant difference in the exponent $\delta$ could be identified. Our algorithm decouples the mobility of the individual species. With that, we could reinforce our analytical claim that the new DGEP universality class is also robust against keeping $C$ individuals immobile for the GEP reaction scheme with diffusing $A$ species.
Taking these results obtained on a system with exclusion constraint together with ours without such, one may faithfully claim the existence of a new universality class that deviates from the well-known GEP universality class. Surprisingly, the values of the exponents of the new DGEP universality class are very similar to those of the GEP universality class. This effect occurs both in the RG calculations as well as the stochastic simulations in two dimensions.

\section{Discussion} \label{sec:conclusion}
In this paper, we explored the impact of mobility on the critical dynamics in epidemics through a combined approach of renormalization group analysis and stochastic simulations.
Specifically, we focused on epidemics with immunization and considered two modifications---called DGEP and CGEP---of the \textit{general epidemic process}~(GEP): DGEP generalizes the GEP model by considering mobile susceptible individuals $A$, while in the CGEP model a new ``caregiving'' reaction~(${B+C \rightarrow 2C}$) is introduced, where mobile recovered individuals~($C$) can aid the recovery of infected individuals~($B$). 
In both models, the infected individuals are coupled to a diffusing species via a non-linear reaction. Diffusivity of the coupled species disrupts a symmetry of the GEP model, the so-called rapidity reversal symmetry.
The universality class of the absorbing state phase transition of the respective models was studied by a one-loop perturbative RG analysis in~$d=6-\epsilon$ dimensions, just below the upper critical dimension. We found that the flow functions of both the CGEP and DGEP models reach the same new RG fixed point, indicating that they belong to a new universality class different from the GEP one. 
Stochastic simulations affirmed these findings, which identified the same set of critical exponents for both systems.  
The critical exponents' values do not fulfill the hyperscaling relation associated with the rapidity reversal symmetry, providing evidence that this symmetry is violated in the new universality class. This new universality class describes the spreading of an epidemic near the extinction threshold, where infected individuals interact with a diffusing species and gain immunization upon recovery.

Based on these analytical and numerical studies, we draw the following three main conclusions. 
Firstly, the emergence of a new universality class shows that mobility plays a crucial role in the critical dynamics of epidemic models,
or, more generally, of non-equilibrium phase transitions. 
This highlights the significance of exploring this phenomenon~\cite{Deng.et.al:2020}. 
Diffusivity is important in other paradigmatic epidemic models for which a complete analytical understanding is still lacking~\cite{Polovnikov.et.al:2022}.
Secondly, in the CGEP model---which couples infected individuals to both a diffusing and a non-diffusing species---the RG flow approaches a fixed point where the coupling to the non-diffusing species vanishes, i.e., the coupling to immobile species becomes increasingly irrelevant approaching the critical regime. 
We hypothesize that this is one example of a more general effect: On large length and time scales and close to criticality, couplings to mobile species dominate the couplings to immobile ones. 
Lastly, we showed that DGEP and CGEP exhibit the same critical behavior, indicating that, though sensitive to the introduction of mobility, universality classes might be surprisingly indifferent to how and with which diffusing species the order parameter is coupled. 
This claim warrants a broader investigation of different ways mobility might be introduced to other non-equilibrium phase transitions.
In our models, one could test the robustness of this effect by adding new reactions, for instance, a vaccination reaction (${A+C \rightarrow 2C}$), coupling the susceptible species $A$ with the recovered $C$, and verify if the critical dynamics is again only determined by the mobility of the healthy species, or if the specific manner in which a healthy species is coupled to the infected one matters.  

The existence of the new DGEP universality class was suggested by the violation of rapidity-reversal symmetry and by the different numbers of absorbing states between GEP and DGEP.
In particular, the GEP model features an infinite number of absorbing states, corresponding to arbitrary arrangements of frozen-in susceptible individuals after the disease has gone extinct. 
In contrast, DGEP exhibits absorbing states of uniformly distributed susceptible individuals~$A$, reached by diffusive currents.
The absorbing states in the CGEP model encompass both immobile and mobile species, making classification based on this criterion challenging. 
However, having shown that the interaction with the immobile species becomes irrelevant at criticality, we can disregard the spatial distribution of the immobile species in the absorbing state and correctly classify the CGEP phase transition into the DGEP universality class. 

To understand why adding mobility to the GEP model affects its critical behavior, we compare our result with previous works
\cite{Tauber.et.al:2005}, which studied the density decay of a two-species annihilation process~(${A+B \rightarrow \emptyset}$).
They found that the corresponding decay rate~$\lambda$ acquires non-trivial RG corrections when both~$A$ and~$B$ are mobile.
As the annihilation reaction is local, the RG corrections can be understood as an altered meeting probability for two diffusing individuals due to stochasticity. 
We believe that similar considerations can be applied to our models. In particular, through a mean-field analysis (section~\ref{sec:mean_field}), we showed that diffusion creates diffusive currents of healthy individuals. 
When stochasticity is included, the effect of these currents can be understood analogously as a change in the probability of encounters between infected and healthy species, which leads to the new DGEP universal class.
We believe that the above explanation in terms of meeting probabilities applies generically, explaining why diffusion can alter the critical behavior and highlighting the robustness of the universality class featuring diffusive individuals. 

Finally, our numerical methods, based on the hybrid next reaction scheme version of the SSA~\cite{Gibson.Bruck:2000,Gillespie:1976, Gillespie:2007,Polovnikov.et.al:2022}, may open avenues for further numerical studies, enhancing our understanding of the role of diffusion in absorbing state phase transitions and non-equilibrium phase transitions in general. 
The stochastic simulations of the DGEP model highlight the importance of decoupling the mobility of different species to allow for a more general analysis of the impact of mobility (see~\cite{Polovnikov.et.al:2022}). 
The crucial step is to abandon the exclusion constraint in the lattice structure. This also aligns with the standard derivation of a field-theoretical 
action in the path-integral formalism that strongly relies on bosonic density fields~\cite{Tauber:2014}.
To conclude, we suggest that our stochastic simulation algorithm holds the potential for exploring the impact of diffusion in a broader range of problems beyond critical dynamics. 
In particular, it could be used to analyze models with anomalous diffusion, for example, extracting anomalous diffusion exponents~\cite{Munoz.et.al:2021} and calculating higher order moments~\cite{Schwarzl.et.al:2017}.

From a broader perspective, our study highlights the importance of mobility for the spreading dynamics of populations close to the extinction threshold. It suggests general mechanisms that may lead to new critical dynamics. Future studies may focus on different types of population dynamics~\cite{Reichenbach.et.al:2007,Knebel.et.al:2020} and non-local interactions between species~\cite{Kolk.et.al:2023, Mukhamadiarov.et.al:2021, Cao.et.al:2022, Grassberger:2013}. 
We also see future potential in investigating the role of different types of individuals mobility like anomalous diffusion realized via Lévy flights~\cite{Homrighausen.et.al:2013,Shapoval.et.al:2022} and long-range dispersal~\cite{Hallatschek.et.al:2014,Villiger.Paulose:2023}.

\ack
We thank Richard Dengler for inspiring discussions and Patrick Wilke and Moritz Striebel for their valuable support in getting started with the numerical code. 
We acknowledge financial support by the German Research Foundation (DFG) through SFB 1032 (Project ID No.~201269156), the Excellence Cluster ORIGINS under Germany’s Excellence Strategy (EXC-2094-390783311), the European Union (ERC, CellGeom, project number 101097810), and the Chan-Zuckerberg Initiative (CZI).
\newpage 

\newpage

\appendix

\section{Derivation of the action}
\label{appendix:A}
In this appendix, we derive a field-theoretical description for our models starting from the general reaction scheme~\eref{eq:chemical.rates1}--\eref{eq:chemical.rates3}.
In section~\ref{sec:mean_field}, we discussed the absorbing state phase transition within the mean-field approximation, where the effect of noise was disregarded.
To account for stochasticity, a field-theoretical analysis is imperative and is developed throughout this section.
We derive a field-theoretical action describing infected individuals' dynamics close to the extinction threshold. 

\subsection*{The Coherent State Path
Integral method.}
We start by reviewing how to map the reaction scheme~\eref{eq:chemical.rates1}--\eref{eq:diffusion.rates3} to a field-theoretical description for the well-mixed case and then generalize to a spatially extended system.
Given the probabilistic nature of the stochastic process, we aim to find the time evolution of the probability~${p(n_A,n_B,n_C,t)=: p(\bi{n},t)}$ that there are $n$ individuals for each given species.
A common way to find it is through the Coherent State Path Integral~(CSPI) method, a specific type of Fock space formulation~\cite{Weber.Frey:2017}.  
In this method, the probability distribution at time~$t$ is marginalized over a Poissonian initial condition~$p(\bi{n},t|\bi{x}_0, t_0)$, with initial mean~${\bi{x}_0=(x_{{0}_A},x_{{0}_B},x_{{0}_C})}$. 
Probability distributions other than a Poissonian distribution could be used for marginalization, but this choice often is the most convenient in calculations~\cite{Weber.Frey:2017}. 
The time evolution of the densities~$\bi{n}(t)$, given by a backward master equation, can be mapped to the time evolution of the Poissonian means~$\bi{x}(t)$ with initial conditions~$\bi{x}_{0}$. 
The different~$x_i(t)$ capture the averaged statistical behavior of the density of each species~$n_i(t)$. Following the standard procedure~\cite{Weber.Frey:2017}, we convert the backward master equation for~$\bi{x}(t)$ into a path integral representation for~$p(\bi{n},t|\bi{x}_0, t_0)$:
\begin{eqnarray}
     p(\bi{n},t|\bi{x}_0,t_0) \propto \int \hspace{1mm}\mathcal{D}[\bi{x}(t'), \bi{q}(t')]
    \exp\left(-\int_{t_0}^{t}\rm{d}t\mathcal{L}[\bi{x}(t'),\bi{q}(t')]\right) \label{eq:marginal.probability}
\end{eqnarray}
The marginalized probability distribution~$p(\bi{n},t|\bi{x}_0,t_0)$ is expressed as a integral over all ``paths" of~$\bi{x}(t)$ and its response field~$\bi{q}(t)$. Those paths are weighted with a stochastic Lagrangian density~$\mathcal{L}$ of the form~\cite{Tauber:2014}
\begin{eqnarray}
    \mathcal{L}[\bi{x}(t),\bi{q}(t)]= \rmi \bi{q}\partial_t \bi{x}(t)-\mathcal{Q}(\bi{x}(t),\rmi\bi{q}(t)+1) ,\label{eq:shifted:lagrangian}
\end{eqnarray}
where~$\rmi \bi{q}\partial_t \bi{x}(t)$ captures the time evolution of the mean~$\bi{x}(t)$ and the Liouville operator~$\mathcal{Q}$ encodes the dependence on the specific reaction scheme. For an illustration, we look at a simple reaction involving three species~$A$,~$B$, and~$C$:
\begin{eqnarray}
    kB+mA  \overset{\lambda}{\rightarrow} lC.
    \label{eq:reaction}
\end{eqnarray}
Calling $A(t)$,~$B(t)$,~$C(t)$ the time evolution of the Poissonian means for individuals~$A$,~$B$,~$C$ and~$\tilde{A}(t)$,~$\tilde{B}(t)$,~$\tilde{C}(t)$ their response fields, the Liouville operator for reaction~\eref{eq:reaction} reads 
\begin{eqnarray}
    \mathcal{Q}= \lambda B^{k}A^{m}(\tilde{C}(t)^{l}-\tilde{B}(t)^{k}\tilde{A}(t)^{m}) \label{eq:standardLiouville}
\end{eqnarray}
When studying a multi-reaction system, the contributions of the different reactions get summed up, as the Liouville operator is linear.

The above procedure can be generalized to a spatially extended system, where the dependence on space can be included by allowing individuals to hop on a lattice with lattice constant~$a$.
To be able to use~\eref{eq:standardLiouville}, individuals' hopping processes between different sites must be formulated as reactions.
This is done by considering individuals on two different sites to belong to different species, coupled through the reaction~$B_i {\rightarrow} B_j$ with the hopping rate specified by $D_B$. 
For a sufficiently large system, a continuous description can be achieved by taking the limit~$a\rightarrow 0$, resulting in the Poissonian means~$\bi{x}(\bi{r},t)$  which are now additionally space-dependent. 
Explicitly for our system, we rename $\bi{x}(\bi{r},t)$ in terms of the different species---e.g.\@ $x_A(\bi{r},t)=A(\bi{r},t)$ and~$q_A(\bi{r},t)=\tilde{A}(\bi{r},t)$. Using the standard form~\eref{eq:standardLiouville} for $\mathcal{Q}$ for the general reaction scheme~\eref{eq:chemical.rates1}--\eref{eq:diffusion.rates3}, we get the following Lagrangian in continuous space:
\begin{eqnarray}
\fl \mathcal{L}= \tilde{A}(\bi{r},t) \big[\partial_{t}-D_{A}\nabla^{2}\big]
A(\bi{r},t) 
+ \tilde{B}
(\bi{r},t)
\big[\partial_{t}-D_{B}\nabla^{2}\big]
B(\bi{r},t) \label{eq:action.from.coherent} \\
+ \tilde{C}(\bi{r},t)
\big[ \partial_{t}-D_{C}\nabla^{2} \big]
C(\bi{r},t) -\zeta B(\bi{r},t)
\Big[ \tilde{C}(\bi{r},t)-\tilde{B}(\bi{r},t)\Big] \nonumber  \\ 
 - \nonumber 
    \sigma A(\bi{r},t)B(\bi{r},t)\Big[ \big( \tilde{B}(\bi{r},t)+1 \big)^2-\big(\tilde{B}(\bi{r},t)+1 \big) \big(\tilde{A}(\bi{r},t)+1 \big) \Big] \\ 
-  \gamma \, C(\bi{r},t)B(\bi{r},t) \Big[ \big(\tilde{C}(\bi{r},t)+1\big)^2 -\big(\tilde{B}(\bi{r},t)+1\big)\big(\tilde{C}(\bi{r},t)+1\big)
\Big] .
\nonumber
\end{eqnarray} 
The first three terms, containing the time and spatial derivatives, correspond to the propagators for the fields~$A$,~$B$, and~$C$. 
The Laplace operator originates from nearest-neighbor hopping. 
The terms proportional to~$\sigma$,~$\zeta$, and~$\gamma$ correspond to the on-site reactions of infection, recovery, and caregiving. 
Those terms arise from the Liouville operators corresponding to each reaction. 
In contrast to the general form for the Liouville operator~\eref{eq:standardLiouville}, all the response fields are shifted by 1, e.g.~${\tilde{A}\rightarrow \tilde{A}+1}$, in accordance to~\eref{eq:shifted:lagrangian}. This shift is known as the ``Doi shift". The same kind of dynamic action functional can also be derived via other formalisms, for example, the Martin-Siggia-Rose-Bausch-Janssen-Wagner-de Dominicis response functional~\cite{Martin.et.al:1973,Janssen:1976, deDominicis:1976, Bausch.et.al:1976, Weber.Frey:2017}. 
The strength of the CSPI approach is that it does not require any assumptions; all the information is contained in the specific form of the reaction scheme. 
However, CSPI characterizes the evolution of fields that represent the statistically averaged behavior of a density (specifically, the Poissonian means of its distribution) rather than the density itself. 
Consequently, some of the system's stochasticity is inherently blurred within this formalism and cannot be captured. An instance of this is ``diffusional noise", also known as conserved noise. 
In a Langevin description, this usually corresponds to a term of the type~$\nabla( \sqrt{\rho} \eta)$, where~$\rho$ is the density and~$\eta$ a white Gaussian noise. To obtain this term in a Fock Space formulation, a description in terms of densities must be recovered by the so-called Cole-Hopf transformation, converting the CSPI action into a Kramers-Moyale type action \cite{Weber.Frey:2017}.
Although diffusional noise cannot be derived within the CSPI formalism, its absence is inconsequential in our context as it is RG irrelevant at our system's upper critical dimension ${d_c=6}$; thus, it can be neglected.

\subsection*{Collecting  relevant terms.}
Close to the extinction threshold, the dynamics predicted by the mean-field equations~\eref{eq:deterministicequations1}--\eref{eq:deterministicequations3} become invalid as fluctuations grow larger in the vicinity of an absorbing state phase transition. 
We aim to calculate the corrections arising from stochasticity with a perturbative renormalization group method. 
The first step is to expand the action around the deterministic critical values for the fields. In particular, we shift the fields
${A(\bi{r},t)\rightarrow A_0 + A(\bi{r},t)}$, and tune~$A_0$ towards the critical~$A_c^{\rm{mf}}$ (see section~\ref{sec:mean_field} ).
This way, the shifted fields represent the deviations from the deterministic critical solution due to noise. 
Via a dimensional analysis, one can determine the naive scaling dimension of the different interactions in~\eref{eq:chemical.rates1}--\eref{eq:chemical.rates3}. 
Thus, one can determine which interactions are \textit{relevant} near criticality and which are \textit{irrelevant}, i.e., their reaction strength goes to zero under the RG flow and can be neglected \cite{Tauber:2014}. 
 To ensure the renormalizability of the action, we have to retain only the relevant terms,  i.e., the ones with a positive momentum dimension.
 The relevant part of action~\eref{eq:action.from.coherent}, where the fields have been already shifted, reads: 
\begin{eqnarray}
\fl  \mathcal{S}= \int \rmd t \rmd^{d}\bi{r} \tilde{A}(\bi{r},t)\big[\partial_{t}-D_{A}\nabla^{2}\big]A(\bi{r},t) + \tilde{B}(\bi{r},t)\big[\partial_{t}-D_{B}\nabla^{2} \big]B(\bi{r},t) \nonumber  \\
 + \tilde{C}(\bi{r},t)[ \partial_{t}-D_{C}\nabla^{2} ]C(\bi{r},t) -\zeta B(\bi{r},t)\Big[ \tilde{C}(\bi{r},t)-\tilde{B}(\bi{r},t)\Big] \nonumber  \\ - 
    \sigma A(\bi{r},t)B(\bi{r},t)  \tilde{B}(\bi{r},t)-\sigma A_0 B(\bi{r},t) \Big[  \tilde{B}^2(\bi{r},t)+\tilde{B}(\bi{r},t)-\tilde{A}(\bi{r},t) \Big] \nonumber  \\ +   \gamma \, C(\bi{r},t)B(\bi{r},t) \tilde{B}(\bi{r},t) \label{eq:ren.action}
\end{eqnarray}
\subsection*{Integrating out the fields~$A$ and~$C$.}
Action~\eref{eq:ren.action} can be simplified by integrating out the fields~$A$ and~$C$, as their corresponding response fields appear only linearly in the interaction terms since the non-linear terms, raising from stochastic noise, are \textit{irrelevant}. 
In particular, we show that the integral over the response fields results in a~$\delta$-function constraining the evolution of~$A(\bi{r},t)$ and~$C(\bi{r},t)$ to obey the linearized part of the deterministic equations~\eref{eq:deterministicequations1}--\eref{eq:deterministicequations3}. Note that this simplification is not necessary since once could equally renormalize the above action and would obtain the same flow functions~\eref{eq:flow.function.CGEP1}--\eref{eq:flow.function.CGEP3}.
We proceed by collecting the terms proportional to~$\tilde{A}$ and~$\tilde{C}$
\begin{eqnarray}
\fl \int \mathcal{D}[B,\tilde{B},A ,C] \exp \left(- \mathcal{S}_{h} \right) \times \nonumber\\
\fl \int \mathcal{D}[ \tilde{A}] \exp  \left(- \int \rmd t \rmd^{d} \bi{r} \tilde{A}(\bi{r},t)\Big[ \partial_{t}A(\bi{r},t)-D_{A}\nabla^{2} A(\bi{r},t) +\sigma A_0 B(\bi{r},t) \Big] \right) \times \nonumber \\
\fl  \int \mathcal{D}[\tilde{C}]\exp   \left(- \int \rmd t \rmd^{d}\bi{r} \tilde{C}(\bi{r},t)\Big[ \partial_{t} C(\bi{r},t)-D_{C}\nabla^{2}C(\bi{r},t) -\zeta B(\bi{r},t)\Big] \right)= \nonumber \\
\fl  \int \mathcal{D}[B,\tilde{B},A,C] \exp  \left(- \mathcal{S}_{h} \right) \times \nonumber
\\ 
   \delta_{\bi{r},t}\left(  \partial_{t}A(\bi{r},t)-D_{A}\nabla^{2} A(\bi{r},t) +\sigma A_0 B(\bi{r},t) \right) \times \nonumber\\
  \delta_{\bi{r},t}\left(  \partial_{t} C(\bi{r},t)-D_{C}\nabla^{2}C(\bi{r},t) -\zeta B(\bi{r},t) \right), \nonumber
\end{eqnarray}
where~$\delta_{\bi{r},t}$ denotes a delta-function imposing its argument to be zero at each point in space and time. To simplify notation, we introduced the action~$\mathcal{S}_{h}$ 
\begin{eqnarray}
    \fl \mathcal{S}_{h}= \int \rmd t \rmd^{d}\bi{r} \Big[ \tilde{B}(\bi{r},t)\left( \partial_{t}-D_{B}\nabla^{2}+ \zeta-A_0 \sigma \right) B(\bi{r},t) 
  -\sigma A_0 B(\bi{r},t)  \tilde{B}^2(\bi{r},t) \nonumber \\ - \sigma A(\bi{r},t)B(\bi{r},t)  \tilde{B}(\bi{r},t)
  +\gamma \,  C(\bi{r},t)B(\bi{r},t) \tilde{B}(\bi{r},t)\Big]. 
\end{eqnarray}
As anticipated, integrating out the fields~$A$ and~$C$ results in a delta-function requiring that the two fields follow the mean-field equations~\eref{eq:deterministicequations1}--\eref{eq:deterministicequations3} linearized around the homogeneous solution~$( A_0, 0, 0 )$. The non-linear part of the mean-field equations do not enter the action~\eref{eq:ren.action} as they are RG irrelevant at ${d_c=6}$.
The solution of the linearized mean-field equations can be easily obtained in Fourier space:
\begin{eqnarray}
    A(\omega,\bi{k})= -\sigma A_0 \frac{ B(\omega,\bi{k})}{\rmi \omega+D_A k^2 } \qquad 
    C(\omega,\bi{k})=\zeta  \frac{B(\omega,\bi{k})}{\rmi \omega+D_C k^2 } \nonumber 
\end{eqnarray}
By inserting the values of~$A$ and~$C$ in the Fourier transformed action~$\mathcal{S}_{h}$, we get the final action
\begin{eqnarray}
\fl \mathcal{S}=\int_{\bi{q}}  \tilde{B}(-\bi{q}) \left(\rmi \omega +D_{B}k^{2}+\zeta-A_0 \sigma  \right)B(\bi{q}) -\int_{\bi{q}_1,\bi{q}_2 }\sigma A_0 B(-\bi{q}_1-\bi{q}_2)  \tilde{B}(\bi{q}_1)\tilde{B}(\bi{q}_2)   \nonumber \\
  + \int_{\bi{q}_1,\bi{q}_2} \Bigg[  \frac{\sigma^2 A_0 B(\bi{q}_1)}{\rmi \omega_1+D_A k_1^2 } +  \frac{\gamma \zeta B(\bi{q}_1)}{\rmi \omega_1+D_C k_1^2 } \Bigg] B(\bi{q}_2)  \tilde{B}(-\bi{q}_1-\bi{q}_2). \label{eq:final.action}
\end{eqnarray}
where we denote frequency~$\omega$ and momentum~$\bi{k}$ (${|\bi{k}|=k}$) by a 4-vector~${\bi{q}=(\omega, \bi{k})}$ and write the integral as~${\int_{\bi{q}}:=\int_{\mathbb{R}^{d+1}}   \rmd \omega \, \rmd^{d}\bi{k}/(2\pi)^{d+1}}$. 
Here,~${B(\bi{q})= B(\omega, \bi{k})}$ denotes the density of the infected individuals in Fourier space, while~${\tilde{B} (\bi{q}) = \tilde{B} (\omega, \bi{k})}$ is its dual response field; for an interpretation on its physical meaning see~\cite{Benitez.et.al:2016,Tauber:2014,Weber.Frey:2017}. 
To recover the right action for the three models discussed in the main text~(GEP, DGEP, and CGEP) we set~${\tau= \zeta-A_0 \sigma}$ in the first term.
The couplings for the non-linear terms and the value of~$D$ differ for each model. 
Only the first non-linearity is the same in all cases~${U=2 \sigma A_0}$. 
In GEP, the caregiving reaction is absent ($\gamma=0$) and~$A$ immobile~($D_A=0$). In the general action~\eref{eq:action} this corresponds to~${V=0, W=\sigma^2 A_0}$. 
Analogously, in DGEP, the caregiving reaction is not present~(${\gamma=0}$), but~${D_A \neq 0}$. 
In the general action~\eref{eq:action} this corresponds to~${V=\sigma^2 A_0, W=0, D=D_A}$. 
In CGEP, all reactions are present, but the susceptible species~$A$ is immobile~${D_A=0}$.
The action of CGEP can be obtained from~\eref{eq:action} setting~${V=\zeta\gamma, W=\sigma^2 A_0, D=D_C}$.

\section{Renormalization}
\label{appendix:RG}
In this appendix, we present how to calculate the RG flow functions~\eref{eq:flow.function.CGEP1}--\eref{eq:flow.function.CGEP3} for the general epidemic model presented in the main text of this paper. 
The flow functions are calculated using the $Z$-factor RG approach and by performing an $\epsilon$-expansion around the upper critical dimension~${d_c=6}$, below which the critical exponents acquire non-trivial values~\cite{ZinnJustin:2002}.
The idea of the $Z$-factor approach is to redefine couplings and fields such that the observables---which seemingly acquire infinite values for ${\epsilon \rightarrow 0}$---are well-defined. Those redefined parameters are termed \textit{renormalized}, whereas the original parameters are referred to as \textit{bare}. Throughout this section, we represent the bare parameters with an overscore ring (e.g.\@~$\mathring{\tau}$ and~$\mathring{D_B}$), while the renormalized parameters are denoted in a standard manner (e.g.\@~$\tau$ and~$D_B$).

Before renormalizing the action~\eref{eq:action}, it is convenient to rescale it in terms of effective couplings, allowing us to express the coefficients $(U,V,W,D)$ in terms of  $(g_2,g_1,g_D)$, otherwise, the resulting equations for the $Z$-factors are underdetermined. In the field-theoretical approach, the fields themselves are ``integration parameters" integrated over each path (see e.g.\@~\eref{eq:marginal.probability}); thus, they can be rescaled by appropriately changing the integration measure. 
To express the action in terms of effective couplings, we rescale the bare fields in the following way: 
\begin{eqnarray}
    \mathring{B}\rightarrow \left( \mathring{U} \mathring{D}_B \mathring{V}^{-1} \right)^{\frac{1}{2}} \mathring{B} \qquad \mathring{\tilde{B}} \rightarrow \left( \mathring{U} \mathring{D}_B \mathring{V}^{-1} \right)^{-\frac{1}{2}} \mathring{\tilde{B}}
\end{eqnarray}
Since the field and response field are inversely rescaled, the determinant of the Jacobian for this change of variable is unitarian, and the general expression for the probability of a density configuration is unchanged~\eref{eq:marginal.probability}.
The rescaled action reads:
\begin{eqnarray}
\fl \mathcal{S}= \int_\bi{q} \mathring{\tilde{B}} (-\bi{q})  \left( \rmi\omega+\mathring{D}_B k^{2}+\mathring{\tau} \right) 
\mathring{B}(\bi{q})-\int_{\bi{q}_1,\bi{q}_2} \frac{(\mathring{g}_2)^{\frac{1}{2}} \mathring{D}_B}{2} \mathring{\tilde{B}}(-\bi{q}_1)\mathring{\tilde{B}}(-\bi{q}_2) \mathring{B}(\bi{q}_2+\bi{q}_1) \nonumber\\
\fl +\int_{\bi{q}_1,\bi{q}_2} \Bigg[ \frac{ \mathring{g}_1  \mathring{D}^2_B }{ (\mathring{g}_2)^{\frac{1}{2}}} \frac{\mathring{B}(\bi{q}_1)}{\rmi\omega_1} 
+ (\mathring{g}_2)^{\frac{1}{2}} \mathring{D}^2_B\frac{\mathring{B}(\bi{q}_1)}{\rmi\omega_1+\mathring{g}_D \mathring{D}^2_B k_1^{2}}  \Bigg] \mathring{B}(\bi{q}_2)  \mathring{\tilde{B}} (\bi{-q}_1-\bi{q}_2) \, , \label{eq:action.reparam}
\end{eqnarray}
where we inserted the value of the bare effective couplings:
\begin{eqnarray}
  \mathring{g}_{1}= \frac{\mathring{U} \mathring{W}}{(4 \mathring{D}_{B} \pi)^{3}},  \qquad
  \mathring{g}_{D}:=\frac{\mathring{D}}{\mathring{D}_{B}},\qquad {\rm{and}} \nonumber \qquad
 \mathring{g}_{2}:= \frac{\mathring{U} \mathring{V}}{(4 \mathring{D}_{B} \pi)^{3}}. \label{eq:effectiv.coupling}
\end{eqnarray}
 In~\eref{eq:action.reparam}, we wrote the integral over a single momentum as ${\int_{\bi{q}}:=\int_{\mathbb{R}^{d+1}}   \rmd \omega \, \rmd^{d}\bi{k}} /(2 \pi)^{d+1}$, but we redefine the integral over the two momenta such that the factor $(4 \pi)^{3}$ appearing from the effective couplings is reabsorbed into the integral measure to lighter the notation: ${\int_{\bi{q}_1, \bi{q}_2 }:=~(4 \pi)^{3/2}\int_{\mathbb{R}^{2d+2}}(\rmd \omega_1 \rmd \omega_2 \rmd^{d}\bi{k}_1 \rmd^{d}\bi{k}_2)/(2\pi)^{2d+2}}$

The dimension of the effective couplings can be found imposing that the action is dimensionless. With respect to an arbitrary momentum scale $\mu$, this gives: 
\begin{eqnarray}
    [\mathring{g}_1]=[\mathring{g}_2]=\mu^{6-d} \qquad {\rm{and}} \qquad [\mathring{g}_D]=\mu^{0}
\end{eqnarray}
The effective couplings are irrelevant for dimensions greater than $d_c=6$ and therefore, flow to 0 under the RG flow. For dimensions smaller than $d_c=6$, instead,  they become relevant and get renormalized. Therefore, the upper critical dimension of action~\eref{eq:action.reparam} is ${d_c=6}$.
 
\subsection*{Reparametrization of the action and minimal subtraction scheme}
We proceed to rewrite the action in terms of renormalized quantities. Those are defined using the $Z$-factors:
\begin{eqnarray}
 \fl \mu^{\epsilon} Z_{g_1} g_1 := \mathring{ g_1}, \quad  Z^{\frac{1}{2}}_{B} B := \mathring{B}, \quad   
    \frac{Z_{\tau}}{\left( Z_{B}Z_{\tilde{B}}\right)^{\frac{1}{2}}}\tau  :=\mathring{\tau}, \quad \frac{\left( Z_{B}Z_{\tilde{B}}\right)^{\frac{1}{2}} }{Z_{D}} g_D := \mathring{ g}_D,   \nonumber \\
    \mu^{\epsilon} Z_{g_2} g_2 := \mathring{ g_2}, 
    \quad
  Z^{{\frac{1}{2}}}_{\tilde{B}} \tilde{B} :=\mathring{\tilde{B}},    \qquad  
     \frac{Z_{D}}{\left( Z_{B}Z_{\tilde{B}}\right)^{\frac{1}{2}} } D_{B}:= \mathring{D}_{B}. 
 \label{eq:Z.factor.definition}
\end{eqnarray}
Note that the renormalized couplings $g_D$ and $D_B$ are defined through the same $Z$-factor, as the diffusion constant $D$ does not renormalize, i.e., ${\mathring{D}=D}$.
Inserting these values in the action~\eref{eq:action.reparam}, we can split it in two parts ${\mathcal{S}=\mathcal{S}_{R}+\mathcal{S}_{\rm{c}}}$. 
The renormalized action $S_{R}$ has the same functional form as the bare one~\eref{eq:action.reparam}, containing renormalized couplings and fields. It reads:
\begin{eqnarray}
\fl \mathcal{S}_{R}=\int_\bi{q} \tilde{B} (-\bi{q})  \left( \rmi\omega+{D}_B k^{2}+\tau \right) 
B(\bi{q})-\int_{\bi{q}_1,\bi{q}_2}  \frac{(g_2)^{\frac{1}{2}} D_B}{2} \tilde{B}(-\bi{q}_1)\tilde{B}(-\bi{q}_2) B(\bi{q}_2+\bi{q}_1) \nonumber\\
\fl +\int_{\bi{q}_1,\bi{q}_2} \Bigg[  \frac{g_1 D^2_B}{(g_2)^{\frac{1}{2}}} \frac{B(\bi{q}_1)}{\rmi\omega_1} 
+ (g_2)^{\frac{1}{2}} D^2_B \frac{B(\bi{q}_1)}{\rmi\omega_1+g_D D_B k_1^{2}} \Bigg] B(\bi{q}_2)  \tilde{B} (\bi{-q}_1-\bi{q}_2). \label{eq:action.reparam.R}
\end{eqnarray}
The extra dependence on the momentum scale $\mu^{\epsilon}$ appearing in the effective couplings $g_1,g_2$ is reabsorbed in the integral measure of the interacting vertices to lighten the notation, i.e., $ {\int_{\bi{q}_1, \bi{q}_1 }:=~(4 \pi)^{3/2}\mu^{\epsilon/2}\int_{\mathbb{R}^{2d+2}}(\rmd \omega_1 \rmd\omega_2 \rmd^{d}\bi{k}_1 \rmd^{d}\bi{k}_2)/(2\pi)^{2d+2}}$. 
This notation will be consistently assumed for integrals over two momenta in what follows.
The action $\mathcal{S}_{\rm{c}}$ is made of ``counter-terms" containing the $Z$-factors:
\begin{eqnarray}
\fl \mathcal{S}_{\rm{c}}= \int_\bi{q} \tilde{B} (-\bi{q})\left( \big[(Z_{\tilde{B}}Z_{B})^{\frac{1}{2}}-1\big]\rmi\omega   
+\big[Z_{D}-1\big]D_B k^{2} +\big[Z_{\tau}-1\big] \tau_{R} \right) B_{R}(\bi{q}) 
 \nonumber\\  -\int_{\bi{q}_1,\bi{q}_2} \Big[Z^{\frac{1}{2}}_{g_2} Z_D Z^{\frac{1}{2}}_{\tilde{B}}-1\Big] \frac{(g_2)^{\frac{1}{2}} D_B}{2} \tilde{B}(-\bi{q}_1)\tilde{B}(-\bi{q}_2) B(\bi{q}_2+\bi{q}_1) \nonumber \\ 
 +\int_{\bi{q}_1,\bi{q}_2} \Big[Z^{-\frac{1}{2}}_{g_2}  Z_{g_1} Z^2_D Z^{-\frac{1}{2}}_{\tilde{B}}-1 \Big] \frac{g_1 D^2_B}{(g_2)^{\frac{1}{2}}\rmi\omega_1} B(\bi{q}_1)  B(\bi{q}_2)  \tilde{B} (\bi{-q}_1-\bi{q}_2) \nonumber \\
 + \int_{\bi{q}_1,\bi{q}_2}  \Big[Z^{\frac{1}{2}}_{g_2} Z^2_D Z^{-\frac{1}{2}}_{\tilde{B}}-1 \Big]\frac{(g_2)^{\frac{1}{2}} D^2_B }{\rmi\omega_1+g_D D_B k_1^{2}} B(\bi{q}_1)
  B(\bi{q}_2)  \tilde{B} (\bi{-q}_1-\bi{q}_2) \, . \nonumber
\end{eqnarray}
In the field-theoretical formalism, all physical observables can be expressed as expectation values of a product of fields~\cite{ZinnJustin:2002}. 
These expectation values correspond to $n$-point correlation functions, or equivalently, vertex functions $\Gamma^{(m,n)}$~\cite{Tauber:2014}. 
The $Z$-factors are defined such that the counter-terms of action $\mathcal{S}_c$ cancel the divergent contributions of the vertex functions $\Gamma_R$ calculated from the action $\mathcal{S}_R$. 
This procedure ensures that observables are divergence-free.

Vertex functions are calculated through a diagrammatic expansion over one-particle irreducible (1PI) diagrams~\cite{Weinberg:1995}, organized by the number $n$ of internal loops of a diagram (or, equivalently, in terms of the deviation from the upper critical dimension $\epsilon^n$). 
The $Z$-factors are perturbatively defined such that the (${n-1}$)-loop order expansion in $\mathcal{S}_{\rm{c}}$ cancels the $n$-loop divergences of the renormalized vertex functions $\Gamma_R$. 
In our case, only $\Gamma_{R}^{(1,1)}$, $\Gamma_{R}^{(2,1)}$ and $\Gamma_{R}^{(1,2)}$ contain divergences. So, the $Z$-factors are determined through the following conditions:  
\begin{eqnarray}
\label{eq:Z.factors.condition} 
 \fl \mathcal{O}(\epsilon^0) \overset{!}{=} \frac{(g_2)^{\frac{1}{2}} D_B( 1-(Z_{g_2}Z^{-1}_{\tilde{B}})^{\frac{1}{2}} Z^2_D  )}{\rmi\omega_1+g_D D_B k_1^2} + 
 \frac{(1- Z^2_D Z_{g_1}(Z_{\tilde{B}}Z_{g_2})^{-\frac{1}{2}} )}{\rmi\omega_1 } \frac{g_1 D_B}{(g_2)^{\frac{1}{2}}} -\Gamma_R^{(2,1)},\label{eq:Zfactorscondition1} \\
\fl \mathcal{O}(\epsilon^0) \overset{!}{=} \left( 1-Z_{\tau} \right) \tau_{R}-(Z_{D}-1)D_{B} k^{2}- \left( (Z_{\tilde{B}}Z_{B})^{{\frac{1}{2}}}-1\right)\rmi\omega-\Gamma_R^{(1,1)},  \label{eq:Zfactorscondition2}  \\
\fl \mathcal{O}(\epsilon^0) \overset{!}{=} \left( Z^{\frac{1}{2}}_{g_2} Z_D Z^{\frac{1}{2}}_{\tilde{B}}-1 \right) (g_2)^{\frac{1}{2}} D^2_B - \Gamma_R^{(1,2)} .\label{eq:Zfactorscondition3}
\end{eqnarray}
To lighten notation, we will consistently use vertex functions rescaled by a factor $\mu^{ \epsilon/2}(4 \pi)^{3/2}$ to compensate the same factor previously reabsorbed in the integral measure of $\int_{\bi{q}_1,\bi{q}_2}$.
Note that the factor $2$ in front of the vertex~$ ((g_2)^{1/2} D_B)/2$ arises from combinatorial factors.

\subsection*{Calculation of the integral of diagram (A)}
Next, we give an example of how to evaluate the divergent part of a Feynman diagram needed to obtain the $Z$-factors. We choose to evaluate diagram (A) since it is responsible for creating an asymmetry in the RG flow functions of $\beta_{g_1}$, and $\beta_{g_2}$~\eref{eq:flow.function.CGEP1}--\eref{eq:flow.function.CGEP3}, making the RG flow approach the new DGEP fixed point. 

\Fref{fig:figureB1} reports the Feynman rules for the renormalized action~\eref{eq:action.reparam.R} together with the diagram (A), which contributes to the vertex function $\Gamma_R^{(2,1)}$. In the diagram, the external frequencies are denoted with $(\omega,\bi{k})$ and the internal momentum over which to integrate by $(\omega_{p},\bi{p})$. The integral corresponding to diagram (A) reads:  
\begin{eqnarray}
\fl (A)= (D_B)^4 \mu^{\frac{3 \epsilon}{2}} (4 \pi)^{\frac{9}{2}} \int \frac{\rmd^{d} \bi{p}}{(2 \pi)^{d}} \frac{\rmd w_{p}}{(2 \pi)}
    \frac{1}{\rmi(w_{2}+w_{p})+D_{B}| \bi{k}_{1}+\bi{p}|^{2}+\tau} \frac{1}{\rmi w_{p}+D_B k^2+\tau} \times           \nonumber\\ 
   \fl \frac{1}{\rmi(w_{1}-w_{p})+D_{B}|\bi{k}_{2}-\bi{p}|^{2}+\tau} 
    \left( \frac{g_1 }{\rmi(g_2)^{\frac{1}{2}}(w_{1}-w_{p})}+ \frac{(g_2)^{\frac{1}{2}}}{\rmi(w_{1}-w_{p})+D|\bi{k}_{1}-\bi{p}|^{2}} \right) \times \nonumber \\
    \left(\frac{g_1 }{\rmi w_{p}}+ \frac{g_2 }{\rmi w_{p}+D p^2} \right).\label{eq:diagram.A}
    \end{eqnarray}
We inserted ${D=g_D D_B}$ to lighten the notation.
Note that the factor $\mu^{3 \epsilon/2} (4 \pi)^{9/2}$ was previously reabsorbed in the measure for integrals over two momenta.
We will show that the only divergence arising from this integral is renormalizing the vertex proportional to $g_1D^2_B\left((g_2)^{1/2}\rmi\omega_1\right)^{-1}$ in~\eref{eq:action.reparam}. 
Therefore, in the following calculation, $\omega_1$ is kept non-zero as it explicitly enters the vertex structure, which we aim to renormalize. At the same time, we can set the other external momenta to zero---i.e., ${\omega_2=\bi{k}_1=\bi{k}_2=0}$---since the divergences cannot depend on their values. 

We first evaluate the integral over $\omega_p$ by using the residue theorem in the complex plane. 
We close the contour in the lower half of the complex plane, where there are three poles $z_i$:
${z_{1} = \omega_1}$, ${z_{2}= \omega_1-\rmi D p^2}$ and ${z_{3}= \omega_1- \rmi (D_B p^2+\tau)}$. We use a forward (It\^{o}) prescription where the pole $z_1$ is included in the lower half plane through the shift ${z_1= \lim_{\delta \rightarrow 0}\omega_1-\rmi \delta}$~\cite{Tauber:2014}.

The integral over momentum $ \bi{p}$ is evaluated just below the upper critical dimension in $d=6-\epsilon$ dimensions with a finite diffusion constant $D\neq 0$. The only divergent contribution of~\eref{eq:diagram.A} in the limit $\epsilon \rightarrow 0$ is the  residue of $z_1$ inserted in the term proportional to $g_1(g_2)^{1/2}$ of~\eref{eq:diagram.A}, which gives rise to the integral $I(D=0)$ with $I(D)$ defined as:
   \begin{eqnarray}
  \fl  I(D)=  \int \frac{\rmd^{d} \bi{p}}{(2\pi)^d} 
    \frac{1}{(D_B-D) \bi{p}^2 +\tau} \, \frac{1}{\big[ \rmi\omega_1 +(D_B+D)\bi{p}^2 +\tau \big]^2} \, \frac{1}{\rmi\omega_1 +2 D \bi{p}^2} \label{eq:niceint}
     \end{eqnarray}
To better understand the limit $D \rightarrow 0$, we also consider the contribution of the residue of $z_2$ inserted in the term proportional to $g_2^{3/2} $, giving rise to the integral $I(D)$. 
Summarizing, diagram (A) gives, among others, the two contributions: 
\begin{eqnarray}
 \fl  \mu^{\frac{3 \epsilon}{2}} (4 \pi)^{\frac{9}{2}}g_2^{\frac{3}{2}} D^4_{B}I(D) \quad {\rm{and}} \qquad \mu^{\frac{3 \epsilon}{2}} (4 \pi)^{\frac{9}{2}}g_1^2g_2^{-\frac{1}{2}} D^4_B I(D=0).
\end{eqnarray}
In the main text, we referred to these contributions as $V D^3_B g_2 I(D)$ and $W D^3_B g_1 I(0)$, as these are the value they would assume in the old parametrization of the action~\eref{eq:action}.
Note that the first term, if divergent, could renormalize the vertex proportional to $(g_2)^{1/2}D_B$ in the second line of~\eref{eq:action.reparam.R} (in which case setting $k_1=0$ would not be justified). In contrast, the second renormalizes the vertex proportional to  $(g_1)^2/(g_2)^{1/2}$. By simple power counting, we expect only $I(0)$ to be divergent so that the two vertices in the second line of~\eref{eq:action.reparam.R} renormalize differently, leading to asymmetric flow functions for $g_1$ and $g_2$. In the following, we evaluate $I(D)$ and $I(0)$ to prove that only the latter is relevant to the renormalization and to illustrate how such integrals are performed.
 We achieve this by first collecting the diffusion constant appearing in the propagators and then using identity~\eref{eq:identiti2}, known as the \textit{Feynman trick}, to bring the integrals in the following form:
\begin{eqnarray}
\fl   I(0)=\frac{\Gamma(3)}{\rmi\omega_1 D^3_B } \int_{0}^{1} \rmd x \, x \int \frac{\rmd^{6-\epsilon} \bi{p}}{(2 \pi)^{d}}  \left( \bi{p}^2+(1-x)\frac{\tau}{D_B}+x\frac{\rmi\omega_1+\tau}{D_B}\right)^{-3} \\
\fl I(D)=\frac{\Gamma(4)}{D(D_B-D)(D_B+D)^2} \times \nonumber\\ \int_0^1 \rmd x  \rmd y \,x \int\frac{\rmd^{6-\epsilon} \bi{p}}{(2 \pi)^{d}} \left( p^2+ \frac{(1-x-y)\tau}{(D_B-D)} + x\frac{\rmi\omega_1+\tau}{D_B+D}+y\frac{\rmi\omega_1}{D}\right)^{-4}
\end{eqnarray}
The integral over momentum $\bi{p}$ can now be expressed in function of the Euler-Gamma function $\Gamma$---the analytical continuation of the factorial function---using identity~\eref{eq:identiti1}:
\begin{eqnarray}
  I(0)=&\frac{1}{\rmi\omega_1 D^3_B } \int_{0}^{1} \rmd x \, x   \frac{\Gamma(\frac{\epsilon}{2}) }{(4 \pi )^{3}} \left((1-x)\frac{\tau}{D_B}+x\frac{\rmi\omega_1+\tau}{D_B}\right)^{-\frac{\epsilon}{2}}
 \label{eq:value.of.I1}\\
 I(D)=&\frac{1}{D(D_B-D)(D_B+D)^2}\int_{0}^{1} \rmd x \rmd y \, x \frac{\Gamma(1+\frac{\epsilon}{2})}{(4\pi )^3} \times \nonumber \\ &\left(  (1-x-y)\frac{\tau}{(D_B-D)} +x \frac{\rmi\omega_1+\tau}{D_B+D}+y\frac{\rmi\omega_1}{D}\right)^{-1-\frac{\epsilon}{2}}
\end{eqnarray}
To obtain the leading order of the integrals $I(0)$ and $I(D)$, we expand in $\epsilon$. In particular, we use
\begin{eqnarray}
    \Gamma(\epsilon)=\frac{1}{\epsilon}+\mathcal{O}(\epsilon^0)\qquad  
{\rm{and}} \qquad \Gamma(1+\epsilon)=\mathcal{O}(\epsilon^0) \nonumber
\end{eqnarray}
such that the divergent part of $I(0)$ in the limit $\epsilon \rightarrow 0$ can be expressed as a $1/ \epsilon$ pole.
 \begin{eqnarray}
  I(0)=&\frac{1}{\rmi\omega_1 D^3_B } \int_{0}^{1} \rmd x \, x   \frac{2 }{\epsilon (4 \pi )^{3}}= \frac{1}{\rmi\omega_1 D^3_B  (4 \pi )^{3}} \frac{1}{\epsilon} 
 \label{eq:value.of.I}\\
 I(D)=&\frac{1}{D(D_B-D)(D_B+D)^2} \int_{0}^{1} \rmd x \rmd y \frac{x}{(4\pi )^3} \times \nonumber \\ 
 &\left(  (1-x-y)\frac{\tau}{(D_B-D)} +x \frac{\rmi\omega_1+\tau}{D_B+D}+y\frac{\rmi\omega_1}{D}\right)^{-1}
\end{eqnarray}
The integral over $x,y$ in $I(D)$ is finite as long as $D\neq0$, as was claimed initially. In the limit $D \rightarrow 0$, the integral over $x,y$ scales as $D \ln(D)$, implying that the integral $I(D \rightarrow 0) \propto \lim_{D\rightarrow 0}\ln(D)$ becomes divergent. On the other hand, the divergent part of $I(0)$ is parameterized in terms of a $1/\epsilon$ pole. This way, we showed that the $1/\epsilon $ pole corresponds to a logarithmic divergence $\lim_{D\rightarrow 0}\ln(D)$ if the limit $D\rightarrow 0$ is taken after the integral is performed. Overall, the only divergent contribution of (A) to the vertex functions in the limit $\epsilon \rightarrow0$ is reported below. To account for the right dimensionality, we multiply by the factor $\mu^{d-d_c}$ to fix the dimensionality of the integral $I(0)$ in~\eref{eq:value.of.I}.
\begin{eqnarray}
   \mu^{3 \epsilon/2} (4 \pi)^{\frac{9}{2}} \frac{(g_1)^2}{(g_2)^{\frac{1}{2}}} D^4_B \mu^{-\epsilon}I(0)= \mu^{ \epsilon/2}(4 \pi)^{\frac{3}{2}} \frac{(g_1)^2 D_B }{(g_2)^{\frac{1}{2}}}\frac{1}{\rmi\omega_1 \epsilon}
\end{eqnarray}

\subsection*{$Z$-factors and flow functions}
\begin{figure}
    \centering
\includegraphics[width=1.02\columnwidth]{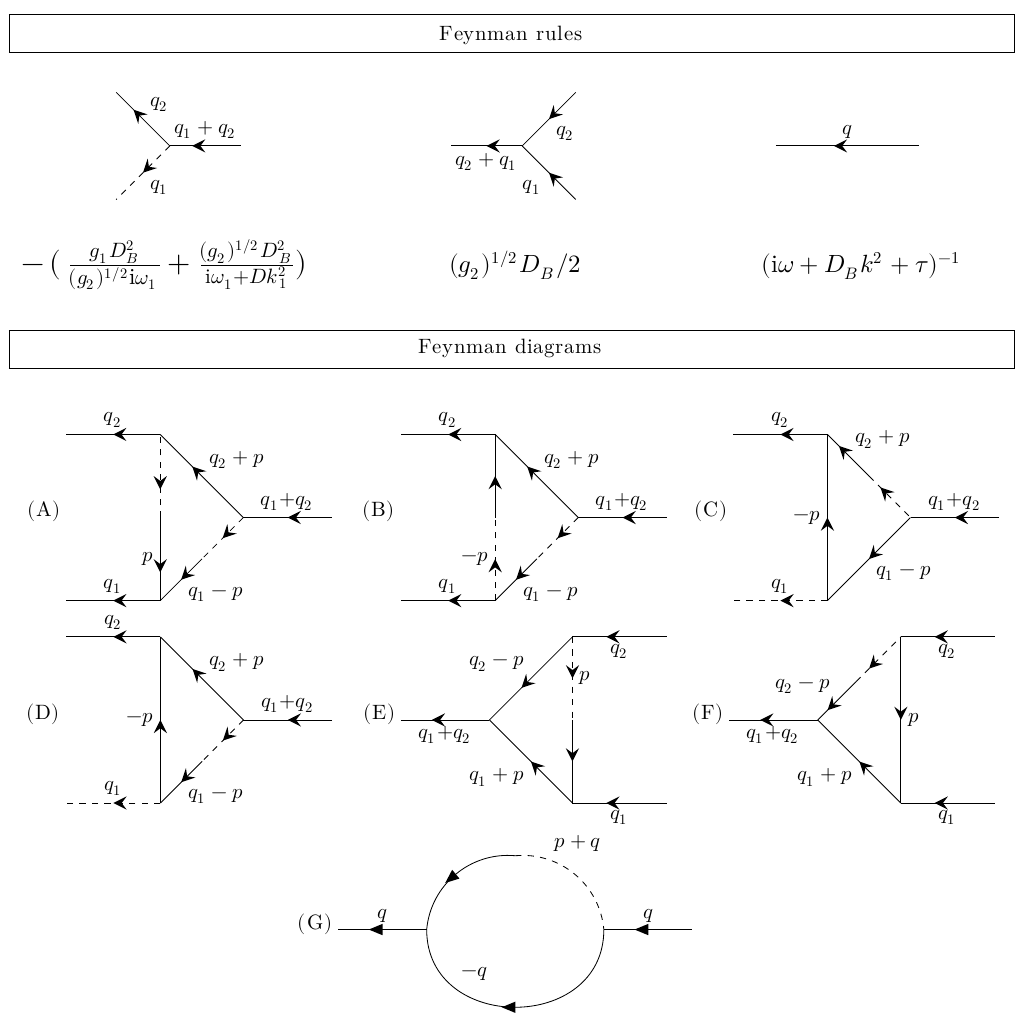}
    \caption{ \textbf{Feynman rules and diagrams}. The Feynman rules and one-loop diagrams contributing to $\Gamma_R^{(2,1)}$ (diagrams (A),(B),(C) and (D)), $\Gamma_R^{(1,2)}$ (diagrams (E) and (F)) and $\Gamma_R^{(1,1)}$ (diagram (G)).}
    \label{fig:figureB1}
\end{figure}
\Fref{fig:figureB1} shows all Feynman diagrams contributing to the vertex functions  $\Gamma_{R}^{(1,1)}$, $\Gamma_{R}^{(2,1)}$ and $\Gamma_{R}^{(1,2)}$. We evaluate the divergent part of those following the same procedure as was done for diagram (A), i.e., the integrals are evaluated at dimension ${d=d_c-\epsilon}$ and solved with the help of the two identities~\eref{eq:identiti1} and~\eref{eq:identiti2}. 
In the limit ${\epsilon \rightarrow 0}$, the divergent contributions appear as poles proportional to $1/\epsilon$.  
The divergent contribution of the Feynman diagrams shown in figure~\ref{fig:figureB1} are reported below: 
\begin{eqnarray}
  (A)=& \mu^{ \frac{\epsilon}{2}}(4 \pi)^{\frac{3}{2}}\frac{g_1 D^2_B}{(g_2)^{\frac{1}{2}} \rmi\omega_1} \frac{g_1}{ \epsilon}\nonumber \\
  (B)=0  \nonumber \\
(C)=&\mu^{ \frac{\epsilon}{2}}(4 \pi)^{\frac{3}{2}} \left( \frac{(g_2)^{\frac{1}{2}} D^2_B} {\rmi\omega_1+ D k_1^2} + \frac{g_1 D^2_B}{(g_2)^{\frac{1}{2}} \rmi\omega_1} \right) \left( \frac{g_1}{\epsilon} \frac{3}{4}+ \frac{g_2}{\epsilon} \frac{3+g_{D}}{4(1+g_{D})^2} \right)\nonumber
 \\
 (D)=&\mu^{ \frac{\epsilon}{2}}(4 \pi)^{\frac{3}{2}}\left( \frac{(g_2)^{\frac{1}{2}} D^2_B} {\rmi\omega_1+ D k_1^2} + \frac{g_1 D^2_B}{(g_2)^{\frac{1}{2}} \rmi\omega_1} \right)  \left( \frac{g_1}{\epsilon}\frac{1}{4}+\frac{g_2}{\epsilon}\frac{1}{4(1+g_{D})} \right) \nonumber
  \\
 (E)=&-\mu^{ \frac{\epsilon}{2}}(4 \pi)^{\frac{3}{2}}(g_2)^{\frac{1}{2}} D_B \left( \frac{g_1}{\epsilon}\frac{1}{4}+\frac{g_2}{\epsilon}\frac{1}{4(1+g_{D})}  \right) \nonumber
\\
   (F)=&- \mu^{ \frac{\epsilon}{2}}(4 \pi)^{\frac{3}{2}}(g_2)^{\frac{1}{2}} D_B\left( \frac{g_1}{\epsilon}\frac{3}{4}+\frac{g_2}{\epsilon}\frac{3+g_{D}}{4(1+g_{D})^2}  \right) \nonumber \\
 (G) =& \mu^{ \frac{\epsilon}{2}}(4 \pi)^{\frac{3}{2}} \Big[\frac{g_1}{\epsilon}\left(\frac{3}{4}\rmi\omega+\tau_R+D_B \frac{k^2}{6}\right)  + \nonumber \\ &\frac{g_2}{\epsilon}\left( \frac{3+g_{D}}{4(1+g_{D})^2}\rmi\omega+\frac{2+g_{D}}{2(1+g_{D})^2} \tau_R + \frac{1+4g_{D}+g_{D}^2}{6(1+g_D)^3}D_B k^2 \right) \Big]\nonumber
\end{eqnarray}
In the above equations, each divergent contribution, i.e., each $1/\epsilon$ pole, proportional to $g_2$ becomes equal to the one proportional to $g_1$ in the limit of ${g_D=0}$, except for the diagram ($A$), where only the term $I(0)$ proportional to $g_1$ is present. 

To obtain the divergent part of the vertex functions, we sum up the corresponding contributions of the Feynman diagrams.
As we mentioned before, we rescale the vertex function by $\mu^{ \epsilon/2}(4 \pi)^{3/2}$ to be consistent with the choice of the integral measure $\int_{\bi{q}_1,\bi{q}_2}$.
\begin{eqnarray}
    \fl \Gamma_{R}^{(2,1)} =&-(C)-(D)-(A)= -\frac{ g_1 D_B}{(g_2)^{\frac{1}{2}} \rmi\omega_1} \left(\frac{2 g_1}{\epsilon}+\frac{g_2}{\epsilon} \frac{g_{D}+2}{2(g_{D}+1)^2} \right) \nonumber \\
   \fl  &-  \frac{ (g_2)^{\frac{1}{2}} D_B }{\rmi\omega_1+ g_D D_B k_1^2} \left( \frac{g_1} {\epsilon}+ \frac{g_2} {\epsilon} \frac{g_{D}+2}{2(g_{D}+1)^2} \right),
    \\
  \fl \Gamma_{R}^{(1,2)} =&-2(E)-2(F)=
  (g_2)^{\frac{1}{2}} D^2_B \left( \frac{2g_1}{\epsilon} + \frac{g_2}{\epsilon}   \frac{g_{D}+2}{(1+g_{D})^2} \right),   \\ 
    \fl \Gamma_{R}^{(1,1)} =&-(G)=
   -\frac{g_1}{\epsilon}\left(\frac{3}{4}\rmi\omega+\tau_R+D_B \frac{k^2}{6}\right) \nonumber \\ \fl &+\frac{g_2}{\epsilon}\left( \frac{3+g_{D}}{4(1+g_{D})^2}\rmi\omega+\frac{2+g_{D}}{2(1+g_{D})^2} \tau_R + \frac{1+4g_{D}+g_{D}^2}{6(1+g_D)^3}D_B k^2) \right) \, ,  
    \end{eqnarray}
where the $2$ in front of (E) and (F) in $\Gamma_{R}^{(1,2)}$ is due to a combinatorial factor.
To find the $Z$-factors, we employ the conditions given by the minimal subtraction scheme~\eref{eq:Zfactorscondition1}--\eref{eq:Zfactorscondition3}. This yields:
\begin{eqnarray}
\fl Z_{g_{1}}=1+\frac{1}{\epsilon}\left(\frac{7}{2}g_{1}+g_{2}\frac{5+5g_{D}+2g_{D}^2}{2(1+g_{D})^3}\right) \qquad 
 &Z_{D} =1+\frac{1}{\epsilon}\left(\frac{g_{1}}{6}+g_{2}\frac{1+4g_{D}+g_{D}^2}{6(1+g_{D})^3}\right)  \nonumber\\
 \fl Z_{g_{2}}=1+\frac{1}{\epsilon}\left(\frac{5}{2}g_{1}+g_{2}\frac{5+5g_{D}+2g_{D}^2}{2(1+g_{D})^3}\right) 
\qquad &Z_{\tau}=1+\frac{1}{\epsilon}\left( g_{1}+g_{2}\frac{2+g_{D}}{2(1+g_{D})^2} \right) 
  \nonumber\\
   \fl Z_{\tilde{B}} =1+\frac{1}{\epsilon} \left( \frac{7}{6}g_1+ g_2{\epsilon}\frac{7+13g_{D}+4g_{D}^2}{6(1+g_{D})^3}\right)  \qquad 
    &Z_{B}=1+\frac{1}{\epsilon} \left(\frac{1}{3} g_1 +\frac{g_2}{\epsilon}\frac{2-g_{D}-g_{D}^2}{6(1+g_{D})^3}\right) \nonumber
\end{eqnarray}
Having calculated the value of the $Z$-factors, we are finally in the position to calculate the flow functions for the general action~\eref{eq:action}. To do so, we rewrite the effective couplings in terms of $Z$-factors and bare couplings and exploit the fact that the bare couplings are independent of the momentum scale $\mu$. Therefore, the derivatives of the bare parameters with respect to $\mu$ have to be zero, e.g.,  $\frac{d}{d \mu} \mathring{g}_2=0$. 
The flow functions at one-loop order read: 
\begin{eqnarray}
 \fl \beta_{g_{1}}:=\mu \frac{d}{d\mu}g_{1}=\mu \frac{d}{d\mu} \left(  \mu^{-\epsilon} Z^{-1}_{g_1} \mathring{g}_{1}\right)=-\epsilon g_{1}+g_{1}\left(\frac{7}{2}g_{1}+g_{2}\frac{5+5g_{D}+2g_{D}^2}{2(1+g_{D})^3} \right)\label{eq:flowfunctions1C}\\
 \fl \beta_{g_{2}}:=\mu \frac{d}{d\mu}g_{2}=\mu \frac{d}{d\mu} \left(  \mu^{-\epsilon}  Z^{-1}_{g_2} \mathring{g}_2\right)=-\epsilon g_{1}+g_{1}\left(\frac{5}{2}g_{1}+g_{2}\frac{5+5g_{D}+2g_{D}^2}{2(1+g_{D})^3}\right)\label{eq:flowfunctions2C}\\
 \fl \beta_{D}:=\mu \frac{d}{d\mu}g_D=\mu \frac{d}{d\mu} \left(    \frac{\left( Z_{B}Z_{\tilde{B}}\right)^{\frac{1}{2}}}{Z_D} \frac{D}{ \mathring{D}_B } \right)=-g_{D}\left(\frac{7}{12}g_{1}+g_{2}\frac{7+4g_{D}+g_{D}^2}{12(1+g_{D})^3}\right) \label{eq:flowfunctions3C}
\end{eqnarray}
A fixed point in the flow function of the effective couplings $\beta_{g_1},\beta_{g_2},\beta_{g_D}$ represents a parameter regime where the system is scale-invariant, as expected from a second-order phase transition. The above flow function posses three different fixed points:  GEP fixed point (${g_1^*=2\epsilon/7,g^*_2=g^*_D=0}$), the  DGEP fixed point (${g_1^*=0,g^*_2=2 \epsilon /5,g^*_D=0}$) and the gaussian fixed line (${g_1^*=g^*_2=0,g^*_D}$) depicted in figure~\ref{fig:3} in the main text. 

It appears unusual that, at the GEP fixed point, the vertex $g_1 / (g_2)^{1/2}$ diverges. This divergence is a consequence of the selected parametrization and does not relate to physical divergences. The problem can be overcome by changing the parametrization of the action so that no divergences arise at the GEP fixed point. One possibility is: 
\begin{eqnarray}
     \mathring{B}\rightarrow \left( \mathring{U} \mathring{D}_B \mathring{W}^{-1} \right)^{\frac{1}{2}} \mathring{B}, \quad \mathring{\tilde{B}} \rightarrow \left( \mathring{U} \mathring{D}_B \mathring{W}^{-1} \right)^{-\frac{1}{2}} \mathring{\tilde{B}} . \nonumber
\end{eqnarray}

\subsection*{Standard integrals}
Lastly, we report two standard identities, which were used to evaluate the divergent part of integrals (A)--(G) through dimensional regularization
\begin{eqnarray}
  \fl  \int_{-\infty}^{\infty} \frac{\rmd^{d} q}{(2 \pi)^{d}} \frac{q^{n}}{(q^2+\tau)^{a}}=\frac{1}{(4 \pi)^{\frac{d}{2}}\Gamma(\frac{d}{2})} \frac{\Gamma(\frac{d+n}{2})\Gamma(a-\frac{d+n}{2})}{\Gamma(a)} \tau^{\frac{d+n}{2}-a} \label{eq:identiti1}
\end{eqnarray}
\begin{eqnarray}
  \fl  \frac{1}{P_1^{a_1}...P_n^{a_n}}=\frac{\Gamma(a_1+...+a_n)}{\Gamma(a_1)...\Gamma(a_n)} \int_{0}^{1} \rmd x_1...\rmd x_n \left( \frac{\delta(1-x_1...-x_n)x_1^{a_1-1}...x_n^{a_n-1}}{(x_1 P_1+...+x_n P_n)^{a_1+...+a_n}}\right) \label{eq:identiti2}
\end{eqnarray}
where $\Gamma$ stands for the Euler-Gamma function, an analytical continuation of the factorial function.

\section{Critical exponents}
  This section aims to derive the value of the critical exponents $z,\nu_{\parallel},\theta,\delta$ describing the scaling of the number of infected individuals, the survival probability, and the mean-squared radius:
  \begin{eqnarray}
    \langle N \rangle \left(t,\sigma \right) 
    &=& t^{\theta} \hat{N}\left( (\sigma-\sigma_c)^{\nu_{\parallel}} t \right), 
    \label{eq:observable1C}\\
    P_{\rm s}(t,\sigma) 
    &=& t^{-\delta} \hat{P}_{\rm s}\left((\sigma-\sigma_c)^{\nu_{\parallel}} t\right) \, , 
    \label{eq:observable2C}
    \\
    \langle R^2 \rangle (t,\sigma) 
    &=& 
    t^{{2}/{z}} \hat{R}^2\left((\sigma-\sigma_c)^{\nu_{\parallel}} t\right)
    \, .
    \label{eq:observable3C}
\end{eqnarray}
 While the $Z$-factors of the effective couplings $(Z_{g_2}, Z_{g_1},Z_{g_D})$ determine the RG flow functions and fixed points, the other $Z$-factors $(Z_{B}, Z_{\tilde{B}}, Z_{\tau}, Z_{D})$ encode the scaling correction of the renormalized parameters as compared to the bare ones and determine the value of the critical exponents. 
 
\subsection*{Anomalous dimension}
By imposing that the action~\eref{eq:action.reparam.R} is dimensionless, the dimension of the bare parameters can be obtained:
\begin{eqnarray}
     [\mathring{B}]=\mu^{(d+2)/2} \quad [\mathring{\tilde{B}}]=\mu^{(d-2)/2} \quad 
     [\mathring{\tau}]=\mu^{2} \quad [\mathring{D}_B]=\mu^{0}
\end{eqnarray}
where dimensionality is measured with respect to the momentum scale $\mu$.
Close to criticality, the scaling of physical parameters, i.e., the renormalized parameters, is not determined only by dimensional analysis; they acquire non-trivial correction termed \textit{anomalous} dimension. To account for these corrections, we give a fictional dimensionality to the $Z$-factors---the anomalous dimensions $W_D, W_\tau, \sigma $ and $\tilde{\sigma}$---:
\begin{eqnarray}
  \fl  Z_{D}:=\mu^{W_D}\qquad  Z_{\tau}:=\mu^{W_\tau} \qquad  Z_{B}:=\mu^{\sigma} \qquad  Z_{\tilde{B}}:=\mu^{\tilde{\sigma}} \label{eq:dimensionZ.factor}
\end{eqnarray}
This way, the scaling of any renormalized parameter is given in terms of dimensional analysis and the corresponding RG corrections, encoded in the scaling of the $Z$ factors~\eref{eq:dimensionZ.factor}. This is done using the definitions for the renormalized parameters~\eref{eq:Z.factor.definition}, such that the correct scaling for the physical parameters is recovered.
To obtain the anomalous dimensions $(W_D,W_{\tau},\sigma,\tilde{\sigma})$, we take the derivative of the $Z$-factors with respect to $\ln(\mu)$, ,i.e., $\mu \frac{ d }{d \mu} Z$. As the $Z$-factors depend only on effective couplings, the derivative $\mu \frac{d}{d \mu}$ can be evaluated using the flow functions~\eref{eq:flowfunctions1C}--\eref{eq:flowfunctions3C}. At a fixed point, the effective couplings stop to flow, and the anomalous dimensions~\eref{eq:dimensionZ.factor} approach a finite constant value, encoding the RG correction to the parameters' scaling. Theories flowing to the same RG fixed point are characterized by parameters with the same scaling behavior, leading to the concept of universality class. Below are calculated the anomalous dimensions at the DGEP ($g^*_2=2\epsilon /5, g^*_1=g^*_D=0$) fixed point: 
\begin{eqnarray}
   W^*_D &= \mu \frac{d}{d \mu} \ln Z_D \bigg|_{FP}= g^*_{2}\frac{1+4g^*_{D}+g_{D}^{*2}}{6(1+g^*_{D})^3}=\frac{1}{15} \epsilon \nonumber\\
    W^*_{\tau} &= \mu \frac{d}{d \mu}Z_{\tau} \bigg|_{FP} =-g^*_{R}\frac{2+D^*}{2(1+D^2)^2}=-\frac{5}{2} \epsilon \nonumber\\
  \sigma^* &= \mu \frac{d}{d \mu } Z_B \bigg|_{FP} = -g^* \frac{2-D^*-(D^*)^2}{6(1+D^*)^3}=-\frac{2}{15} \epsilon \hspace{0,8cm} \nonumber \\ 
  \tilde{\sigma}^* &= \mu \frac{d}{d \mu } Z_{\tilde{B}}\bigg|_{FP}=- g_2^* \frac{7+13g_{D}^*+4g_{D}^{*2}}{6(1+g_{D}^*)^3}= -\frac{7}{15}\epsilon \hspace{0,3cm} \nonumber 
\end{eqnarray}
where the asterisk denotes that the anomalous dimensions
are evaluated at the DGEP fixed point.
From the knowledge of the scaling correction from the $Z$-factors, we calculate the critical scaling of the renormalized fields and parameters, inferred from~\eref{eq:Z.factor.definition}:
\begin{eqnarray}
     \lbrack B\rbrack =[\mathring{B}]Z_{B}^{-\frac{1}{2}}=\mu^{\frac{d+2-\sigma^{*}}{2}} \label{appendixC:scaling1} \\  
 \lbrack \tilde{B} \rbrack=
    [\mathring{\tilde{B}}]Z_{\tilde{B}}^{-\frac{1}{2}}
    =\mu^{\frac{d-2-\tilde{\sigma}^{*}}{2}} \label{appendixC:scaling2} \\ 
    \lbrack\tau \rbrack=[\mathring{\tau}]
    Z_\tau^{-1}Z_B^{\frac{1}{2}}
    Z_{\tilde{B}}^{\frac{1}{2}}
    =\mu^{2-W^*_{\tau}+\frac{\sigma+\tilde{\sigma}}{2}} \label{appendixC:scaling3}\\
 \lbrack D_B \rbrack=[\mathring{D}_B]Z_D^{-1}Z_B^{\frac{1}{2}}Z_{\tilde{B}}^{\frac{1}{2}}=\mu^{-W^{*}_D+\frac{\sigma+\tilde{\sigma}}{2}} \label{appendixC:scaling4}
\end{eqnarray}
\subsection*{Scaling of Green's functions}
At criticality, the scaling of the renormalized parameters changes the scaling exponents of the observables~\eref{eq:observable1C}--\eref{eq:observable3C} as compared with their scaling away from criticality, given by simple dimensional analysis. Using the field-theoretical formalism, any physical observable can be expressed as an expectation value of fields. Those expectation values are the connected Green functions $G^{(m,n)}_R$ of the theory:
\begin{eqnarray}
    G^{(m,n)}_R(\tau,\{\bi{q}_i\},{g_i})=\Bigl \langle B(\bi{q}_1)...B(\bi{q}_m) \tilde{B}(\bi{q}_{m+1})...\tilde{B}(\bi{q}_{m+n}) \Bigr \rangle
 \nonumber
\end{eqnarray}
where $\langle \cdot \rangle$ denotes a path integral expectation value weighted with the action $S_R$. 
The scaling of the Green functions can be found using the method of characteristics and equations~\eref{appendixC:scaling1}--\eref{appendixC:scaling4}
This procedure is called the Callan-Symanzik formalism~\cite{Tauber:2014} and allows the expression of the scaling of the renormalized Green functions at different momentum scales $\mu$. 
Here we state the final expression for $\mu \rightarrow 0$, i.e., the IR critical regime:
\begin{eqnarray}
  \ G^{(m,n)}_R(\tau,\bi{r},t,\{g_{i}\})= ( t)^{-n\frac{d-2+\tilde{\sigma}^*}{2z}-m\frac{d+2+\sigma}{2z}}\Phi^{(m,n)} \Bigl(\frac{\bi{x}}{\xi} , \frac{t}{\xi^{z}} \Bigr) \label{eq:Green.function.scaling}
\end{eqnarray}
where $\Phi^{(m,n)}$ is a generic non-linear scaling function just like $\hat{P_s}, \hat{N}, \langle \hat{R}^2 \rangle  $ in~\eref{eq:observable1C}--\eref{eq:observable3C}. The correlation length $\xi$ and the critical exponent $z$ appear in its argument.
On the left of the expression, the power of time comes from the scaling of the renormalized fields $B_R$ and $\tilde{B}$, given by~\eref{appendixC:scaling1} and~\eref{appendixC:scaling2}.
Here, we will present a heuristic argument of how the scaling of $\xi$ and the critical exponent $z$ can be found. For a formal derivation, please refer to~\cite{Tauber:2014}. 

Requiring that all terms in the propagator of the field $B$ have the same dimension, e.g. $[D_B k^2]=[\omega]$, and using \eref{appendixC:scaling1}--\eref{appendixC:scaling3}, one can infer
\begin{eqnarray}
[\omega]:= \mu^{z}\qquad \rm{with}\qquad  z=2-W^{*}_D+\frac{\sigma^*+\tilde{\sigma}^*}{2}.   
\end{eqnarray}
A similar argument can be made for the scaling of the correlation length $\xi$. The correlation length of the mean-field system can be expressed in terms of bare parameter (by simple dimensional analysis): $\mathring{\xi}=(\mathring{\tau}/\mathring{D}_B)^{1/2}$. 
 Assuming that the scaling of $\xi$ is corrected by \textit{anomalous} dimensions, while its functional form remains the same at criticality, we write $\xi=(\tau / D_B)^{\nu}$ where $\nu$ is the critical exponent. One can determine the value $\nu$ by requiring that the renormalized correlation length has the right dimension: 
\begin{eqnarray}
[\tau/ D_B]^{\nu}=\mu^{-1} \rightarrow \nu=(2+W_D-W_{\tau})^{-1}.   
\end{eqnarray}
\subsection*{Scaling of physical observables}
To complete the derivation of the scaling form of the observables~\eref{eq:observable1C}--\eref{eq:observable3C}, we have to express them in terms of $G^{(m,n)}_R$. Here, we only report the final results. For a derivation of the expression for the survival probability $P_s$ see~\cite{Janssen:2005}, for the derivation of $\langle N \rangle$ and $\langle R^2 \rangle$ see~\cite{Janssen.Tauber:2005}
\begin{eqnarray}
  P_s (t,\tau) \propto \langle \tilde{B} (\bi{r}=0,-t) \rangle =G^{(0,1)}_R(\tau,\bi{r}=0,t) \label{appendixC:scaling.of.P} \\
 \langle N  \rangle (t,\tau) = \int \rmd^d \bi{r} \langle \tilde{B}(0,0) B(\bi{r},t) \rangle =\int \rmd^d \bi{r} G^{(1,1)}_R(\tau,\bi{r},t) \label{appendixC:scaling.of.N} \\
 \langle R^2 \rangle (t,\tau) = \frac{1}{\int \rmd^d \bi{r} \langle B(\bi{r},t) \rangle} \int \rmd^d \bi{r} \, r^2 \langle B(\bi{r},t) \rangle \nonumber \\ 
\textcolor{white}{\langle R^2 \rangle (t,\tau)} =\frac{1}{\int \rmd^d \bi{r} G^{(1,0)}_R( \tau,\bi{r},t) } \int \rmd^d \bi{r} \, r^2 G^{(1,0)}_R(\tau,\bi{r},t) \label{appendixC:scaling.of.R}
\end{eqnarray}
The critical exponents are derived by inserting the scaling of the Green functions~\eref{eq:Green.function.scaling} into the expression for the observables~\eref{appendixC:scaling.of.P}--\eref{appendixC:scaling.of.R}. The critical exponent $\delta$ can be easily obtained from the $G^{(0,1)}_R$ scaling. The scaling form of $\langle N  \rangle (t,\tau)$ is obtained by changing the integration variable in the last expression of~\eref{appendixC:scaling.of.N} to $\bi{r}/ \xi$ and relating the length scale $\xi$ to a time scale, i.e., $\xi \propto t^{1/ z}$. At last, the critical exponent $\nu_{\parallel}$ is obtained by inserting the correlation length $\xi \propto (\tau)^{\nu} $ in the second argument of the scaling function of the Green function~\eref{eq:Green.function.scaling} and using that $\tau$ encodes how far we are from criticality, i.e.,  $\tau \propto (\sigma-\sigma_c) $.
The final values for the critical exponents read:
\begin{eqnarray}
   z-2=W_D-\frac{\sigma+\tilde{\sigma}}{2}=-\frac{7}{30}\epsilon \\
    \delta=\frac{d-2+\tilde{\sigma}}{2z}= 1-\frac{1}{4}\epsilon \\ 
   \theta=-\frac{\sigma+\tilde{\sigma}}{2z}=\frac{3}{20}\epsilon \qquad \qquad \\ \nu_{\parallel}=z \nu=1-\frac{1}{20}\epsilon
    \end{eqnarray}

\section{Choice of the time intervals}
\label{appendix:time.interval}

This section provides additional details regarding the choice of the time interval~${\Delta t = 10}$ in the averaged local slope equation~\eref{eq:local.slope.theta}~\cite{Grassberger:1979, Grassberger:1989}. 
The averaged local slope technique explained in Sec.~\ref{sec:numerics} of the main text, aims to approximate the derivative in every point via the notion of a difference quotient. 
However, the underlying data is inherently noisy as it results from averaging over a finite number of stochastic realizations. 
Therefore, one has to strike a balance between a more accurate approximation of a derivative (i.e., smallest possible~$\Delta t$) and not disproportionately considering noise-induced slope changes (i.e., large~$\Delta t$).
By testing various values of~$\Delta t$, we find that the choice of~${\Delta t=10}$ is a sensible middle ground.
This value is comparable to values used in previous studies~\cite{Grassberger:1989, Jensen.et.al:1990, Jensen:1993, Voigt.Ziff:1997, Deng.et.al:2020}. 
Most importantly, our estimates of the critical exponents are stable against varying the value of~$\Delta t$ by up to~$40 \%$.
While previous work~\cite{Deng.Odor:2023} on the DGEP model chose a substantially smaller~$\Delta t$, they additionally employ a moving average to smooth out stochastic fluctuations. 
This is similar to choosing a larger time interval in the sense that both methods mitigate what are assumed to be noise-induced slope changes. Increasing the time interval achieves this by a coarser difference quotient, while the moving average per definition results in fewer deviations between neighboring points.
Furthermore, a direct comparison of~$\Delta t$ between the Gillespie algorithm~\cite{Polovnikov.et.al:2022} used in this paper with a clear notion of time and one where time is defined through the number of updating steps~\cite{Deng.Odor:2023} should be done with caution. 
One additional constraint for the time interval is that it has to be compatible with the recording interval of the simulation algorithm; otherwise, the local slope cannot be calculated. Our recording interval is chosen to be~$1$ in terms of our base units. 
Hence,~$\Delta t$ has to be chosen as an integer in our case.

\section{Error estimation}
\label{appendix:error.estimation}

This section summarizes our systematic approaches to estimating the errors of the critical exponents obtained from the numerical data of stochastic simulations. 
We present several complementary approaches for error estimation and use the one that gives the largest value.

\subsection*{Error estimation for the averaged local slope method}
\fulltable{ 
\textbf{ Error estimation methods.} List of methods used for the error estimation for critical exponents. The parameters used for each of the methods are listed as well. In particular, for the sampling methods, we report the fitting interval and the size of the random samples, while for the fitting range method, the range of fitting intervals is listed.}
\label{table:E1} \\
\br
Model & Exponent & Error estimation method & Parameters\\[1.5mm]

\mr
GEP & $\theta, \delta, z$ & Non-averaged min-max & 1/$t_{\mathrm{min}}=2 \times 10^{-5}$\\
\mr
DGEP & $\theta$ & Sampling  &1/$t_{\mathrm{min}}=2 \times 10^{-5}$, 5\% of the data per repetition \\
&  $\delta$& Sampling &1/$t_{\mathrm{min}}=2 \times 10^{-5}$, 5\% of the data per repetition \\
& $z$ & Sampling & 1/$t_{\mathrm{min}}=2 \times 10^{-5}$, 5\% of the data per repetition \\
\mr
CGEP & $\theta$ & Sampling &1/$t_{\mathrm{min}}=2 \times 10^{-5}$, 5\% of the data per repetition \\
&  $\delta$& Sampling & 1/$t_{\mathrm{min}}=2 \times 10^{-5}$, 5\% of the data per repetition \\
& $z$ & Fitting Range& 1/ $t_{\mathrm{min}} \in [5\times 10^{-6}, 4\times 10^{-5}] $ \\

\endfulltable
In summary, the estimation of critical exponents and their errors from numerical data using the \textit{averaged local slope method} is susceptible to the following partially subjective factors: The choice of data points used to fit to condition~\eref{eq:asymptotic.local.slope} and the method used to estimate the impact of deviations from the resulting linear fit.
Here, we can confidently present results and assert strong statements about universality classes by choosing relatively large errors derived from a range of error estimation methods.

\begin{figure}[t]
    \centering
    \includegraphics[width=\columnwidth]{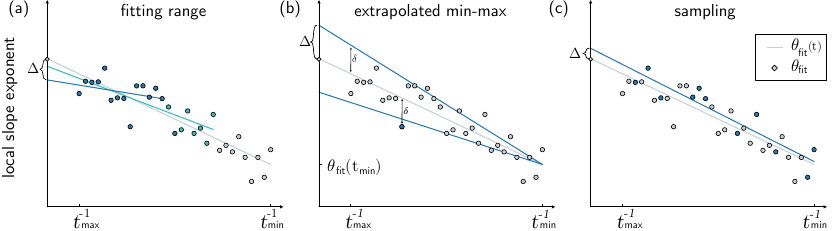}
    \caption{
    {\bf Error estimations.}
    Visualization of three different methods of error estimations on the example of the critical exponent~$\theta$.
    The filled dots represent artificial numerical data of the local slope~$\theta_{\rm data}(t)$, the gray straight line the corresponding fit~$\theta_{\rm fit}(t)$ and~$\Delta$ the resulting error estimates.
    (a) \textit{Fitting range error} obtained by varying the fitting range. For the next smaller fitting range, only the data points colored in light blue and dark blue are taken into account, and for the smallest fitting range, only the data points colored in dark blue are taken into account. The respective linear extrapolations are shown as light blue and dark blue straight lines.
    (b) Example of the \textit{extrapolated min-max error}. The data point with the largest deviation~$\delta$ from~$\theta_{\rm fit}(t)$ is shown in dark blue. The new linear extrapolations~$\tilde\theta_{\rm fit}(t)$ obtained by shifting~$\theta_{\rm fit}(t_{\rm max})$ are shown in light blue.
    (c) Example of the \textit{sampling error} obtained from selecting a random sample (shown in dark blue) and its corresponding linear fit is shown in light blue.
    }
    \label{fig:figureE1}
\end{figure}

As we discussed in the main text, a common method to extract critical exponents from numerical data---here described in the example of~$\theta$---is to first obtain its time-dependent approximation~$\theta(t)$. 
This can be achieved by calculating the time-dependent  \textit{averaged local slope} of the corresponding observable---$\langle N\rangle (t)$ in the case of~$\theta$---which amounts to locally approximating the slope of~$\langle N\rangle (t)$ in a double logarithmic plot by a discretized derivative (see~\eref{eq:local.slope.theta}).
At criticality, $\theta(t)$ is expected to asymptotically approach the actual value of the critical exponent~$\theta$ as~\cite{Grassberger:1989}
\begin{eqnarray}
\lim_{t\rightarrow\infty}\theta(t)=\theta+\frac{a}{t}, \label{eq:asymptotic.local.slope}
\end{eqnarray}
with some value~$a$ for the \textit{local correction} to the actual critical exponent.
Hence, the critical exponent~$\theta$ can be obtained by plotting~$\theta(1/t)$, estimating~$a$ by approximating the slope of~$\theta(t)$ in the limit~$1/t\rightarrow 0$ and calculating the intercept of the ensuing straight line with the $1/t=0$ axis.
For this procedure to work, one additionally has to find the critical value of the tuning parameter (in our case, the critical infection rate~$\sigma_c$), which is obtained by requiring that the condition~\eref{eq:asymptotic.local.slope} holds, i.e, by requiring that~$\theta(1/t)$ approaches a straight line close to $1/t=0$.
In summary, we take the following steps to estimate critical exponents from numerical data: Estimate the critical infection rate as described above and determine the best fit of the corresponding dataset of local slopes~$\theta_{\rm data}(t)$ (shown as bullets in \fref{fig:figureE1}) to the expected asymptotic behavior~$\theta_{\rm fit}(t) = \theta_{\rm fit} + a_{\rm fit}/t$ (shown as a gray straight line in \fref{fig:figureE1}). 
Importantly, only data points inside a certain \textit{fitting range}~$I=[1/t_{\rm max}, 1/t_{\rm min}]$---where $t_{\rm min}$ is approximately the smallest time where we see the asymptotic behavior~\eref{eq:asymptotic.local.slope} and $t_{\max}$ is given by the maximal simulation time---are taken into account for the above fit.
The resulting $\theta_{\rm fit}$, which is equal to the intercept of $\theta_{\rm fit}(t)$ with the $1/t=0$ axis, is then our best estimate for the critical exponent. The same procedure is subsequently applied to the remaining critical exponents.
In the following paragraphs, all mentioned linear fits and straight lines refer to the estimated value of~$a_{\rm fit}$ and the resulting straight lines in the vicinity of~$1/t=0$.

Even though this method is widely used, there is no single gold standard for estimating the errors of critical exponents obtained via the above procedure~\cite{Henkel.et.al:2008, Hinrichsen:2000}.
Nevertheless, two guidelines for the error estimation of critical exponents exist: 
Firstly, slopes of straight lines should not be estimated via fitting methods like $\chi^2$--error estimates~\cite{Henkel.et.al:2008}. 
Secondly, the error of critical exponent estimates is usually one order of magnitude larger than the error estimate of the critical value of the tuning parameter~\cite{Henkel.et.al:2008}.
In particular, there is no established approach to determine the interval over which to fit the data---thereby obtaining the extrapolated linear fit, its intercept with the $1/t=0$ axis and thus the estimated exponent.
This choice, however, can have a significant influence on the results. 
Due to stochastic fluctuations in the data and corrections to the described asymptotic behavior at finite times, the estimated value of a critical exponents is susceptible to this choice of the fitting range~$I$.
Given these challenges, we explored three distinct methods to assess the exponents and their associated errors. 
Opting for the approach that produces the widest range of possible values for the exponents, we derive a highly conservative estimate of errors.

We are interested in the critical dynamics on longer time scales, where asymptotic scaling behavior is to be expected. 
The procedure detailed above, used to extract this asymptotic behavior from numerical data, however, depends on the choice of~$t_{\rm min}$. 
This choice corresponds to the point where the system is close to fulfilling~\eref{eq:asymptotic.local.slope} and is, at least partially, subjective.
A possible error estimation, therefore, involves varying~$t_{\rm min}$ within a reasonable range and, in accordance with the asymptotic behavior~\eref{eq:asymptotic.local.slope}, fitting a straight line to the corresponding data points, obtained by calculating the averaged local slope as described in~\eref{eq:local.slope.theta}.
The error is then estimated as the difference between the largest and smallest resulting intercept (see \fref{fig:figureE1}(a)).
We term this the \textit{fitting range error}.

Another error estimation can be obtained by a method we term the \textit{extrapolated min-max error}, explained on the example of the critical exponent~$\theta$. To calculate this error, we proceed as follows: We first pick a sensible value for~$t_{\rm min}$, resulting in the fitting range~${I=[1/t_{\rm max}, 1/t_{\rm min}]}$.
Subsequently, the local correction~$a_{\rm fit}$ and the critical exponent~$\theta_{\rm fit}$ are again estimated by fitting the averaged local slope to condition~\eref{eq:asymptotic.local.slope}.
We then calculated the maximal deviation from the resulting straight line, i.e.,
\begin{eqnarray}
    \delta = \max_{1/t\in I}|\theta_{\rm fit}(t) - \theta_{\rm data}(t)| = \max_{1/t\in I} |\theta_{\rm fit}(t_{\rm min}) + \frac{a}{t} - \theta_{\rm data}(t)|.
\end{eqnarray}
Subsequently, a new straight line~$\tilde\theta_{\rm fit}(t)$ is obtained by adding (or subtracting)~$\delta$ to~$\theta_{\rm fit}(t_{\rm max})$ and connecting this to~$\theta_{\rm fit}(t_{\rm min})$.
Note that we add~$\delta$ to~$\theta_{\rm fit}(t_{\rm max})$ instead of directly using the data point of the largest deviation to draw a new straight line to avoid error overestimations in cases where the largest deviation is observed close to $1/t_{\rm min}$.
The \textit{extrapolated min-max error} is then the distance between the original estimate~$\theta_{\rm}$ and the intercept of~$\tilde\theta_{\rm fit}(t)$ with the $1/t=0$ axis; it is thus given by
\begin{eqnarray}
    \Delta_{\rm min-max} = \frac{\delta}{1-{t_{\rm min}}/{t_{\rm max}}}.
\end{eqnarray}
A visualization of the above procedure is shown in \fref{fig:figureE1}(b).

Yet another approach is to test how stable the fit obtained through condition~\eref{eq:asymptotic.local.slope} is against removing data points.
To achieve this, we randomly sampled between~1--5$\%$ of the data points within a given fitting interval and fitted a straight line to each sample. 
This process was repeated $5,000$ times. The ensuing \textit{sampling error} was then determined as the maximum difference between the intercept of the full fit and the intercepts of the sampled fits (see \fref{fig:figureE1}(c)).

We calculated all three types of errors for each exponent and chose the largest one for the final estimate. This approach led to conservative error estimates in comparison to other studies~\cite{Grassberger:1989, Jensen.et.al:1990, Jensen:1993, Voigt.Ziff:1997, Deng.et.al:2020}. 
Table~\ref{table:E1} provides a summary of the chosen error estimates and fitting ranges for the respective exponents across all studied epidemic models. 
For the GEP model, however, the situation is more involved. 
Despite investing significant computational resources, two possible candidates for the critical infection rate---${\sigma_c=1.376}$ and~${\sigma_c=1.377}$---remained. 
Resolving this ambiguity would require longer simulation times, which are currently impractical with the employed algorithm.
Lacking such studies, we calculated a point-wise mean; for example, ${\bar\theta(t)=(\theta_-(t) + \theta_+(t))/2}$, where $\theta_\pm$ correspond to data obtained from simulations with~${\sigma=1.376}$ and~${\sigma=1.377}$, respectively. 
The resulting averaged data set was subsequently used to obtain an estimate for the critical exponents.
This approach, however, results in an underestimation of the error when applying all of the above error estimation methods to the averaged data.
Thus, we did not use the averaged data ($\bar\theta(t)$) but rather calculated the \textit{extrapolated min-max error} for both candidate data sets~($\theta_-(t)$ and $\theta_+(t)$) and chose the largest deviation from the critical exponents obtained from the averaged data set as the \textit{non-averaged min-max error}.
The \textit{non-averaged min-max error} is a more conservative approach to estimating the error and, therefore a reasonable substitute.

\subsection*{Error estimation for data collapses}
\begin{figure}[htb]
\centering
\includegraphics[width=\columnwidth]{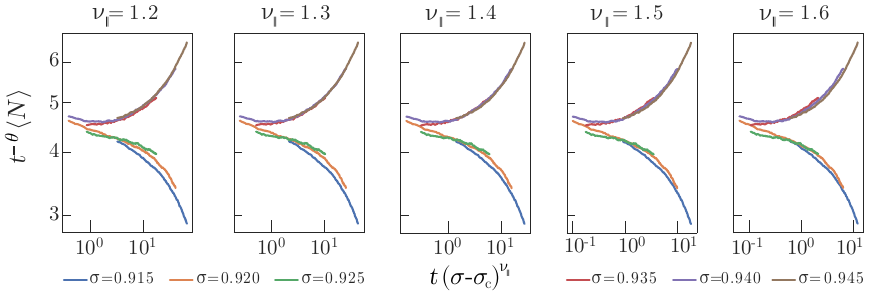}
\caption{
{\bf Data collapse error estimation for DGEP.} Data collapses for the average number of infected individuals for the DGEP model at different values of $\nu_{\parallel}$ close to our final estimate of~$\nu_{\parallel}=1.4$, as indicated by the graph titles. Apart from~$\nu_\parallel$, the parameters and data used in these figures are identical. The individual curves are rescaled as explained in the main text in section~\ref{sec:numerics}. They are obtained from simulations for different values of the infection rate~$\sigma$ as indicated below the figures; the critical value~$\sigma_c=0.93$ and the exponent~$\theta=0.54$ were chosen for the rescaling.
}
    \label{fig:figureE2}
\end{figure}

\begin{figure}[htb]
\centering
\includegraphics[width=\columnwidth]{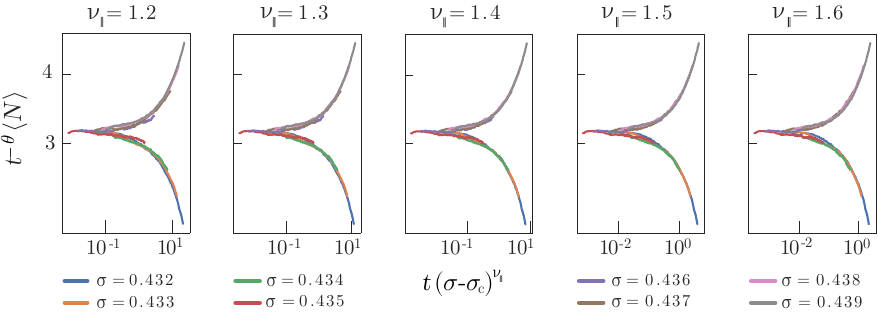}
\caption{
{\bf Data collapse error estimation for CGEP.} Data collapses for the average number of infected individuals for the CGEP model at different values of $\nu_{\parallel}$ close to our final estimate of~$\nu_{\parallel}=1.4$, as indicated by the graph titles. Apart from $\nu_\parallel$, the parameters and data used in these figures are identical. The individual curves are rescaled as explained in the main text in section~\ref{sec:numerics}. They are obtained from simulations for different values of the infection rate~$\sigma$ as indicated below the figures; the critical value~$\sigma_c=0.43537$ and the exponent~$\theta=0.55$ were chosen for the rescaling.
}
    \label{fig:figureE3}
\end{figure}
There are two equally valid approaches to estimating the error of critical exponents obtained through a data collapse. 
One way is to define a loss function for the quality of the fit and then let an optimization algorithm determine the best collapse and the critical exponents with their errors~\cite{Sorge:2015, Melchert:2009, Houdayer.Hartmann:2004, Nelder.Mead:1965, Bhattacharjee.Seno:2001}. Alternatively, one can employ a manual approach and determine the best data collapse visual inspection~\cite{Henkel.et.al:2008, Polovnikov.et.al:2022}.

However, employing a loss-function approach introduces some inherent bias.
First of all, initial values of the critical exponents for the algorithm are required, and the results change depending on this choice. 
The choice of loss function also plays a significant role in the final estimates. 
Hence, we decided to adhere to a manual approach to mitigate these potential biases.
The best fit is determined by visual inspection, and the error margins are then estimated from the breakdown of a reasonable data collapse~\cite{Henkel.et.al:2008, Polovnikov.et.al:2022}. 
In particular, we determine the value of~$\nu_\parallel$ for which the appropriate rescaling (which can be determined from~\eref{eq:observable1}) leads to only two distinct curves: one for all data obtained from simulations with infection rates larger than the critical infection rate~$\sigma_c$ and one for data obtained for infection rates smaller than~$\sigma_c$.
Figures~\ref{fig:figureE2} and~\ref{fig:figureE3} show the data collapses for different values of~$\nu_\parallel$ of the DGEP and CGEP data, respectively. 
In both cases, we determine that the best collapse is achieved for~${\nu_\parallel=1.4}$. 
To estimate the error, we further determine for which values the collapse is still ``reasonably good'' and for which this is no longer the case. It should, however, be noted that this choice is at least partially subjective.
In figure~\ref{fig:figureE2}, one can nevertheless clearly observe that the curves for larger infection rates---${\sigma=0.935,\ 0.940}$ and~$0.945$---
diverge from each other for~${\nu_\parallel=1.2}$ and~${\nu_\parallel=1.6}$.
Similarly, in figure~\ref{fig:figureE3}, both the curves for larger and smaller infection rates diverge from each other for~${\nu_\parallel=1.4}$ and the curves for~${\sigma=0.432}$--$0.435$ also clearly separate for~${\nu_\parallel=1.6}$.
We thus conclude that in both cases, the value of~$\nu_\parallel$ is given by~${\nu_\parallel=1.4\pm 0.1}$.

\section{Additional numerical data}
\label{appendix:num.data}
\begin{figure}[t]
\centering
\includegraphics[width=\columnwidth]{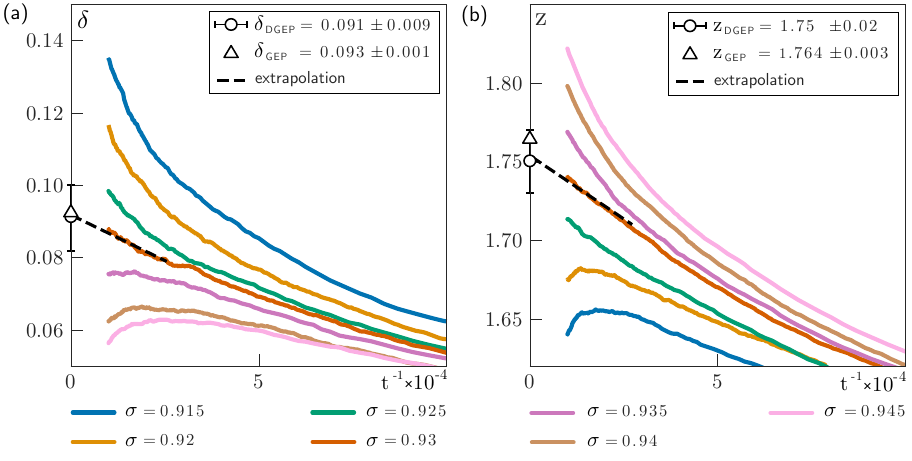}
\caption{
{\bf Critical behavior of the DGEP model's other observables.} \\
(a) Averaged local slope for the survival probability of the infected individuals for the DGEP model, for a set of values for the infection rate~$\sigma$ indicated below the graph.
Based on these results and the critical value from figure~\ref{fig:5}(a), we estimate the value of~${\delta=0.091 \pm 0.009}$ as indicated by the circular marker with error bars. The triangle shows our estimate for GEP for comparison, where we omitted the error bars because they are smaller than the marker.
(b) Averaged local slope for the mean-squared radius of the infected cluster for the DGEP model for a set of values for the infection rate~$\sigma$ indicated below the graph and identical to the values of $\sigma$ used for (a).
Based on these results and the critical value from figure~\ref{fig:5}(a), we estimate the value of~${z=1.75 \pm 0.02}$ as indicated by the circular marker with error bars. The triangle shows our estimate for GEP for comparison, where we omitted the error bars because they are smaller than the marker.
The parameters for both figures are identical to those used for figure~\ref{fig:5}(a) and are summarized in table~\ref{table:F1}.
}
    \label{fig:figureF1}
\end{figure}

\begin{figure}[!htb]
\centering
\includegraphics[width=\columnwidth]{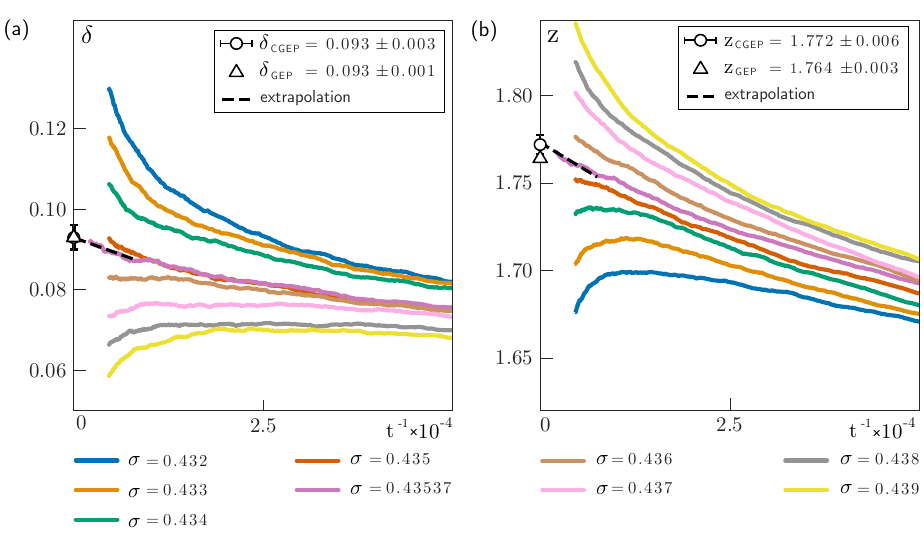}
\caption{
{\bf Critical behavior of the CGEP model's other observables.} \\
(a) Averaged local slope for the survival probability of the infected individuals for the CGEP model, for a set of values for the infection rate~$\sigma$ indicated below the graph.
Based on these results and the critical value from figure~\ref{fig:5}(b), we estimate the value of~${\delta=0.093 \pm 0.003}$ as indicated by the circular marker with error bars. The triangle shows our estimate for GEP for comparison, where we omitted the error bars because they are smaller than the marker.
(b) Averaged local slope for the mean-squared radius of the infected cluster for the CGEP model for a set of values for the infection rate~$\sigma$ indicated below the graph and identical to the values of $\sigma$ used for (a).
Based on these results and the critical value from figure~\ref{fig:5}(a), we estimate the value of~${z=1.772 \pm 0.006}$ as indicated by the circular marker with error bars. The triangle shows our estimate for GEP for comparison, where we omitted the error bars because they are smaller than the marker.
The parameters for both figures are identical to those used for figure~\ref{fig:5}(b) and are summarized in table~\ref{table:F1}.
}
    \label{fig:figureF2}
\end{figure}

This section presents additional data we omitted in the main text, as well as the exact definitions of the observables from a numerical point of view. 
We define the survival probability as the number of realizations with at least one infected individual divided by the total number of realizations. The average number of infected individuals is calculated as an average over all realizations, including those where no infected individuals remain. The mean squared cluster radius is obtained by squaring the difference between the right- and leftmost infected individual in the x- or y-direction, whichever is larger. The average is taken over all realizations with at least one infected individual.
Figure~\ref{fig:figureF1} shows the survival probability and the mean-squared cluster radius of the DGEP models. It is evident from the indicated resulting critical exponents that we cannot distinguish them from the GEP ones with statistical significance. The same is true for figure~\ref{fig:figureF2}, where the corresponding observables are shown for the CGEP model. In table~\ref{table:F1}, we summarize the simulation parameters used for each model under study.

\fulltableh{{\bf Simulation parameters.} Parameters used for the simulations of GEP, DGEP and CGEP. We list the number of realizations of the stochastic process taken into account for computing the observables' average; the length of the 2D square-shaped domain measured in units of the lattice spacing; the maximal time measured in units of the diffusion constants (see discussion in section~\ref{sec:numerics}); the number of infected individuals in the initial conditions~$N_0$; the diffusion constants of all species; the reaction rates~$\zeta$ for recovery and~$\gamma$. In addition, we provide the figures to which the respective parameters apply for easy reference.
} \label{table:F1}\\
\br
Model & Realizations &  Length & Time & $N_0$ & $D_A$ & $D_B$ & $D_C$ & $\zeta$ & $\gamma$ & Figures\\
\mr
GEP & 24,000 & 10,000 & 50,000 & 10 & 0 & 0.1 & 0 & 0.1 & 0 & \ref{fig:4} \\
DGEP & 20,000 & 2,600 & 10,000 & 10 & 0.1 & 0.1 & 0 & 0.1 & 0 & \ref{fig:5}(a), \ref{fig:6}(a), \ref{fig:figureF1}, \ref{fig:figureE2}\\
CGEP\\
\hspace{3mm}critical & 38,000 & 8,800 & 40,000 & 10 & 0 & 0.1 & 0.1 & 0.05 & 0.1 & \ref{fig:5}(b), \ref{fig:6}(b), \ref{fig:figureF2}, \ref{fig:figureE3} \\
\hspace{3mm}off-critical & 34,000 & 4,400 & 20,000 & 10 & 0 & 0.1 & 0.1 & 0.05 & 0.1 & \ref{fig:5}(b), \ref{fig:6}(b), \ref{fig:figureF2}, \ref{fig:figureE3}\\
\endfulltableh

\FloatBarrier 
\section*{References}
\providecommand{\newblock}{}


\begin{thebibliography}{100}
\expandafter\ifx\csname url\endcsname\relax
  \def\url#1{{\tt #1}}\fi
\expandafter\ifx\csname urlprefix\endcsname\relax\def\urlprefix{URL }\fi
\providecommand{\eprint}[2][]{\url{#2}}

\bibitem{Chinazzi.et.al:2020}
Chinazzi M, Davis J~T, Ajelli M, Gioannini C, Litvinova M, Merler S, y~Piontti A~P, Mu K, Rossi L, Sun K, Viboud C, Xiong X, Yu H, Halloran M~E, Longini I~M and Vespignani A 2020 {\em Science\/} {\bf 368} 395--400 ISSN 0036-8075

\bibitem{Kissler.et.al:2020}
Kissler S~M, Tedijanto C, Goldstein E, Grad Y~H and Lipsitch M 2020 {\em Science\/} {\bf 368} 860--868 ISSN 0036-8075

\bibitem{Arenas.et.al:2020}
Arenas A, Cota W, Gómez-Gardeñes J, Gómez S, Granell C, Matamalas J~T, Soriano-Paños D and Steinegger B 2020 {\em Physical Review\/} X {\bf 10} 041055

\bibitem{Bauer.et.al:2021}
Bauer S, Contreras S, Dehning J, Linden M, Iftekhar E, Mohr S~B, Olivera-Nappa A and Priesemann V 2021 {\em PLoS Computational Biology\/} {\bf 17} e1009288 ISSN 1553-734X

\bibitem{Dehning.et.al:2020}
Dehning J, Zierenberg J, Spitzner F~P, Wibral M, Neto J~P, Wilczek M and Priesemann V 2020 {\em Science\/} {\bf 369} eabb9789 ISSN 0036-8075

\bibitem{Ferretti.et.al:2020}
Ferretti L, Wymant C, Kendall M, Zhao L, Nurtay A, Abeler-Dörner L, Parker M, Bonsall D and Fraser C 2020 {\em Science\/} {\bf 368} eabb6936 ISSN 0036-8075

\bibitem{Gros.et.al:2021}
Gros C, Valenti R, Schneider L, Valenti K and Gros D 2021 {\em Scientific Reports\/} {\bf 11} 6848

\bibitem{Muley.et.al:2020}
Muley D, Shahin M, Dias C and Abdullah M 2020 {\em Sustainability\/} {\bf 12} 7367

\bibitem{Epstein.et.al:2008}
Epstein J~M, Parker J, Cummings D and Hammond R~S 2008 {\em PLoS ONE\/} {\bf 3} e3955

\bibitem{Weitz.et.al:2020}
Weitz J~S, Park S~W, Eksin C and Dushoff J 2020 {\em Proceedings of the National Academy of Sciences\/} {\bf 117} 32764--32771 ISSN 0027-8424

\bibitem{Lloyd.et.al:1996}
Lloyd A~L and May R~M 1996 {\em Journal of Theoretical Biology\/} {\bf 179} 1--11 ISSN 0022-5193

\bibitem{Pastor-Satorras.et.al:2015}
Pastor-Satorras R, Castellano C, Mieghem P~V and Vespignani A 2015 {\em Reviews of Modern Physics\/} {\bf 87} 925--979 ISSN 0034-6861

\bibitem{Holley.Liggett:1975}
Holley R~A and Liggett T~M 1975 {\em The Annals of Probability\/} {\bf 3} ISSN 0091-1798

\bibitem{Castellano.et.al:2009}
Castellano C, Fortunato S and Loreto V 2009 {\em Reviews of Modern Physics\/} {\bf 81} 591--646 ISSN 0034-6861

\bibitem{Drossel.Schwabl:1992}
Drossel B and Schwabl F 1992 {\em Physical Review Letters\/} {\bf 69} 1629--1632 ISSN 0031-9007

\bibitem{Albano:1995}
Albano E~V 1995 {\em Physica\/} A {\bf 216} 213--226 ISSN 0378-4371

\bibitem{Ziff.et.al:1986}
Ziff R~M, Gulari E and Barshad Y 1986 {\em Physical Review Letters\/} {\bf 56} 2553--2556 ISSN 0031-9007

\bibitem{Jensen.et.al:1990}
Jensen I, Fogedby H~C and Dickman R 1990 {\em Physical Review\/} A {\bf 41} 3411--3414 ISSN 1050-2947

\bibitem{Marro.Dickman:1999}
Marro J and Dickman R 1999 {\em Nonequilibrium Phase Transitions in Lattice Models\/} (Cambridge University Press) ISBN 9780511524288

\bibitem{Odor:2004}
Ódor G 2004 {\em Reviews of Modern Physics\/} {\bf 76} 663--724 ISSN 0034-6861

\bibitem{Tauber:2014}
Tauber U~C 2014 {\em Critical Dynamics\/} (Cambridge, United Kingdom: Press, Cambridge University)

\bibitem{Tauber.et.al:2005}
Täuber U~C, Howard M and Vollmayr-Lee B~P 2005 {\em Journal of Physics\/} A {\bf 38} R79 ISSN 0305-4470

\bibitem{Henkel.et.al:2008}
Henkel M, Hinrichsen H and Lübeck S 2008 {\em {Absorbing Phase Transitions}\/} vol~1 {\em Non-Equilibrium Phase Transitions\/} (Heidelberg, Germany: Springer Netherlands)

\bibitem{Hinrichsen:2000}
Hinrichsen H 2000 {\em Advances in Physics\/} {\bf 49} 815--958 ISSN 0001-8732

\bibitem{Janssen:1981}
Janssen H~K 1981 {\em Zeitschrift für Physik\/} B {\bf 42} 151

\bibitem{Janssen.Tauber:2005}
Janssen H~K and Tauber U~C 2005 {\em Annals of Physics\/} {\bf 315} 147--192 ISSN 0003-4916

\bibitem{Grassberger:1983}
Grassberger P 1983 {\em Mathematical Biosciences\/} {\bf 63} 157--172 ISSN 0025-5564

\bibitem{Janssen:1985}
Janssen H~K 1985 {\em Zeitschrift für Physik\/} B {\bf 58} 311--317 ISSN 0722-3277

\bibitem{Harris:1974}
Harris T~E 1974 {\em The Annals of Probability\/} {\bf 2} ISSN 0091-1798

\bibitem{Kermack.McKendrick:1927}
Kermack W~O and McKendrick A~G 1927 {\em Proceedings of the Royal Society of London. Series A, Containing Papers of a Mathematical and Physical Character\/} {\bf 115} 700--721 ISSN 0950-1207

\bibitem{Grassberger:1997}
Grassberger P, Chaté H and Rousseau G 1997 {\em Physical Review\/} E {\bf 55} 2488--2495 ISSN 1539-3755

\bibitem{Janssen.Stenull:2008}
Janssen H~K and Stenull O 2008 {\em Physical Review\/} E {\bf 78} 061117 ISSN 1539-3755

\bibitem{Kolk.et.al:2023}
van~der Kolk J, Raßhofer F, Swiderski R, Haldar A, Basu A and Frey E 2023 {\em Physical Review Letters\/} {\bf 131} ISSN 0031-9007

\bibitem{Jensen:1993}
Jensen I 1993 {\em Physical Review Letters\/} {\bf 70} 1465--1468 ISSN 0031-9007

\bibitem{Munoz.et.al.1998}
Muñoz M~A, Grinstein G and Dickman R 1998 {\em Journal of Statistical Physics\/} {\bf 91} 541--569 ISSN 0022-4715

\bibitem{Silva.Dickman:1999}
da~Silva J~K~L and Dickman R 1999 {\em Physical Review\/} E {\bf 60} 5126--5129 ISSN 1539-3755

\bibitem{Grassberger:1982}
Grassberger P 1982 {\em Zeitschrift für Physik\/} B {\bf 47} 365--374 ISSN 0722-3277

\bibitem{Howard.Tauber:1997}
Howard M~J and Täuber U~C 1997 {\em Journal of Physics\/} A {\bf 30} 7721 ISSN 0305-4470

\bibitem{Henkel.Hinrichsen:2004}
Henkel M and Hinrichsen H 2004 {\em Journal of Physics\/} A {\bf 37} R117 ISSN 0305-4470

\bibitem{Deng.et.al:2020}
Deng S, Li W and Täuber U~C 2020 {\em Physical Review\/} E {\bf 102} 042126 ISSN 2470-0045

\bibitem{Deng.Odor:2023}
Deng S and Ódor G 2023 {\em Physical Review\/} E {\bf 107} 014303 ISSN 2470-0045

\bibitem{Jensen.Dickman:1999}
Jensen I and Dickman R 1999 {\em Journal of Physics\/} A {\bf 26} L151 ISSN 0305-4470

\bibitem{Wijland.et.al:1998}
van Wijland F, Oerding K and Hilhorst H~J 1998 {\em Physica\/} A {\bf 251} 179--201

\bibitem{Doussal.Wiese:2015}
Doussal P~L and Wiese K~J 2015 {\em Physical Review Letters\/} {\bf 114} 110601 ISSN 0031-9007

\bibitem{Janssen.Stenull:2016}
Janssen H~K and Stenull O 2016 {\em Physical Review\/} E {\bf 94} 042138 ISSN 2470-0045

\bibitem{Kree.et.al:1989}
Kree R, Schaub B and Schmittmann B 1989 {\em Physical Review\/} A  2214--2221 ISSN 1050-2947

\bibitem{Janssen:2001}
Janssen H~K 2001 {\em Physical Review\/} E {\bf 64} 058101 ISSN 1539-3755

\bibitem{Tarpin.et.al:2017}
Tarpin M, Benitez F, Canet L and Wschebor N 2017 {\em Physical Review\/} E {\bf 96} 022137 ISSN 2470-0045

\bibitem{Oerding.et.al:2000}
Oerding K, Wijland F, Leroy J~P and Hilhorst H~J 2000 {\em Journal of Statistical Physics\/} {\bf 99} 1365--1395 ISSN 0022-4715

\bibitem{Maia.Dickman:2007}
Maia D~S and Dickman R 2007 {\em Journal of Physics: Condensed Matter\/} {\bf 19} 065143 ISSN 0953-8984

\bibitem{Dickman.Maia:2008}
Dickman R and Maia D~S 2008 {\em Journal of Physics A: Mathematical and Theoretical\/} {\bf 41} 405002 ISSN 1751-8121

\bibitem{Argolo.et.al:2019}
Argolo C, Tenório V and Lyra M~L 2019 {\em Physica A: Statistical Mechanics and its Applications\/} {\bf 517} 422--430 ISSN 0378-4371

\bibitem{Polovnikov.et.al:2022}
Polovnikov B, Wilke P and Frey E 2022 {\em Physical Review Letters\/} {\bf 128} 078302 ISSN 0031-9007

\bibitem{Kermack.et.al:1927}
Kermack W~O and McKendrick A~G 1927 {\em Proceedings of the Royal Society of London. Series A, Containing Papers of a Mathematical and Physical Character\/} {\bf 115} 700--721 ISSN 0950-1207

\bibitem{Cardy:1983}
Cardy J~L 1983 {\em Journal of Physics\/} A {\bf 16} L709 ISSN 0305-4470

\bibitem{Cardy.Grassberger:1985}
Cardy J~L and Grassberger P 1985 {\em Journal of Physics\/} A {\bf 18} L267 ISSN 0305-4470

\bibitem{Tome.Ziff:2010}
Tomé T and Ziff R~M 2010 {\em Physical Review\/} E {\bf 82} 051921 ISSN 1539-3755

\bibitem{Frey.Brauns:2018}
Frey E and Brauns F 2018 {Self-organization of Protein Patterns} {\em Active Matter and Nonequilibrium Statistical Physics\/} ({\em Lecture Notes of the Les Houches Summer School\/} vol 112) (Oxford, United Kingdom: Oxford University Press) ISBN 9780192858313

\bibitem{COMSOL}
COMSOL M 1998 {\em COMSOL Multiphysics, Burlington, MA, accessed Feb\/} {\bf 9} 2018

\bibitem{Weber.Frey:2017}
Weber M~F and Frey E 2017 {\em Reports on Progress in Physics\/} {\bf 80} 046601 ISSN 0034-4885

\bibitem{Wilson:1975}
Wilson K~G 1975 {\em Reviews of Modern Physics\/} {\bf 47} 773--840 ISSN 0034-6861

\bibitem{Wilson.Fisher:1972}
Wilson K~G and Fisher M~E 1972 {\em Physical Review Letters\/} {\bf 28} 240--243 ISSN 0031-9007

\bibitem{Fisher:1974}
Fisher M~E 1974 {\em Reviews of Modern Physics\/} {\bf 46} 597--616 ISSN 0034-6861

\bibitem{Frey.Tauber:1994}
Frey E and T\"auber U~C 1994 {\em Physical Reviews\/} E {\bf 50}(2) 1024--1045

\bibitem{Weinberg:1995}
Weinberg S 1995 {\em The Quantum Theory of Fields\/} (Cambridge, United Kingdom: Cambridge University Press)

\bibitem{Costa.et.al:2007}
da~Costa N~V, Fulco U~L, Lyra M~L and Gléria I~M 2007 {\em Physical Review\/} E {\bf 75} 031112 ISSN 1539-3755

\bibitem{Munoz.et.al:1999}
Muñoz M~A, Dickman R, Vespignani A and Zapperi S 1999 {\em Physical Review\/} E {\bf 59} 6175--6179 ISSN 1539-3755

\bibitem{Pastor-Satorras.Vespignani:2000}
Pastor-Satorras R and Vespignani A 2000 {\em Physical Review\/} E {\bf 62} R5875--R5878 ISSN 1539-3755

\bibitem{Janssen:1997}
Janssen H~K 1997 {\em Physical Review\/} E {\bf 55} 6253--6256 ISSN 1539-3755

\bibitem{Grassberger:1989}
Grassberger P 1989 {\em Journal of Physics\/} A {\bf 22} 3673--3679

\bibitem{Argolo.et.al:2011}
Argolo C, Quintino Y, Gleria I and Lyra M~L 2011 {\em Physica\/} A {\bf 390} 1433--1439 ISSN 0378-4371

\bibitem{Souza.et.al:2011}
de~Souza D~R, Tom{\'{e}} T and Ziff R~M 2011 {\em Journal of Statistical Mechanics: Theory and Experiment\/} {\bf 2011} P03006

\bibitem{Grassberger:1979}
Grassberger P and de~la Torre A 1979 {\em Annals of Physics\/} {\bf 122} 373--396

\bibitem{Gillespie:1976}
Gillespie D~T 1976 {\em Journal of Computational Physics\/} {\bf 22} 403--434

\bibitem{Gillespie:2007}
Gillespie D~T 2007 {\em Annual Review of Physical Chemistry\/} {\bf 58} 35--55

\bibitem{Gibson.Bruck:2000}
Gibson M~A and Bruck J 2000 {\em The Journal of Physical Chemistry\/} A {\bf 104} 1876--1889

\bibitem{Voigt.Ziff:1997}
Voigt C~A and Ziff R~M 1997 {\em Physical Review\/} E {\bf 56} R6241--R6244

\bibitem{Munoz.et.al:2021}
et~al G~M~G 2021 {\em Nature Communications\/} {\bf 12}

\bibitem{Schwarzl.et.al:2017}
Schwarzl M, Godec A and Metzler R 2017 {\em Scientific Reports\/} {\bf 7} ISSN 2045-2322

\bibitem{Reichenbach.et.al:2007}
Reichenbach T, Mobilia M and Frey E 2007 {\em Nature\/} {\bf 448} 1046–1049 ISSN 1476-4687

\bibitem{Knebel.et.al:2020}
Knebel J, Geiger P~M and Frey E 2020 {\em Physical Review Letters\/} {\bf 125} ISSN 1079-7114

\bibitem{Mukhamadiarov.et.al:2021}
Mukhamadiarov R~I, Deng S, Serrao S~R, Priyanka, Nandi R, Yao L~H and T\"{a}uber U~C 2021 {\em Scientific Reports\/} {\bf 11} ISSN 2045-2322

\bibitem{Cao.et.al:2022}
Cao X, Doussal P~L and Rosso A 2022 {\em Physical Review Letters\/} {\bf 129} ISSN 1079-7114

\bibitem{Grassberger:2013}
Grassberger P 2013 {\em Journal of Statistical Physics\/} {\bf 153} 289–311 ISSN 1572-9613

\bibitem{Homrighausen.et.al:2013}
Homrighausen I, Winkler A and Frey E 2013 {\em Physical Review E\/} {\bf 88} 012111 ISSN 1539-3755

\bibitem{Shapoval.et.al:2022}
Shapoval D, Blavatska V and Dudka M 2022 {\em Journal of Physics A: Mathematical and Theoretical\/} {\bf 55} 455002 ISSN 1751-8121

\bibitem{Hallatschek.et.al:2014}
Hallatschek O and Fisher D~S 2014 {\em Proceedings of the National Academy of Sciences\/} {\bf 111} ISSN 1091-6490

\bibitem{Villiger.Paulose:2023}
Villiger N and Paulose J 2023 {\em G3 Genes|Genomes|Genetics\/} ISSN 2160-1836

\bibitem{Martin.et.al:1973}
Martin P~C, Siggia E and Rose H~A 1973 {\em Phys. Rev. A\/} {\bf 8}(1) 423--437

\bibitem{Janssen:1976}
Janssen H~K 1976 {\em Zeitschrift für Physik B Condensed Matter and Quanta\/} {\bf 23} 377–380 ISSN 1434-6036

\bibitem{deDominicis:1976}
de~Dominicis C 1976 Techniques de renormalisation de la th{\'e}orie des champs et dynamique des ph{\'e}nomenes critiques {\em J. Phys., Colloq\/} vol~37 p 247

\bibitem{Bausch.et.al:1976}
Bausch R, Janssen H~K and Wagner H 1976 {\em Zeitschrift für Physik B Condensed Matter\/} {\bf 24} 113--127 ISSN 0722-3277

\bibitem{Benitez.et.al:2016}
Benitez F, Duclut C, Chaté H, Delamotte B, Dornic I and Muñoz M~A 2016 {\em Physical Review Letters\/} {\bf 117} 100601 ISSN 0031-9007

\bibitem{ZinnJustin:2002}
Zinn-Justin J 2002 {\em Quantum Field Theory and Critical Phenomena\/} (Oxford, United Kingdom: Oxford University Press) ISBN 9780198509233

\bibitem{Janssen:2005}
Janssen H~K 2005 {\em Journal of Physics: Condensed Matter\/} {\bf 17} S1973 ISSN 0953-8984

\bibitem{Sorge:2015}
Sorge A 2015 Pyfssa 0.7.6

\bibitem{Melchert:2009}
Melchert O 2009 autoscale.py - a program for automatic finite-size scaling analyses: A user's guide

\bibitem{Houdayer.Hartmann:2004}
Houdayer J and Hartmann A 2004 {\em Physical Review\/} B {\bf 70} 014418

\bibitem{Nelder.Mead:1965}
Nelder J~A and Mead R 1965 {\em The Computer Journal\/} {\bf 7} 308--313

\bibitem{Bhattacharjee.Seno:2001}
Bhattacharjee S~M and Seno F 2001 {\em Journal of Physics\/} A {\bf 34} 6375--6380

\end{thebibliography}
\end{document}